\documentclass[prl,twocolumn, aps,amssymb,longbibliography,showpacs,superscriptaddress,nofootinbib,floatfix]{revtex4-1}

\usepackage{graphicx}
\usepackage{float}
\usepackage{dcolumn}
\usepackage{siunitx}
\usepackage{float}
\usepackage{hyperref}
\usepackage{mathtools}
\usepackage{svg}
\usepackage[english]{babel}
\usepackage{physics}
\normalsize
\usepackage{booktabs,makecell,tabularx}
\usepackage{color}
\usepackage{comment}
\usepackage{dsfont}
\usepackage[export]{adjustbox} 
\usepackage{multirow}
\usepackage[normalem]{ulem}
\newcommand{\expct}[1]{\langle #1 \rangle}

\begin{document}

\title{Universality of Shallow Global Quenches in Critical Spin Chains}
\author{Julia Wei}
\thanks{These authors contributed equally to this work.}
\affiliation{Department of Physics, Harvard University, Cambridge, MA 02138 USA}

\author{Méabh Allen}
\thanks{These authors contributed equally to this work.}
\affiliation{Department of Physics, University of California, Berkeley, California 94720 USA}

\author{Jack Kemp}
\thanks{These authors contributed equally to this work.}
\affiliation{Department of Physics, Harvard University, Cambridge, MA 02138 USA}
\affiliation{TCM Group, University of Cambridge, JJ Thomson Avenue, Cambridge, CB3 0US UK}

\author{Chenbing Wang}
\affiliation{Department of Physics, Harvard University, Cambridge, MA 02138 USA}

\author{Zixia Wei}
\affiliation{Department of Physics, Harvard University, Cambridge, MA 02138 USA}

\author{Joel E. Moore}
\affiliation{Department of Physics, University of California, Berkeley, California 94720 USA}
\affiliation{Materials Science Division, Lawrence Berkeley National Laboratory, Berkeley, CA 94720, USA}

\author{Norman Y. Yao}
\affiliation{Department of Physics, Harvard University, Cambridge, MA 02138 USA}


\begin{abstract}
Measuring universal data in the strongly correlated regime of quantum critical points remains a fundamental objective for quantum simulators. In foundational work, Calabrese and Cardy demonstrated how this data governs the dynamics of certain global quenches to 1+1-dimensional conformal field theories. 
While the quasiparticle picture they introduce has been widely successful in both theory and experiment, their seminal prediction that the critical exponents are simply encoded in the relaxation rates of local observables is more challenging to investigate experimentally; in particular,  the specific initial state required for their analysis is generated via imaginary time evolution. 
In this work, we examine the critical quench dynamics of local observables from two types of readily-accessible initial conditions: ground states and finite-temperature ensembles. 
We identify universal scaling collapses and scaling functions in both cases, utilizing a combination of conformal perturbation theory and tensor network numerics. 
For the finite-temperature quenches, we determine a regime in which the conformal field theory results are recovered, thereby allowing universal quantum critical data to be extracted from realistic quenches.
\end{abstract}

\maketitle

The advent of analog quantum simulators has enabled the precision investigation of 
unitary quantum many-body dynamics. 
This has revealed a rich landscape of emergent behavior, from anomalously slow thermalization~\cite{smith:2016, bernien:2017, bluvstein:2021, adler:2024, wang:2025, jin:2025} to anomalously fast spin transport~\cite{jepsen:2020, joshi:2022, wei:2022b, rosenberg:2024a}. 
A natural regime for the exploration of such correlated phenomena is the vicinity of quantum phase transitions, where  universality dictates that systems with dramatically disparate microscopic details nevertheless exhibit the same macroscopic physics. 
To probe this physics, two dynamical strategies are often used: sweeping through criticality to exploit the quantum Kibble-Zurek mechanism \cite{zurek2005dynamics,keesling2019quantum,dupont2022quantum, andersen:2025} or direct preparation of the quantum critical ground state \cite{dborin:2022,anand:2023, haghshenas:2024, fang2024probing}.

One might naturally wonder if a more straightforward strategy exists.
In particular, can one simply perform quench dynamics under the critical Hamiltonian in order to extract universal properties of the critical point?
In addition to being especially well-suited for near-term quantum simulators, the use of such critical quenches at \emph{high energies} has recently led to the discovery of novel dynamical phase transitions~\cite{heyl:2013, heyl:2014, heyl:2015, heyl:2016, zunkovic:2016, karrasch:2017, titum:2019, titum:2020, halimeh:2020,  rossini:2020, dag:2023, dag:2023a, robertson:2023, bandyopadhyay:2023}. 
\begin{figure}
\centering
\includegraphics[width=0.5\textwidth]{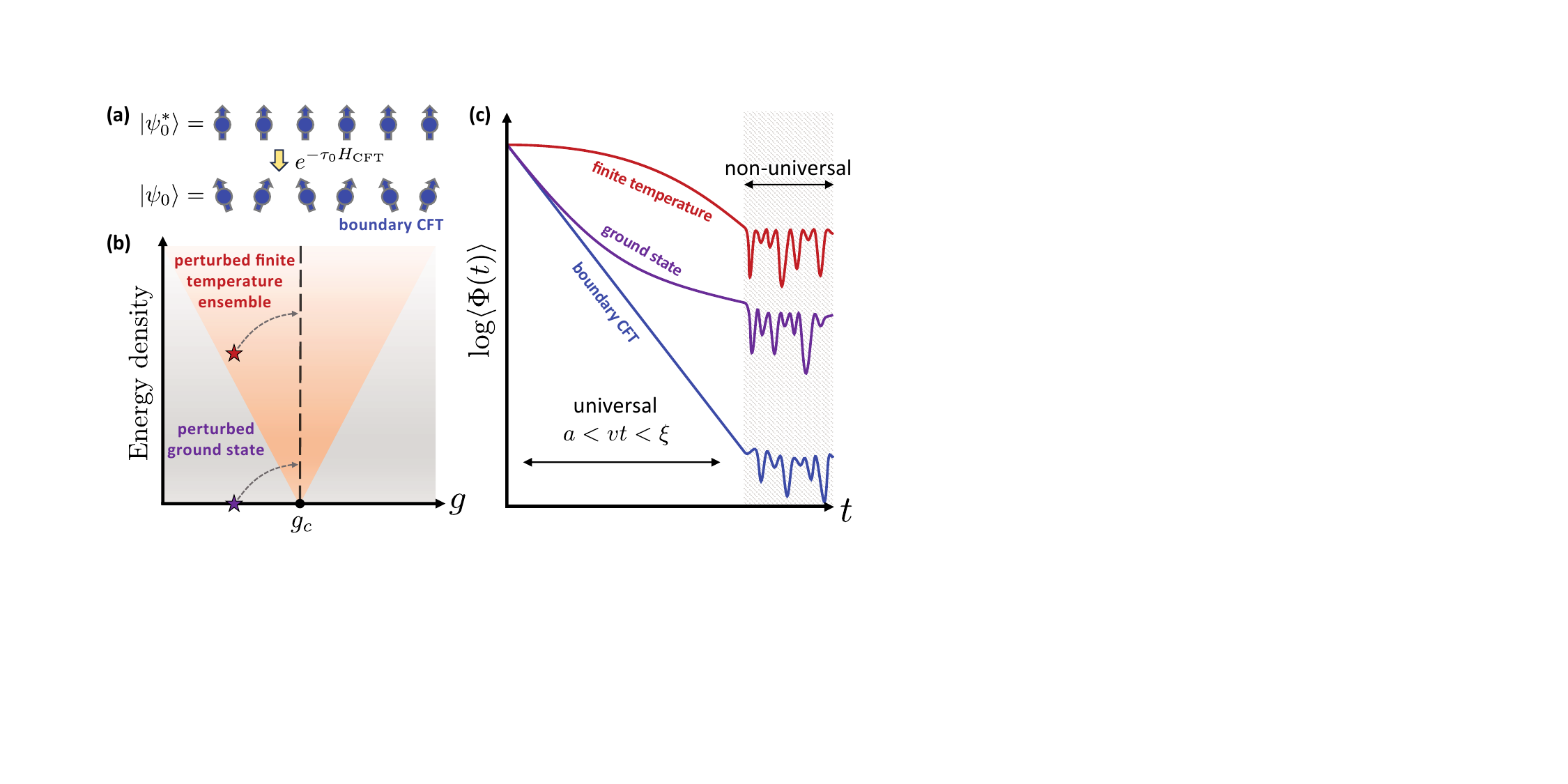}
\caption{(a) Schematic depiction of the initial state $\ket{\psi_0}$ considered by Calabrese and Cardy~\cite{calabrese2016quantum} for a quantum critical quench, generated by deforming a conformally invariant boundary state, $\ket{\psi_0^*}$, via imaginary time evolution for time $\tau_0$. 
(b) Two alternative initial conditions:  a ground state (purple) and a finite-temperature ensemble (red), produced by  perturbing the Hamiltonian away from the quantum critical point, $g_c$.  
The orange region depicts the quantum critical fan, where quantum fluctuations  dominate the system's behavior at finite temperatures.
(c) Starting from these three different initial conditions, the critical quench dynamics of a local observable, $\langle \Phi(t) \rangle$, are shown.
All three initial states exhibit universal dynamics in the shallow quench limit, albeit with different scaling forms. 
The universal regime is bounded by early-time dynamics set by the lattice spacing, $a$, and quasiparticle velocity, $v$; and late-time dynamics arising from differences between the CFT and the lattice model~\cite{SM}. 
}

\label{fig:schematic}
\end{figure}

The broad question of whether critical quenches at \emph{low energies} can, instead, extract the universal properties of the ground-state phase transition, was explored in a set of seminal papers by Calabrese and Cardy (Fig.~\ref{fig:schematic})~\cite{calabrese2005evolution,calabrese2006time,Calabrese_Cardy_2007,calabrese2016quantum,cardy2016furtherresults}. In addition to introducing an intuitive and widely applicable~\cite{kim:2013, bonnes:2014, cotler:2016, alba:2018, bastianello:2018, bertini:2018, cevolani:2018, vonkeyserlingk:2018a, najafi:2018, bertini:2019, cao:2019, calabrese:2020, klobas:2021, schneider:2021, rottoli:2025}  quasiparticle picture since confirmed in experiments~\cite{cheneau:2012, jurcevic:2014, kaufman:2016, kormos:2017, vovrosh:2021}, this work gave a prescription to obtain much more information about the critical theory than appears in the quantum Kibble-Zurek process.
The prescription is based on quenching from a specific initial state [Fig.~\ref{fig:schematic}(a)], obtained by evolving a conformally invariant boundary state in imaginary time, which is amenable to the powerful machinery of boundary conformal field theory (CFT) in 1+1D (see Appendix~A).
In this context, Calabrese and Cardy showed that, under a critical quench, primary fields exhibit a simple exponential decay. Moreover, the ratio of relaxation times for any two such primary fields is \emph{universal} and given by the ratio of their scaling dimensions. Unfortunately, the specific initial state considered is not practical to realize in experiment, as it requires imaginary time evolution to prepare~\cite{nimag}.
We are thus led to ask the following: are there experimentally accessible classes of initial states for which critical quench dynamics can extract the same critical information as in the full Calabrese-Cardy prescription?  And if not, are selected critical properties still manifest in those quenches?

\begin{table}[t]
    \begin{tabular}{lccc}
    \toprule
     Initial State & Scaling & U. Ratio & Exp.  \\
    \midrule
    $|\psi_0\rangle \propto e^{-\tau_0 H_{\textrm{c}}}|\psi_0^*\rangle$ &$\tau_0$& \checkmark & \checkmark \\
    $|\psi_{\textrm{GS}}\rangle \propto \lim\limits_{\tau \to \infty} e^{-\tau( H_{\textrm{c}}+g H_\Psi)} |\psi_0^*\rangle$ & $g^{-\nu}$& $\cross$ & $\cross$ \\
    \multirow{2}{*}{ $\rho_{\beta, g_\Psi} \propto e^{-\beta(H_\mathrm{c} + g_\Psi H_\Psi)}$ } & $\beta$& $\checkmark$ & \makecell[c]{$x_\Psi =1$: $\checkmark$ \\$x_\Psi \neq 1$: $\cross$ }
    \\
    \bottomrule
    \end{tabular}
\caption{\label{tab:forms} 
Summarizes three key properties of the critical quench dynamics starting from  different initial states: (i) the scaling collapse parameter, (ii) whether the ratio of decay times of primary fields is universal ratio~\cite{calabrese2016quantum}, and (iii) whether the decay is generically given by a simple exponential. 
We consider only relevant perturbations.
}
\label{table:initial_states}
\end{table}
In this Letter, we resolve this question for two of the most physically relevant types of initial conditions (Table~\ref{table:initial_states}): ground states and finite-temperature ensembles [Fig.~\ref{fig:schematic}(b)]. 
Using a combination of  conformal perturbation theory, free fermion analytics and large-scale tensor network numerics~\cite{hauschild:2024,barthelOptimizedLieTrotter2020,SCHOLLWOCK201196,karraschReducingNumericalEffort2013}, we explore the relaxation of local observables after instantaneous, ``shallow" global quenches to criticality, where the low-energy limit is taken before the late-time [Fig.~\ref{fig:schematic}(c)].
To ensure generality, we perform an extensive investigation of three distinct models (Table~\ref{table:models}): (i) the transverse-field Ising model (TFIM), (ii) the three-state Potts model, and (iii) the axial next-nearest-neighbor Ising (ANNNI) model. 

Our main results are as follows. We begin by computing critical quench dynamics for both classes of initial conditions across all three models.
One might naively expect, via renormalization group arguments, that both of these initial conditions should exhibit scaling dynamics similar to that observed for the boundary-CFT initial condition~\cite{nintegr}. 
Although we do observe a scaling collapse for the relaxation of the primary fields in both cases, we instead uncover a distinct scaling form for the critical quench dynamics. 
For the ground-state initial condition, we find that the dynamics collapse  under rescaling by the energy gap.
However, the ratio of relaxation times is no longer given by the ratio of scaling dimensions (Table~\ref{table:initial_states}).

For the finite-temperature initial condition, conformal perturbation theory allows us to explicitly determine the scaling function.
Inspection then reveals that the ratio of relaxation times matches the boundary-CFT expectation at late times (Table~\ref{table:initial_states}).
However, at early times, the extraction of universal quantities from the ratio of relaxation times is complicated by the presence of additional terms in the scaling function. 
Interestingly, these additional terms drop out in the specific case of integer scaling dimension, where one recovers the simple exponential decay predicted by the boundary-CFT analysis.
Finally, we describe a pair of experimental, critical quench protocols aimed at observing the  different predicted scaling behaviors (Fig.~\ref{fig:schematic}).

 \begin{table}[t]
    \begin{tabular}{lllll}
    \toprule
     & $\nu$ & $\Phi$& $x_{\Phi}$ & Lattice Operator \\
    \midrule
    \multirow{2}{*}{Ising}
    &\multirow{2}{*}{1}& $\sigma_\text{I}$ & 1/8 & $Z_i$ \\
    && $\epsilon_\text{I}$ & 1 & $Z_i Z_{i+1} - X_i$ \\
    \midrule
    \multirow{4}{*}{Potts } & \multirow{4}{*}{5/6\,}& $\sigma_\text{P}$ & 2/15 \, & $s_i$\\
    && $\epsilon_\text{P}$ & 4/5 & $(s_is_{i+1}^\dagger - \tau_i) + \mathrm{h.c.}$ \\
    && $\psi \bar \psi \,$ & 4/3 & \makecell[l]{ $s_i^\dagger(2 - 3 \omega^2 \tau_i - 3\omega \tau_i^2)$\\ - $2(s_{i-1}s_i + s_i s_{i+1})$ }\\
    \bottomrule
    \end{tabular}
\caption{\label{tab:primary} Critical exponent $\nu$, primary fields $\Phi$, and scaling dimensions $x_{\Phi}$ for the 1+1D Ising and Potts conformal field theories. In the three-state Potts model, $\omega \equiv e^{2\pi i/3}$. 
Note that both the TFI and ANNNI models are governed by the Ising critical point. 
The final column depicts the lattice-to-CFT dictionaries for the critical TFI, ANNNI, and Potts models~\cite{zou2020conformal, mong2014parafermionic,SM}.}
\label{table:models}
\end{table}

\textit{Shallow quenches to CFTs}---Let us start with the setting. 
Consider a system prepared in an initial state which is close to an equilibrium state of a quantum critical Hamiltonian, $H_\textrm{c}$. 
Here, we study the quench dynamics at the quantum critical points of three paradigmatic spin chains.
First, we consider the transverse-field Ising model,
\begin{equation}
H_{\mathrm{TFI}} = -J\sum_i Z_i Z_{i+1} + (1-g)X_i,
\label{eqn:tfi}
\end{equation}
where $Z_i$, $X_i$ are Pauli matrices. 
The TFIM is a non-interacting model, which allows us to take advantage of a wealth of exact results~\cite{barouch1970statistical, li:2009, degrandi:2010,Calabrese_Essler_Fagotti_2011, calabrese2012quantum, calabrese2012quantum2, delfino:2014, Essler_Fagotti_2016,delfino:2017,das:2017, hodsagi:2018,Granet_Fagotti_Essler_2020, granet:2022}.
Next, we consider the three-state Potts model, an interacting integrable spin chain,
\begin{equation}
H_{\mathrm{Potts}} = -J\sum_i s_i^\dagger s_{i+1} + (1-g)\tau_i + \mathrm{h.c.},
\end{equation} where $s_i$ and $\tau_i$ are the $\mathbb{Z}_3$ clock and shift operators, respectively~\cite{SM}. 
Finally, we consider the self-dual ANNNI  model, a non-integrable spin chain obtained by adding integrability-breaking terms to the TFIM~\cite{surace},
\begin{equation}
H_{\mathrm{ANNNI}} = H_{\mathrm{TFI}} -J \gamma\sum_i X_iX_{i+1} + Z_i Z_{i+2}.
\end{equation}   
At $g = 0$ and low energies, the TFI, Potts, and ANNNI (for $-0.3 < \gamma < 250$~\cite{rahmani2015mzm,milsted2015annni,milsted2017extraction}) models are each characterized by an emergent 1+1D conformal field theory. 
In order to study the critical quench dynamics of the primary fields, $\Phi$, with scaling dimensions, $x_\Phi$, we measure the corresponding local lattice operators, as shown in Table~\ref{table:models}.

\begin{figure*}
\centering
\includegraphics[width=1.0\textwidth]{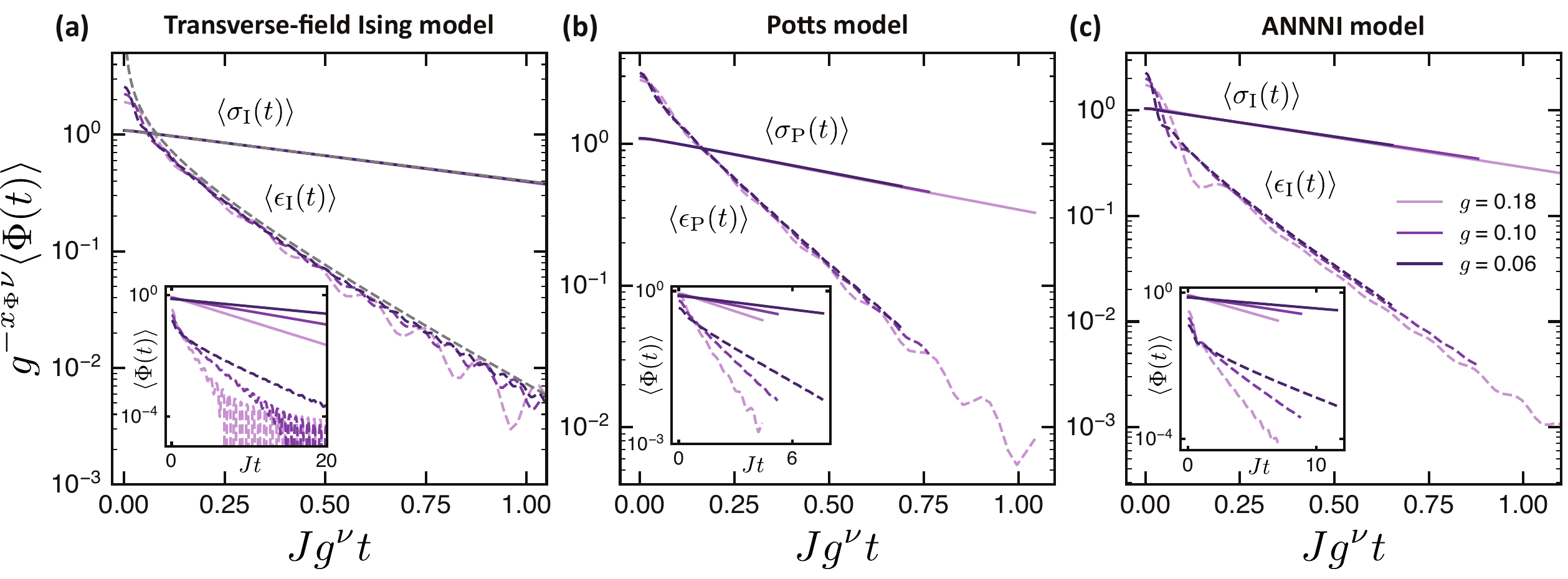}
\caption{Scaling collapse of the critical primary-field dynamics of (a) TFI (system size $L = 500$), (b) Potts ($L = 500$, bond dimension $\chi = 512$), and (c) ANNNI ($L = 2000$, $\gamma = 0.24$, $\chi = 384$) open spin chains.
The initial state for the shallow quench is a ground state in the ordered phase (perturbed away from the critical point). 
The analytical scaling limits are shown in
dotted gray lines for the TFIM. 
For $\epsilon_\mathrm{I}(t)$ in the TFIM, an additional dynamical regime is observable: at fixed $g$, coherent oscillations $\sim (Jt)^{-3/2}\cos(8J t + \pi/4)$ dominate when the long-time limit is taken before the low-energy limit, a free-fermion lattice effect~\cite{calabrese2006time,barouch1970statistical}. 
Generically, this limit will be governed by non-universal physics~\cite{Essler_Fagotti_2016,SM}. 
When the primary fields are non-Hermitian (e.g.~$\sigma_\mathrm{P}$ in the Potts model), the real part  is plotted.
Insets: Depict primary-field dynamics before rescaling. 
}
\label{fig:groundstate}
\end{figure*}

\textit{Ground state critical quenches}---
For each of the three models, we consider critical quenches starting from a nearby ground state on the ordered side of the transition ($0<g\ll 1$) (Fig.~\ref{fig:groundstate}).
In each case, the $\sigma$ primary field exhibits exponential decay with a relaxation time that depends on the distance, $g$, from the critical point (insets, Fig.~\ref{fig:groundstate}). 
After an initial transient, the $\epsilon$ primary field also seems to exhibit exponential decay with oscillations arising at late times (insets, Fig.~\ref{fig:groundstate}). 
 This is suggestive of the behavior from the boundary-CFT prediction, which one might naively expect to hold, on the basis that at large length- and time-scales the dynamics should be insensitive to the details of short-range correlations. However, a careful analysis immediately reveals that the ratio of decay times for any two primary fields does not match the ratio of scaling dimensions (for any of the three models)~\cite{SM}. 

A hint of the underlying issue is provided by the analytic tractability of the TFIM. 
To uncover the behavior of the underlying critical point, we take the scaling limit, $g \rightarrow 0$ \cite{calabrese2012quantum}, \emph{before} the late-time limit. 
This yields the intermediate-time, exponentially decaying behavior seen in the inset of Fig. \ref{fig:groundstate}(a).
Indeed, one finds that the exact forms of $\langle\sigma_\mathrm{I}(t)\rangle$~\cite{calabrese2012quantum} and $\langle \epsilon_\mathrm{I}(t) \rangle$~\cite{barouch1970statistical, SM} are given by:
\begin{equation}
  \langle\sigma_\mathrm{I}(t)\rangle  \simeq (2g)^{1/8}e^{-J g t}, \hspace{2mm}
    \langle \epsilon_\mathrm{I}(t)\rangle \simeq \sqrt{\frac{g}{2\pi}}e^{-4Jgt}\frac{1}{\sqrt{Jt} }.
     \label{eqn:gs_quench_epsilon}
\end{equation}
One immediately notes two distinctions from the boundary-CFT case: (i) the decay of the $\epsilon_\mathrm{I}$ primary field is not a simple exponential, and (ii) even asymptotically, the ratio of relaxation times is $1/4$ as opposed to $x_\sigma / x_\epsilon  = 1/8$ (Table~\ref{table:models}).

Two remarks are in order.
First, the analytic results above explicitly demonstrate that the boundary-CFT predictions do not hold in the TFIM when starting the critical quench from a ground-state initial condition.
More generally, for all three models,  deviations from the boundary-CFT prediction can be understood as a consequence of the emergent integrability of the underlying CFTs~\cite{calabrese2012quantum2, calabrese2016quantum}. 
Crucially, this integrability naturally implies a sensitivity to the starting state of the critical quench, since the subsequent relaxation dynamics will be constrained by additional local conserved quantities beyond the energy.
Indeed, for an arbitrary initial state, the system will relax to a generalized Gibbs ensemble (GGE) distinct from the critical Gibbs ensemble associated with the boundary-CFT initial condition~\cite{cardy2016furtherresults, surace:2020, robertson:2022}. 

\begin{figure*}[t]
\centering
\includegraphics[width=1.\textwidth]{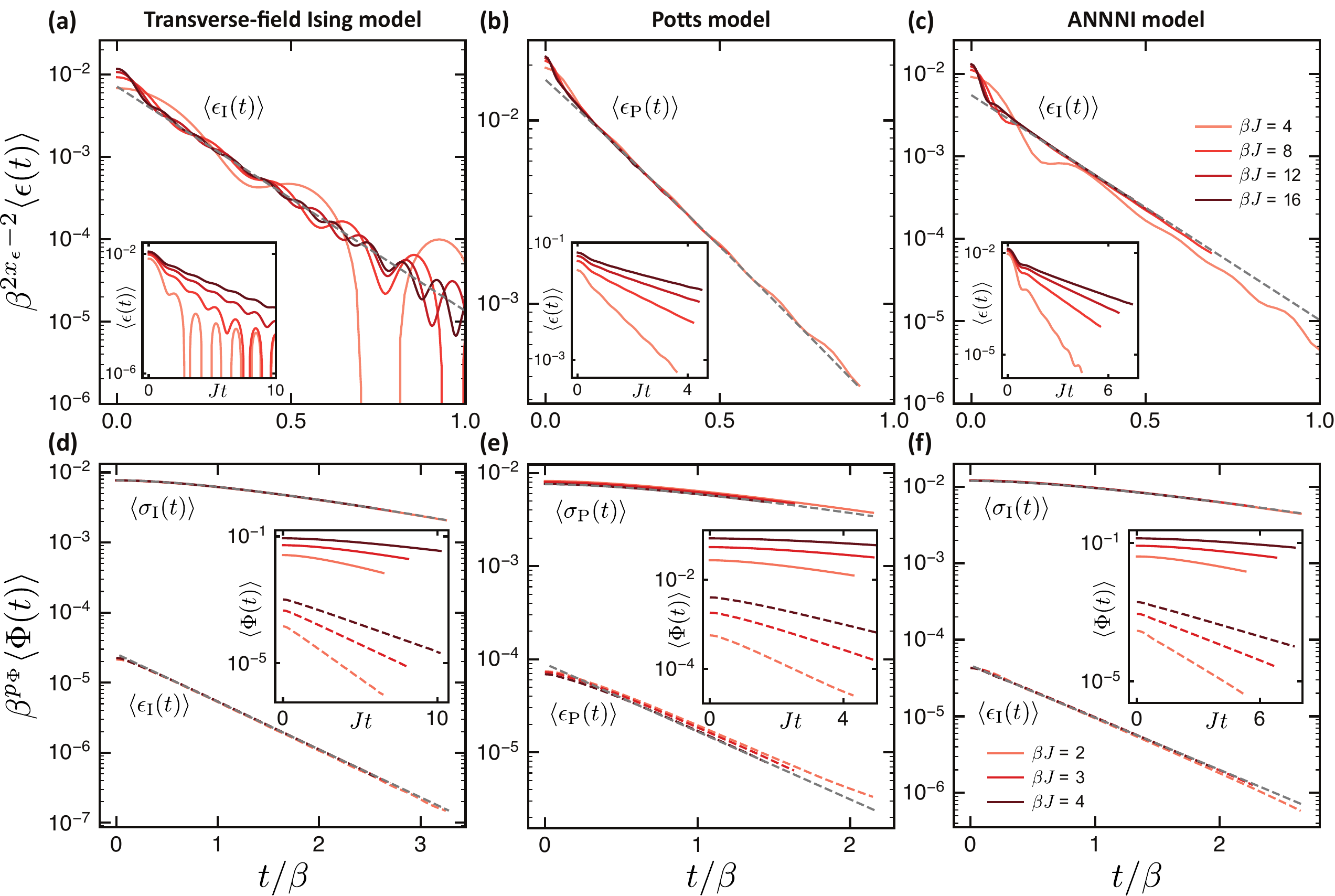}
\caption{Scaling collapse of the critical primary-field dynamics after a transverse [longitudinal] field quench from a thermal ensemble with perturbation sizes $g_\epsilon=5.62\times 10^{-3}$ and $g_\sigma = 1.78\times 10^{-3}$, respectively. 
(a)[(d)] Depicts the TFIM ($\chi=384$), (b)[(e)] the Potts model ($\chi=384$), and (c)[(f)] the ANNNI model ($\chi=256$). 
For (d-f), the rescaling factor $p_{\Phi}$ corresponds to $p_\sigma = 2 - 2x_{\sigma}$, and $p_\epsilon = 4 - 2x_\sigma - x_\epsilon$. 
All models have length $L=400$. 
 The dotted gray lines represent the comparison to conformal perturbation theory, which matches the analytical scaling limit seen in panel (a) for the TFIM~\cite{SM}.
In the TFIM, the non-universal oscillations observed at late times for fixed $\beta$ are analogous to those seen in ground state quenches for fixed $g$ [Fig. \ref{fig:groundstate}(a)]~\cite{SM}. 
Insets: Depict primary-field dynamics before rescaling. 
\label{fig:finitetemp}}
\end{figure*}

Second, although the universal ratio of relaxation times is lost, one might hope that the dynamics of local observables nevertheless collapse onto an observable-dependent scaling function.
A natural guess for the functional form of such a nonequilibrium scaling ansatz is as follows:
\begin{equation}
 \langle \psi_{\mathrm{GS}}|\Phi(t) |\psi_{\mathrm{GS}}\rangle \simeq  \mathcal{A}^{\Phi}_{\mathrm{GS}}g^{x_\Phi \nu} \mathcal{F}_{\Phi}^{\mathrm{GS}}(|\Delta|t).
 \label{eqn:scaling_gs}
\end{equation}
The dependence on $g$ is based upon the scaling hypotheses for spinless energy and magnetization-like operators
~\cite{ginsparg1988applied}, while the dynamic scaling function, $\mathcal{F}_{\Phi}^{\mathrm{GS}}$, depends on the only relevant energy scale in the protocol, namely, the gap of the initial state, $|\Delta| \sim J|g|^{\nu}$~\cite{sachdevQuantumPhaseTransitions2011}. $A_{\mathrm{GS}}^{\Phi}$ is a constant that may vanish depending on the symmetries of the ground state and $\Phi$.
As depicted in Fig.~\ref{fig:groundstate}, rescaling according to the ansatz in Eqn.~\ref{eqn:scaling_gs} yields a collapse of the data across all models and all primary fields~\cite{SM}.

\textit{Finite-temperature critical quenches}---The fact that the critical quench dynamics depend sensitively on the initial state suggests that a more careful choice is required in order to observe the universal ratios of decay times.
A particularly natural setting is to consider shallow quenches from finite-temperature initial conditions perturbed slightly away from the critical Gibbs ensemble:
\begin{equation}
    \rho_{\beta, g} \propto e^{-\beta(H_\mathrm{c} + g H')},
    \label{eq:initialgibbs}
\end{equation}
with $ g \ll 1 / (\beta J)$.
Although one can consider an arbitrary perturbation, for simplicity, we will focus on perturbations, $H' =  H_\Psi$, where $H_\Psi$ is a sum of lattice operators corresponding to the primary field $\Psi$\cite{nOPE}.
Specifically, we will consider two cases: (i) an $\epsilon$ field perturbation with $H_\epsilon = -J \sum_i X_i$ for the TFI/ANNNI models and $H_\epsilon = -J \sum_i ( \tau_i + \tau_{i}^\dagger$) for the Potts model;  and (ii) a $\sigma$ field perturbation with $H_\sigma = -J \sum_i Z_i$ for the TFI/ANNNI models and $H_\sigma = -J \sum_i (s_i + s_{i}^\dagger)$ for the Potts model~\cite{nsigmaquench}. 

Let us begin by considering the critical quench dynamics of a finite-temperature ensemble, $\rho_{\beta,g_\epsilon}$, perturbed by the $\epsilon$ field.
Working with a series of low temperatures in the scaling limit, $\beta J  \gg 1$, we find that the relaxation dynamics of the $ \epsilon$ field seem to exhibit an exponential decay in all three models [insets, Fig. \ref{fig:finitetemp}(a-c)]. 
Not only is this similar to the boundary-CFT prediction, but for the TFI/ANNNI models, the exponent itself matches that of the boundary-CFT quench which thermalizes to a Gibbs ensemble at the same temperature~\cite{ntau0}.
Unfortunately, one cannot test the ratio of decay times since the $\sigma$ field is identically zero by symmetry.

Turning to the critical quench dynamics of the finite-temperature ensemble, $\rho_{\beta,g_\sigma}$, perturbed by the $\sigma$ field [insets, Fig. \ref{fig:finitetemp}(d-f)],  one immediately observes two distinctions from the boundary-CFT prediction: first,  the relaxation dynamics of the $\sigma$ field is not exponential, and second, the relaxation dynamics of the $\epsilon$ field are significantly slower than expected for a quench thermalizing to the same temperature. 
This raises the following question: Despite the initial ensembles, $\{ \rho_{\beta,g_\epsilon}$, $\rho_{\beta,g_\sigma} \}$, both being \emph{perturbatively} close to the critical Gibbs state, why are the dynamics so qualitatively distinct? 
And indeed, how and why does the functional form of the relaxation depend on both the nature of the perturbation and the observable?

To shed light on this discrepancy, we turn to conformal perturbation theory for analytical insight into the scaling forms of the relaxation dynamics after a finite-temperature critical quench~\cite{Zamolodchikov_1991,Mussardo_2010}.
In this framework, we treat the deviation from the critical Gibbs state as a perturbation to the underlying CFT and obtain the dynamics of the primary fields via a perturbative expansion over correlation functions (see Appendix~B)~\cite{SM}.

For a perturbing field $\Psi$ and an observable $\Phi$, the leading order, $\mathcal{O}(g)$, critical quench dynamics are given by:
\begin{equation}
\begin{aligned}
    \langle\Phi (t )\rangle_{1}&\sim g \delta_{x_\Psi x_\Phi} \beta^{2-2x_\Phi}\left(A_{x_\Phi} + B_{x_\Phi}t/\beta \right)e^{-2\pi x_\Phi t/\beta},
    \label{eqn:g^1}
\end{aligned}
\end{equation}
with $A_{x_\Phi}\sim  1 + \frac{\sin(\pi x_\Phi)\left[x_\Phi^{-1}-\gamma_E -\pi \csc(\pi x_\Phi) -\psi^{\prime}(1-x_\Phi)\right]}{\pi}$ and $B_{x_\Phi} \sim 2\sin(\pi x_\Phi )$ universal constants~\cite{ndga,ndims}.

We note that when $\Psi$ and $\Phi$ have different scaling dimensions, the dynamics are zero via symmetry.
This necessitates going to second order, $\mathcal{O}(g^2)$, to obtain the leading behavior when $x_\Psi \neq x_\Phi$: 
\begin{align}
    &\langle\Phi (t )\rangle_{2} \sim g^2 C_{\Phi\Psi\Psi} \beta^{4-2x_\Psi-x_\Phi}\Big[D_{x_{\Phi},x_\Psi} e^{-4\pi  x_\Psi t/\beta}  +\label{eqn:g^2} \\
    & \quad \big( E_{x_{\Phi},x_\Psi} +  F_{x_{\Phi},x_\Psi}t/\beta+G_{x_{\Phi},x_\Psi}t^2/\beta^2 \big)e^{-2\pi  x_{\Phi} t/\beta}\Big]\nonumber,
\end{align}
where $C_{\Phi\Psi\Psi} $ is the three-point structure constant of the CFT, and $D_{x_{\Phi},x_\Psi} ,E_{x_{\Phi},x_\Psi} ,F_{x_{\Phi},x_\Psi} $, and $G_{x_{\Phi},x_\Psi} $ are universal constants. 

Remarkably, the scaling form predicted by the conformal perturbation theory immediately yields a nearly perfect collapse of the numerical data across all models, perturbations, and observables (Fig.~\ref{fig:finitetemp}). 
Crucially, this implies that one can use the analytics to understand the origin of the observed discrepancies from the boundary-CFT predictions.
For example, the linear in $t$ term in $\langle\Phi (t)\rangle_1$ is precisely the reason why the relaxation dynamics of the $\sigma$ field are not observed to be exponential in the insets of Figs.~\ref{fig:finitetemp}(d-f).
Moreover, from the functional form of $\langle\Phi (t)\rangle_2$, one sees that, when $x_\Psi \neq x_\Phi$,  
the asymptotic relaxation dynamics depends on the smaller of $2 x_\Phi$ versus $x_\Psi$; this explains our previously observed slower decay timescale for the $\epsilon$ field [insets, Fig.~\ref{fig:finitetemp}(d-f)].

A few remarks are in order. 
First, we note that in the scaling limit, the relaxation timescale for our finite-temperature ensemble (Eqn.~\ref{eqn:g^1}) 
matches the boundary-CFT initial condition (Eqn.~\ref{eqn:cardydecay}). This means that, in principle, at late times, one should be able to  recover the universal ratio of relaxation times by taking a finite-temperature ensemble perturbed by two primary fields and 
measuring the critical quench dynamics of the same fields. 
Second, in the specific case where the scaling dimension of the (relevant) observable is an integer, $x_\Phi = 1$, we find 
that $B\left(x_\Phi\right) \sim \sin \left(\pi x_\Phi \right) = 0$, implying that the linear in $t$ term drops out of the dynamics (Eqn.~\ref{eqn:g^1}). 
Finally, our conformal perturbation theory enables us to gain physical insight into the critical quench dynamics.
For example, the two terms in $ \langle\Phi (t)\rangle_1$ arise from the horizon effect of a perturbation to the CFT.
In particular, the initial thermal ensemble emits quasiparticles which are correlated over distances set by their velocity; the build-up of correlations inside this light-cone leads to the linear in $t$ term in Eqn.~\ref{eqn:g^1}, distinct from the contribution due to spatially-decaying correlations.

\emph{Outlook}---Our results suggest two possible experimental protocols utilizing quantum quenches to probe universal scaling. 
First, adiabatically preparing ground states near the critical point at different perturbations strengths, $g$, and then quenching is sufficient to measure the critical exponent $\nu$, analogous to a Kibble-Zurek sweep. 
Second, and more powerfully, if one performs an initial quench in an integrability-broken model, the reduced density matrix of a subsystem should ``thermalize" to an effective Gibbs ensemble~\cite{Deutsch_1991,Srednicki_1994,Rigol_Dunjko_Olshanii_2008}.
Crucially, using this ``thermalized'' initial condition for the critical quench enables one to take advantage of our finite-temperature results~\cite{nltc}. 

On the theoretical front, our work opens the door to two intriguing directions. 
First, in 1+1D, it would be interesting to test whether our framework generalizes to other CFTs currently out of reach of classical simulation, as well as those with continuously-varying critical exponents such as the easy-plane XXZ spin chain~\cite{bonnes:2014}.
Second, a natural extension of our work is to higher dimensions, where significantly fewer exact results are available on the nature of physical conformal field theories.  Since interacting conformal field theories are no longer integrable in higher dimensions, we expect the phenomenology to be qualitatively distinct.

\textit{Acknowledgements} We gratefully acknowledge the insights of and discussions with P. Calabrese, J. Cardy, P. Zoller, M. P. Zaletel, C. Kokail, T. Zache, S. Sachdev, M. Bintz, and S. Anand.
This work was supported in part by the Air Force Office of Scientific Research via the MURI program (FA9550-21-1-0069) and by the NSF QLCI program through Grant No. OMA-2016245.
J.W. acknowledges support from the Department of Energy Computational Science Graduate Fellowship under award number DE-SC0022158. 
Z.W. is supported by the Society of Fellows at Harvard University.
J.K. acknowledges support from EPSRC Grant No. EP/V062654/1. 
N.Y.Y. acknowledges support from a Simons Investigator award. 

\nocite{*}
\bibliography{sources}

\begin{thebibliography}{124}%
\makeatletter
\providecommand \@ifxundefined [1]{%
 \@ifx{#1\undefined}
}%
\providecommand \@ifnum [1]{%
 \ifnum #1\expandafter \@firstoftwo
 \else \expandafter \@secondoftwo
 \fi
}%
\providecommand \@ifx [1]{%
 \ifx #1\expandafter \@firstoftwo
 \else \expandafter \@secondoftwo
 \fi
}%
\providecommand \natexlab [1]{#1}%
\providecommand \enquote  [1]{``#1''}%
\providecommand \bibnamefont  [1]{#1}%
\providecommand \bibfnamefont [1]{#1}%
\providecommand \citenamefont [1]{#1}%
\providecommand \href@noop [0]{\@secondoftwo}%
\providecommand \href [0]{\begingroup \@sanitize@url \@href}%
\providecommand \@href[1]{\@@startlink{#1}\@@href}%
\providecommand \@@href[1]{\endgroup#1\@@endlink}%
\providecommand \@sanitize@url [0]{\catcode `\\12\catcode `\$12\catcode `\&12\catcode `\#12\catcode `\^12\catcode `\_12\catcode `\%12\relax}%
\providecommand \@@startlink[1]{}%
\providecommand \@@endlink[0]{}%
\providecommand \url  [0]{\begingroup\@sanitize@url \@url }%
\providecommand \@url [1]{\endgroup\@href {#1}{\urlprefix }}%
\providecommand \urlprefix  [0]{URL }%
\providecommand \Eprint [0]{\href }%
\providecommand \doibase [0]{http://dx.doi.org/}%
\providecommand \selectlanguage [0]{\@gobble}%
\providecommand \bibinfo  [0]{\@secondoftwo}%
\providecommand \bibfield  [0]{\@secondoftwo}%
\providecommand \translation [1]{[#1]}%
\providecommand \BibitemOpen [0]{}%
\providecommand \bibitemStop [0]{}%
\providecommand \bibitemNoStop [0]{.\EOS\space}%
\providecommand \EOS [0]{\spacefactor3000\relax}%
\providecommand \BibitemShut  [1]{\csname bibitem#1\endcsname}%
\let\auto@bib@innerbib\@empty
\bibitem [{\citenamefont {Smith}\ \emph {et~al.}(2016)\citenamefont {Smith}, \citenamefont {Lee}, \citenamefont {Richerme}, \citenamefont {Neyenhuis}, \citenamefont {Hess}, \citenamefont {Hauke}, \citenamefont {Heyl}, \citenamefont {Huse},\ and\ \citenamefont {Monroe}}]{smith:2016}%
  \BibitemOpen
  \bibfield  {author} {\bibinfo {author} {\bibfnamefont {J.}~\bibnamefont {Smith}}, \bibinfo {author} {\bibfnamefont {A.}~\bibnamefont {Lee}}, \bibinfo {author} {\bibfnamefont {P.}~\bibnamefont {Richerme}}, \bibinfo {author} {\bibfnamefont {B.}~\bibnamefont {Neyenhuis}}, \bibinfo {author} {\bibfnamefont {P.~W.}\ \bibnamefont {Hess}}, \bibinfo {author} {\bibfnamefont {P.}~\bibnamefont {Hauke}}, \bibinfo {author} {\bibfnamefont {M.}~\bibnamefont {Heyl}}, \bibinfo {author} {\bibfnamefont {D.~A.}\ \bibnamefont {Huse}}, \ and\ \bibinfo {author} {\bibfnamefont {C.}~\bibnamefont {Monroe}},\ }\bibfield  {title} {\enquote {\bibinfo {title} {Many-body localization in a quantum simulator with programmable random disorder},}\ }\href {\doibase 10.1038/nphys3783} {\bibfield  {journal} {\bibinfo  {journal} {Nature Physics}\ }\textbf {\bibinfo {volume} {12}},\ \bibinfo {pages} {907--911} (\bibinfo {year} {2016})}\BibitemShut {NoStop}%
\bibitem [{\citenamefont {Bernien}\ \emph {et~al.}(2017)\citenamefont {Bernien}, \citenamefont {Schwartz}, \citenamefont {Keesling}, \citenamefont {Levine}, \citenamefont {Omran}, \citenamefont {Pichler}, \citenamefont {Choi}, \citenamefont {Zibrov}, \citenamefont {Endres}, \citenamefont {Greiner}, \citenamefont {Vuleti{\'c}},\ and\ \citenamefont {Lukin}}]{bernien:2017}%
  \BibitemOpen
  \bibfield  {author} {\bibinfo {author} {\bibfnamefont {Hannes}\ \bibnamefont {Bernien}}, \bibinfo {author} {\bibfnamefont {Sylvain}\ \bibnamefont {Schwartz}}, \bibinfo {author} {\bibfnamefont {Alexander}\ \bibnamefont {Keesling}}, \bibinfo {author} {\bibfnamefont {Harry}\ \bibnamefont {Levine}}, \bibinfo {author} {\bibfnamefont {Ahmed}\ \bibnamefont {Omran}}, \bibinfo {author} {\bibfnamefont {Hannes}\ \bibnamefont {Pichler}}, \bibinfo {author} {\bibfnamefont {Soonwon}\ \bibnamefont {Choi}}, \bibinfo {author} {\bibfnamefont {Alexander~S.}\ \bibnamefont {Zibrov}}, \bibinfo {author} {\bibfnamefont {Manuel}\ \bibnamefont {Endres}}, \bibinfo {author} {\bibfnamefont {Markus}\ \bibnamefont {Greiner}}, \bibinfo {author} {\bibfnamefont {Vladan}\ \bibnamefont {Vuleti{\'c}}}, \ and\ \bibinfo {author} {\bibfnamefont {Mikhail~D.}\ \bibnamefont {Lukin}},\ }\bibfield  {title} {\enquote {\bibinfo {title} {Probing many-body dynamics on a 51-atom quantum simulator},}\ }\href {\doibase 10.1038/nature24622} {\bibfield
  {journal} {\bibinfo  {journal} {Nature}\ }\textbf {\bibinfo {volume} {551}},\ \bibinfo {pages} {579--584} (\bibinfo {year} {2017})}\BibitemShut {NoStop}%
\bibitem [{\citenamefont {Bluvstein}\ \emph {et~al.}(2021)\citenamefont {Bluvstein}, \citenamefont {Omran}, \citenamefont {Levine}, \citenamefont {Keesling}, \citenamefont {Semeghini}, \citenamefont {Ebadi}, \citenamefont {Wang}, \citenamefont {Michailidis}, \citenamefont {Maskara}, \citenamefont {Ho}, \citenamefont {Choi}, \citenamefont {Serbyn}, \citenamefont {Greiner}, \citenamefont {Vuleti{\'c}},\ and\ \citenamefont {Lukin}}]{bluvstein:2021}%
  \BibitemOpen
  \bibfield  {author} {\bibinfo {author} {\bibfnamefont {D.}~\bibnamefont {Bluvstein}}, \bibinfo {author} {\bibfnamefont {A.}~\bibnamefont {Omran}}, \bibinfo {author} {\bibfnamefont {H.}~\bibnamefont {Levine}}, \bibinfo {author} {\bibfnamefont {A.}~\bibnamefont {Keesling}}, \bibinfo {author} {\bibfnamefont {G.}~\bibnamefont {Semeghini}}, \bibinfo {author} {\bibfnamefont {S.}~\bibnamefont {Ebadi}}, \bibinfo {author} {\bibfnamefont {T.~T.}\ \bibnamefont {Wang}}, \bibinfo {author} {\bibfnamefont {A.~A.}\ \bibnamefont {Michailidis}}, \bibinfo {author} {\bibfnamefont {N.}~\bibnamefont {Maskara}}, \bibinfo {author} {\bibfnamefont {W.~W.}\ \bibnamefont {Ho}}, \bibinfo {author} {\bibfnamefont {S.}~\bibnamefont {Choi}}, \bibinfo {author} {\bibfnamefont {M.}~\bibnamefont {Serbyn}}, \bibinfo {author} {\bibfnamefont {M.}~\bibnamefont {Greiner}}, \bibinfo {author} {\bibfnamefont {V.}~\bibnamefont {Vuleti{\'c}}}, \ and\ \bibinfo {author} {\bibfnamefont {M.~D.}\ \bibnamefont {Lukin}},\ }\bibfield  {title} {\enquote
  {\bibinfo {title} {Controlling quantum many-body dynamics in driven {{Rydberg}} atom arrays},}\ }\href {\doibase 10.1126/science.abg2530} {\bibfield  {journal} {\bibinfo  {journal} {Science}\ }\textbf {\bibinfo {volume} {371}},\ \bibinfo {pages} {1355--1359} (\bibinfo {year} {2021})}\BibitemShut {NoStop}%
\bibitem [{\citenamefont {Adler}\ \emph {et~al.}(2024)\citenamefont {Adler}, \citenamefont {Wei}, \citenamefont {Will}, \citenamefont {Srakaew}, \citenamefont {Agrawal}, \citenamefont {Weckesser}, \citenamefont {Moessner}, \citenamefont {Pollmann}, \citenamefont {Bloch},\ and\ \citenamefont {Zeiher}}]{adler:2024}%
  \BibitemOpen
  \bibfield  {author} {\bibinfo {author} {\bibfnamefont {Daniel}\ \bibnamefont {Adler}}, \bibinfo {author} {\bibfnamefont {David}\ \bibnamefont {Wei}}, \bibinfo {author} {\bibfnamefont {Melissa}\ \bibnamefont {Will}}, \bibinfo {author} {\bibfnamefont {Kritsana}\ \bibnamefont {Srakaew}}, \bibinfo {author} {\bibfnamefont {Suchita}\ \bibnamefont {Agrawal}}, \bibinfo {author} {\bibfnamefont {Pascal}\ \bibnamefont {Weckesser}}, \bibinfo {author} {\bibfnamefont {Roderich}\ \bibnamefont {Moessner}}, \bibinfo {author} {\bibfnamefont {Frank}\ \bibnamefont {Pollmann}}, \bibinfo {author} {\bibfnamefont {Immanuel}\ \bibnamefont {Bloch}}, \ and\ \bibinfo {author} {\bibfnamefont {Johannes}\ \bibnamefont {Zeiher}},\ }\bibfield  {title} {\enquote {\bibinfo {title} {Observation of {{Hilbert}} space fragmentation and fractonic excitations in {{2D}}},}\ }\href {\doibase 10.1038/s41586-024-08188-0} {\bibfield  {journal} {\bibinfo  {journal} {Nature}\ }\textbf {\bibinfo {volume} {636}},\ \bibinfo {pages} {80--85} (\bibinfo {year}
  {2024})}\BibitemShut {NoStop}%
\bibitem [{\citenamefont {Wang}\ \emph {et~al.}(2025)\citenamefont {Wang}, \citenamefont {Shi}, \citenamefont {Sun}, \citenamefont {Chen}, \citenamefont {Wang}, \citenamefont {Zhao}, \citenamefont {Liu}, \citenamefont {Ma}, \citenamefont {Wang}, \citenamefont {Li}, \citenamefont {Zhang}, \citenamefont {Liu}, \citenamefont {Deng}, \citenamefont {Li}, \citenamefont {He}, \citenamefont {Liu}, \citenamefont {Peng}, \citenamefont {Song}, \citenamefont {Xue}, \citenamefont {Yu}, \citenamefont {Huang}, \citenamefont {Xiang}, \citenamefont {Zheng}, \citenamefont {Xu},\ and\ \citenamefont {Fan}}]{wang:2025}%
  \BibitemOpen
  \bibfield  {author} {\bibinfo {author} {\bibfnamefont {Yong-Yi}\ \bibnamefont {Wang}}, \bibinfo {author} {\bibfnamefont {Yun-Hao}\ \bibnamefont {Shi}}, \bibinfo {author} {\bibfnamefont {Zheng-Hang}\ \bibnamefont {Sun}}, \bibinfo {author} {\bibfnamefont {Chi-Tong}\ \bibnamefont {Chen}}, \bibinfo {author} {\bibfnamefont {Zheng-An}\ \bibnamefont {Wang}}, \bibinfo {author} {\bibfnamefont {Kui}\ \bibnamefont {Zhao}}, \bibinfo {author} {\bibfnamefont {Hao-Tian}\ \bibnamefont {Liu}}, \bibinfo {author} {\bibfnamefont {Wei-Guo}\ \bibnamefont {Ma}}, \bibinfo {author} {\bibfnamefont {Ziting}\ \bibnamefont {Wang}}, \bibinfo {author} {\bibfnamefont {Hao}\ \bibnamefont {Li}}, \bibinfo {author} {\bibfnamefont {Jia-Chi}\ \bibnamefont {Zhang}}, \bibinfo {author} {\bibfnamefont {Yu}~\bibnamefont {Liu}}, \bibinfo {author} {\bibfnamefont {Cheng-Lin}\ \bibnamefont {Deng}}, \bibinfo {author} {\bibfnamefont {Tian-Ming}\ \bibnamefont {Li}}, \bibinfo {author} {\bibfnamefont {Yang}\ \bibnamefont {He}}, \bibinfo {author}
  {\bibfnamefont {Zheng-He}\ \bibnamefont {Liu}}, \bibinfo {author} {\bibfnamefont {Zhen-Yu}\ \bibnamefont {Peng}}, \bibinfo {author} {\bibfnamefont {Xiaohui}\ \bibnamefont {Song}}, \bibinfo {author} {\bibfnamefont {Guangming}\ \bibnamefont {Xue}}, \bibinfo {author} {\bibfnamefont {Haifeng}\ \bibnamefont {Yu}}, \bibinfo {author} {\bibfnamefont {Kaixuan}\ \bibnamefont {Huang}}, \bibinfo {author} {\bibfnamefont {Zhongcheng}\ \bibnamefont {Xiang}}, \bibinfo {author} {\bibfnamefont {Dongning}\ \bibnamefont {Zheng}}, \bibinfo {author} {\bibfnamefont {Kai}\ \bibnamefont {Xu}}, \ and\ \bibinfo {author} {\bibfnamefont {Heng}\ \bibnamefont {Fan}},\ }\bibfield  {title} {\enquote {\bibinfo {title} {Exploring {{Hilbert-Space Fragmentation}} on a {{Superconducting Processor}}},}\ }\href {\doibase 10.1103/PRXQuantum.6.010325} {\bibfield  {journal} {\bibinfo  {journal} {PRX Quantum}\ }\textbf {\bibinfo {volume} {6}},\ \bibinfo {pages} {010325} (\bibinfo {year} {2025})}\BibitemShut {NoStop}%
\bibitem [{\citenamefont {Jin}\ \emph {et~al.}(2025)\citenamefont {Jin}, \citenamefont {Jiang}, \citenamefont {Zhu}, \citenamefont {Bao}, \citenamefont {Shen}, \citenamefont {Wang}, \citenamefont {Zhu}, \citenamefont {Xu}, \citenamefont {Song}, \citenamefont {Chen}, \citenamefont {Tan}, \citenamefont {Wu}, \citenamefont {Zhang}, \citenamefont {Gao}, \citenamefont {Wang}, \citenamefont {Zou}, \citenamefont {Zhang}, \citenamefont {Li}, \citenamefont {Zhong}, \citenamefont {Cui}, \citenamefont {Han}, \citenamefont {He}, \citenamefont {Wang}, \citenamefont {Yang}, \citenamefont {Wang}, \citenamefont {Shen}, \citenamefont {Liu}, \citenamefont {Deng}, \citenamefont {Dong}, \citenamefont {Zhang}, \citenamefont {Li}, \citenamefont {Yuan}, \citenamefont {Lu}, \citenamefont {Sun}, \citenamefont {Li}, \citenamefont {Zhang}, \citenamefont {Song}, \citenamefont {Wang}, \citenamefont {Guo}, \citenamefont {Machado}, \citenamefont {Kemp}, \citenamefont {Iadecola}, \citenamefont {Yao}, \citenamefont {Wang},\ and\
  \citenamefont {Deng}}]{jin:2025}%
  \BibitemOpen
  \bibfield  {author} {\bibinfo {author} {\bibfnamefont {Feitong}\ \bibnamefont {Jin}}, \bibinfo {author} {\bibfnamefont {Si}~\bibnamefont {Jiang}}, \bibinfo {author} {\bibfnamefont {Xuhao}\ \bibnamefont {Zhu}}, \bibinfo {author} {\bibfnamefont {Zehang}\ \bibnamefont {Bao}}, \bibinfo {author} {\bibfnamefont {Fanhao}\ \bibnamefont {Shen}}, \bibinfo {author} {\bibfnamefont {Ke}~\bibnamefont {Wang}}, \bibinfo {author} {\bibfnamefont {Zitian}\ \bibnamefont {Zhu}}, \bibinfo {author} {\bibfnamefont {Shibo}\ \bibnamefont {Xu}}, \bibinfo {author} {\bibfnamefont {Zixuan}\ \bibnamefont {Song}}, \bibinfo {author} {\bibfnamefont {Jiachen}\ \bibnamefont {Chen}}, \bibinfo {author} {\bibfnamefont {Ziqi}\ \bibnamefont {Tan}}, \bibinfo {author} {\bibfnamefont {Yaozu}\ \bibnamefont {Wu}}, \bibinfo {author} {\bibfnamefont {Chuanyu}\ \bibnamefont {Zhang}}, \bibinfo {author} {\bibfnamefont {Yu}~\bibnamefont {Gao}}, \bibinfo {author} {\bibfnamefont {Ning}\ \bibnamefont {Wang}}, \bibinfo {author} {\bibfnamefont {Yiren}\
  \bibnamefont {Zou}}, \bibinfo {author} {\bibfnamefont {Aosai}\ \bibnamefont {Zhang}}, \bibinfo {author} {\bibfnamefont {Tingting}\ \bibnamefont {Li}}, \bibinfo {author} {\bibfnamefont {Jiarun}\ \bibnamefont {Zhong}}, \bibinfo {author} {\bibfnamefont {Zhengyi}\ \bibnamefont {Cui}}, \bibinfo {author} {\bibfnamefont {Yihang}\ \bibnamefont {Han}}, \bibinfo {author} {\bibfnamefont {Yiyang}\ \bibnamefont {He}}, \bibinfo {author} {\bibfnamefont {Han}\ \bibnamefont {Wang}}, \bibinfo {author} {\bibfnamefont {Jianan}\ \bibnamefont {Yang}}, \bibinfo {author} {\bibfnamefont {Yanzhe}\ \bibnamefont {Wang}}, \bibinfo {author} {\bibfnamefont {Jiayuan}\ \bibnamefont {Shen}}, \bibinfo {author} {\bibfnamefont {Gongyu}\ \bibnamefont {Liu}}, \bibinfo {author} {\bibfnamefont {Jinfeng}\ \bibnamefont {Deng}}, \bibinfo {author} {\bibfnamefont {Hang}\ \bibnamefont {Dong}}, \bibinfo {author} {\bibfnamefont {Pengfei}\ \bibnamefont {Zhang}}, \bibinfo {author} {\bibfnamefont {Weikang}\ \bibnamefont {Li}}, \bibinfo {author}
  {\bibfnamefont {Dong}\ \bibnamefont {Yuan}}, \bibinfo {author} {\bibfnamefont {Zhide}\ \bibnamefont {Lu}}, \bibinfo {author} {\bibfnamefont {Zheng-Zhi}\ \bibnamefont {Sun}}, \bibinfo {author} {\bibfnamefont {Hekang}\ \bibnamefont {Li}}, \bibinfo {author} {\bibfnamefont {Junxiang}\ \bibnamefont {Zhang}}, \bibinfo {author} {\bibfnamefont {Chao}\ \bibnamefont {Song}}, \bibinfo {author} {\bibfnamefont {Zhen}\ \bibnamefont {Wang}}, \bibinfo {author} {\bibfnamefont {Qiujiang}\ \bibnamefont {Guo}}, \bibinfo {author} {\bibfnamefont {Francisco}\ \bibnamefont {Machado}}, \bibinfo {author} {\bibfnamefont {Jack}\ \bibnamefont {Kemp}}, \bibinfo {author} {\bibfnamefont {Thomas}\ \bibnamefont {Iadecola}}, \bibinfo {author} {\bibfnamefont {Norman~Y.}\ \bibnamefont {Yao}}, \bibinfo {author} {\bibfnamefont {H.}~\bibnamefont {Wang}}, \ and\ \bibinfo {author} {\bibfnamefont {Dong-Ling}\ \bibnamefont {Deng}},\ }\href {\doibase 10.48550/arXiv.2501.04688} {\enquote {\bibinfo {title} {Observation of topological prethermal strong
  zero modes},}\ } (\bibinfo {year} {2025}),\ \Eprint {http://arxiv.org/abs/2501.04688} {arXiv:2501.04688 [quant-ph]} \BibitemShut {NoStop}%
\bibitem [{\citenamefont {Jepsen}\ \emph {et~al.}(2020)\citenamefont {Jepsen}, \citenamefont {{Amato-Grill}}, \citenamefont {Dimitrova}, \citenamefont {Ho}, \citenamefont {Demler},\ and\ \citenamefont {Ketterle}}]{jepsen:2020}%
  \BibitemOpen
  \bibfield  {author} {\bibinfo {author} {\bibfnamefont {Paul~Niklas}\ \bibnamefont {Jepsen}}, \bibinfo {author} {\bibfnamefont {Jesse}\ \bibnamefont {{Amato-Grill}}}, \bibinfo {author} {\bibfnamefont {Ivana}\ \bibnamefont {Dimitrova}}, \bibinfo {author} {\bibfnamefont {Wen~Wei}\ \bibnamefont {Ho}}, \bibinfo {author} {\bibfnamefont {Eugene}\ \bibnamefont {Demler}}, \ and\ \bibinfo {author} {\bibfnamefont {Wolfgang}\ \bibnamefont {Ketterle}},\ }\bibfield  {title} {\enquote {\bibinfo {title} {Spin transport in a tunable {{Heisenberg}} model realized with ultracold atoms},}\ }\href {\doibase 10.1038/s41586-020-3033-y} {\bibfield  {journal} {\bibinfo  {journal} {Nature}\ }\textbf {\bibinfo {volume} {588}},\ \bibinfo {pages} {403--407} (\bibinfo {year} {2020})}\BibitemShut {NoStop}%
\bibitem [{\citenamefont {Joshi}\ \emph {et~al.}(2022)\citenamefont {Joshi}, \citenamefont {Kranzl}, \citenamefont {Schuckert}, \citenamefont {Lovas}, \citenamefont {Maier}, \citenamefont {Blatt}, \citenamefont {Knap},\ and\ \citenamefont {Roos}}]{joshi:2022}%
  \BibitemOpen
  \bibfield  {author} {\bibinfo {author} {\bibfnamefont {M.~K.}\ \bibnamefont {Joshi}}, \bibinfo {author} {\bibfnamefont {F.}~\bibnamefont {Kranzl}}, \bibinfo {author} {\bibfnamefont {A.}~\bibnamefont {Schuckert}}, \bibinfo {author} {\bibfnamefont {I.}~\bibnamefont {Lovas}}, \bibinfo {author} {\bibfnamefont {C.}~\bibnamefont {Maier}}, \bibinfo {author} {\bibfnamefont {R.}~\bibnamefont {Blatt}}, \bibinfo {author} {\bibfnamefont {M.}~\bibnamefont {Knap}}, \ and\ \bibinfo {author} {\bibfnamefont {C.~F.}\ \bibnamefont {Roos}},\ }\bibfield  {title} {\enquote {\bibinfo {title} {Observing emergent hydrodynamics in a long-range quantum magnet},}\ }\href {\doibase 10.1126/science.abk2400} {\bibfield  {journal} {\bibinfo  {journal} {Science}\ }\textbf {\bibinfo {volume} {376}},\ \bibinfo {pages} {720--724} (\bibinfo {year} {2022})}\BibitemShut {NoStop}%
\bibitem [{\citenamefont {Wei}\ \emph {et~al.}(2022)\citenamefont {Wei}, \citenamefont {{Rubio-Abadal}}, \citenamefont {Ye}, \citenamefont {Machado}, \citenamefont {Kemp}, \citenamefont {Srakaew}, \citenamefont {Hollerith}, \citenamefont {Rui}, \citenamefont {Gopalakrishnan}, \citenamefont {Yao}, \citenamefont {Bloch},\ and\ \citenamefont {Zeiher}}]{wei:2022b}%
  \BibitemOpen
  \bibfield  {author} {\bibinfo {author} {\bibfnamefont {David}\ \bibnamefont {Wei}}, \bibinfo {author} {\bibfnamefont {Antonio}\ \bibnamefont {{Rubio-Abadal}}}, \bibinfo {author} {\bibfnamefont {Bingtian}\ \bibnamefont {Ye}}, \bibinfo {author} {\bibfnamefont {Francisco}\ \bibnamefont {Machado}}, \bibinfo {author} {\bibfnamefont {Jack}\ \bibnamefont {Kemp}}, \bibinfo {author} {\bibfnamefont {Kritsana}\ \bibnamefont {Srakaew}}, \bibinfo {author} {\bibfnamefont {Simon}\ \bibnamefont {Hollerith}}, \bibinfo {author} {\bibfnamefont {Jun}\ \bibnamefont {Rui}}, \bibinfo {author} {\bibfnamefont {Sarang}\ \bibnamefont {Gopalakrishnan}}, \bibinfo {author} {\bibfnamefont {Norman~Y.}\ \bibnamefont {Yao}}, \bibinfo {author} {\bibfnamefont {Immanuel}\ \bibnamefont {Bloch}}, \ and\ \bibinfo {author} {\bibfnamefont {Johannes}\ \bibnamefont {Zeiher}},\ }\bibfield  {title} {\enquote {\bibinfo {title} {Quantum gas microscopy of {{Kardar-Parisi-Zhang}} superdiffusion},}\ }\href {\doibase 10.1126/science.abk2397} {\bibfield
  {journal} {\bibinfo  {journal} {Science}\ }\textbf {\bibinfo {volume} {376}},\ \bibinfo {pages} {716--720} (\bibinfo {year} {2022})}\BibitemShut {NoStop}%
\bibitem [{\citenamefont {Rosenberg}\ \emph {et~al.}(2024)\citenamefont {Rosenberg}, \citenamefont {Andersen}, \citenamefont {Samajdar}, \citenamefont {Petukhov}, \citenamefont {Hoke}, \citenamefont {Abanin}, \citenamefont {Bengtsson}, \citenamefont {Drozdov}, \citenamefont {Erickson}, \citenamefont {Klimov}, \citenamefont {Mi}, \citenamefont {Morvan}, \citenamefont {Neeley}, \citenamefont {Neill}, \citenamefont {Acharya}, \citenamefont {Allen}, \citenamefont {Anderson}, \citenamefont {Ansmann}, \citenamefont {Arute}, \citenamefont {Arya}, \citenamefont {Asfaw}, \citenamefont {Atalaya}, \citenamefont {Bardin}, \citenamefont {Bilmes}, \citenamefont {Bortoli}, \citenamefont {Bourassa}, \citenamefont {Bovaird}, \citenamefont {Brill}, \citenamefont {Broughton}, \citenamefont {Buckley}, \citenamefont {Buell}, \citenamefont {Burger}, \citenamefont {Burkett}, \citenamefont {Bushnell}, \citenamefont {Campero}, \citenamefont {Chang}, \citenamefont {Chen}, \citenamefont {Chiaro}, \citenamefont {Chik}, \citenamefont
  {Cogan}, \citenamefont {Collins}, \citenamefont {Conner}, \citenamefont {Courtney}, \citenamefont {Crook}, \citenamefont {Curtin}, \citenamefont {Debroy}, \citenamefont {Barba}, \citenamefont {Demura}, \citenamefont {Di~Paolo}, \citenamefont {Dunsworth}, \citenamefont {Earle}, \citenamefont {Faoro}, \citenamefont {Farhi}, \citenamefont {Fatemi}, \citenamefont {Ferreira}, \citenamefont {Burgos}, \citenamefont {Forati}, \citenamefont {Fowler}, \citenamefont {Foxen}, \citenamefont {Garcia}, \citenamefont {Genois}, \citenamefont {Giang}, \citenamefont {Gidney}, \citenamefont {Gilboa}, \citenamefont {Giustina}, \citenamefont {Gosula}, \citenamefont {Dau}, \citenamefont {Gross}, \citenamefont {Habegger}, \citenamefont {Hamilton}, \citenamefont {Hansen}, \citenamefont {Harrigan}, \citenamefont {Harrington}, \citenamefont {Heu}, \citenamefont {Hill}, \citenamefont {Hoffmann}, \citenamefont {Hong}, \citenamefont {Huang}, \citenamefont {Huff}, \citenamefont {Huggins}, \citenamefont {Ioffe}, \citenamefont {Isakov},
  \citenamefont {Iveland}, \citenamefont {Jeffrey}, \citenamefont {Jiang}, \citenamefont {Jones}, \citenamefont {Juhas}, \citenamefont {Kafri}, \citenamefont {Khattar}, \citenamefont {Khezri}, \citenamefont {Kieferov{\'a}}, \citenamefont {Kim}, \citenamefont {Kitaev}, \citenamefont {Klots}, \citenamefont {Korotkov}, \citenamefont {Kostritsa}, \citenamefont {Kreikebaum}, \citenamefont {Landhuis}, \citenamefont {Laptev}, \citenamefont {Lau}, \citenamefont {Laws}, \citenamefont {Lee}, \citenamefont {Lee}, \citenamefont {Lensky}, \citenamefont {Lester}, \citenamefont {Lill}, \citenamefont {Liu}, \citenamefont {Locharla}, \citenamefont {Mandr{\`a}}, \citenamefont {Martin}, \citenamefont {Martin}, \citenamefont {McClean}, \citenamefont {McEwen}, \citenamefont {Meeks}, \citenamefont {Miao}, \citenamefont {Mieszala}, \citenamefont {Montazeri}, \citenamefont {Movassagh}, \citenamefont {Mruczkiewicz}, \citenamefont {Nersisyan}, \citenamefont {Newman}, \citenamefont {Ng}, \citenamefont {Nguyen}, \citenamefont {Nguyen},
  \citenamefont {Niu}, \citenamefont {O'Brien}, \citenamefont {Omonije}, \citenamefont {Opremcak}, \citenamefont {Potter}, \citenamefont {Pryadko}, \citenamefont {Quintana}, \citenamefont {Rhodes}, \citenamefont {Rocque}, \citenamefont {Rubin}, \citenamefont {Saei}, \citenamefont {Sank}, \citenamefont {Sankaragomathi}, \citenamefont {Satzinger}, \citenamefont {Schurkus}, \citenamefont {Schuster}, \citenamefont {Shearn}, \citenamefont {Shorter}, \citenamefont {Shutty}, \citenamefont {Shvarts}, \citenamefont {Sivak}, \citenamefont {Skruzny}, \citenamefont {Smith}, \citenamefont {Somma}, \citenamefont {Sterling}, \citenamefont {Strain}, \citenamefont {Szalay}, \citenamefont {Thor}, \citenamefont {Torres}, \citenamefont {Vidal}, \citenamefont {Villalonga}, \citenamefont {Heidweiller}, \citenamefont {White}, \citenamefont {Woo}, \citenamefont {Xing}, \citenamefont {Yao}, \citenamefont {Yeh}, \citenamefont {Yoo}, \citenamefont {Young}, \citenamefont {Zalcman}, \citenamefont {Zhang}, \citenamefont {Zhu},
  \citenamefont {Zobrist}, \citenamefont {Neven}, \citenamefont {Babbush}, \citenamefont {Bacon}, \citenamefont {Boixo}, \citenamefont {Hilton}, \citenamefont {Lucero}, \citenamefont {Megrant}, \citenamefont {Kelly}, \citenamefont {Chen}, \citenamefont {Smelyanskiy}, \citenamefont {Khemani}, \citenamefont {Gopalakrishnan}, \citenamefont {Prosen},\ and\ \citenamefont {Roushan}}]{rosenberg:2024a}%
  \BibitemOpen
  \bibfield  {author} {\bibinfo {author} {\bibfnamefont {E.}~\bibnamefont {Rosenberg}}, \bibinfo {author} {\bibfnamefont {T.~I.}\ \bibnamefont {Andersen}}, \bibinfo {author} {\bibfnamefont {R.}~\bibnamefont {Samajdar}}, \bibinfo {author} {\bibfnamefont {A.}~\bibnamefont {Petukhov}}, \bibinfo {author} {\bibfnamefont {J.~C.}\ \bibnamefont {Hoke}}, \bibinfo {author} {\bibfnamefont {D.}~\bibnamefont {Abanin}}, \bibinfo {author} {\bibfnamefont {A.}~\bibnamefont {Bengtsson}}, \bibinfo {author} {\bibfnamefont {I.~K.}\ \bibnamefont {Drozdov}}, \bibinfo {author} {\bibfnamefont {C.}~\bibnamefont {Erickson}}, \bibinfo {author} {\bibfnamefont {P.~V.}\ \bibnamefont {Klimov}}, \bibinfo {author} {\bibfnamefont {X.}~\bibnamefont {Mi}}, \bibinfo {author} {\bibfnamefont {A.}~\bibnamefont {Morvan}}, \bibinfo {author} {\bibfnamefont {M.}~\bibnamefont {Neeley}}, \bibinfo {author} {\bibfnamefont {C.}~\bibnamefont {Neill}}, \bibinfo {author} {\bibfnamefont {R.}~\bibnamefont {Acharya}}, \bibinfo {author} {\bibfnamefont
  {R.}~\bibnamefont {Allen}}, \bibinfo {author} {\bibfnamefont {K.}~\bibnamefont {Anderson}}, \bibinfo {author} {\bibfnamefont {M.}~\bibnamefont {Ansmann}}, \bibinfo {author} {\bibfnamefont {F.}~\bibnamefont {Arute}}, \bibinfo {author} {\bibfnamefont {K.}~\bibnamefont {Arya}}, \bibinfo {author} {\bibfnamefont {A.}~\bibnamefont {Asfaw}}, \bibinfo {author} {\bibfnamefont {J.}~\bibnamefont {Atalaya}}, \bibinfo {author} {\bibfnamefont {J.~C.}\ \bibnamefont {Bardin}}, \bibinfo {author} {\bibfnamefont {A.}~\bibnamefont {Bilmes}}, \bibinfo {author} {\bibfnamefont {G.}~\bibnamefont {Bortoli}}, \bibinfo {author} {\bibfnamefont {A.}~\bibnamefont {Bourassa}}, \bibinfo {author} {\bibfnamefont {J.}~\bibnamefont {Bovaird}}, \bibinfo {author} {\bibfnamefont {L.}~\bibnamefont {Brill}}, \bibinfo {author} {\bibfnamefont {M.}~\bibnamefont {Broughton}}, \bibinfo {author} {\bibfnamefont {B.~B.}\ \bibnamefont {Buckley}}, \bibinfo {author} {\bibfnamefont {D.~A.}\ \bibnamefont {Buell}}, \bibinfo {author} {\bibfnamefont
  {T.}~\bibnamefont {Burger}}, \bibinfo {author} {\bibfnamefont {B.}~\bibnamefont {Burkett}}, \bibinfo {author} {\bibfnamefont {N.}~\bibnamefont {Bushnell}}, \bibinfo {author} {\bibfnamefont {J.}~\bibnamefont {Campero}}, \bibinfo {author} {\bibfnamefont {H.-S.}\ \bibnamefont {Chang}}, \bibinfo {author} {\bibfnamefont {Z.}~\bibnamefont {Chen}}, \bibinfo {author} {\bibfnamefont {B.}~\bibnamefont {Chiaro}}, \bibinfo {author} {\bibfnamefont {D.}~\bibnamefont {Chik}}, \bibinfo {author} {\bibfnamefont {J.}~\bibnamefont {Cogan}}, \bibinfo {author} {\bibfnamefont {R.}~\bibnamefont {Collins}}, \bibinfo {author} {\bibfnamefont {P.}~\bibnamefont {Conner}}, \bibinfo {author} {\bibfnamefont {W.}~\bibnamefont {Courtney}}, \bibinfo {author} {\bibfnamefont {A.~L.}\ \bibnamefont {Crook}}, \bibinfo {author} {\bibfnamefont {B.}~\bibnamefont {Curtin}}, \bibinfo {author} {\bibfnamefont {D.~M.}\ \bibnamefont {Debroy}}, \bibinfo {author} {\bibfnamefont {A.~Del~Toro}\ \bibnamefont {Barba}}, \bibinfo {author} {\bibfnamefont
  {S.}~\bibnamefont {Demura}}, \bibinfo {author} {\bibfnamefont {A.}~\bibnamefont {Di~Paolo}}, \bibinfo {author} {\bibfnamefont {A.}~\bibnamefont {Dunsworth}}, \bibinfo {author} {\bibfnamefont {C.}~\bibnamefont {Earle}}, \bibinfo {author} {\bibfnamefont {L.}~\bibnamefont {Faoro}}, \bibinfo {author} {\bibfnamefont {E.}~\bibnamefont {Farhi}}, \bibinfo {author} {\bibfnamefont {R.}~\bibnamefont {Fatemi}}, \bibinfo {author} {\bibfnamefont {V.~S.}\ \bibnamefont {Ferreira}}, \bibinfo {author} {\bibfnamefont {L.~Flores}\ \bibnamefont {Burgos}}, \bibinfo {author} {\bibfnamefont {E.}~\bibnamefont {Forati}}, \bibinfo {author} {\bibfnamefont {A.~G.}\ \bibnamefont {Fowler}}, \bibinfo {author} {\bibfnamefont {B.}~\bibnamefont {Foxen}}, \bibinfo {author} {\bibfnamefont {G.}~\bibnamefont {Garcia}}, \bibinfo {author} {\bibfnamefont {{\'E}.}~\bibnamefont {Genois}}, \bibinfo {author} {\bibfnamefont {W.}~\bibnamefont {Giang}}, \bibinfo {author} {\bibfnamefont {C.}~\bibnamefont {Gidney}}, \bibinfo {author} {\bibfnamefont
  {D.}~\bibnamefont {Gilboa}}, \bibinfo {author} {\bibfnamefont {M.}~\bibnamefont {Giustina}}, \bibinfo {author} {\bibfnamefont {R.}~\bibnamefont {Gosula}}, \bibinfo {author} {\bibfnamefont {A.~Grajales}\ \bibnamefont {Dau}}, \bibinfo {author} {\bibfnamefont {J.~A.}\ \bibnamefont {Gross}}, \bibinfo {author} {\bibfnamefont {S.}~\bibnamefont {Habegger}}, \bibinfo {author} {\bibfnamefont {M.~C.}\ \bibnamefont {Hamilton}}, \bibinfo {author} {\bibfnamefont {M.}~\bibnamefont {Hansen}}, \bibinfo {author} {\bibfnamefont {M.~P.}\ \bibnamefont {Harrigan}}, \bibinfo {author} {\bibfnamefont {S.~D.}\ \bibnamefont {Harrington}}, \bibinfo {author} {\bibfnamefont {P.}~\bibnamefont {Heu}}, \bibinfo {author} {\bibfnamefont {G.}~\bibnamefont {Hill}}, \bibinfo {author} {\bibfnamefont {M.~R.}\ \bibnamefont {Hoffmann}}, \bibinfo {author} {\bibfnamefont {S.}~\bibnamefont {Hong}}, \bibinfo {author} {\bibfnamefont {T.}~\bibnamefont {Huang}}, \bibinfo {author} {\bibfnamefont {A.}~\bibnamefont {Huff}}, \bibinfo {author} {\bibfnamefont
  {W.~J.}\ \bibnamefont {Huggins}}, \bibinfo {author} {\bibfnamefont {L.~B.}\ \bibnamefont {Ioffe}}, \bibinfo {author} {\bibfnamefont {S.~V.}\ \bibnamefont {Isakov}}, \bibinfo {author} {\bibfnamefont {J.}~\bibnamefont {Iveland}}, \bibinfo {author} {\bibfnamefont {E.}~\bibnamefont {Jeffrey}}, \bibinfo {author} {\bibfnamefont {Z.}~\bibnamefont {Jiang}}, \bibinfo {author} {\bibfnamefont {C.}~\bibnamefont {Jones}}, \bibinfo {author} {\bibfnamefont {P.}~\bibnamefont {Juhas}}, \bibinfo {author} {\bibfnamefont {D.}~\bibnamefont {Kafri}}, \bibinfo {author} {\bibfnamefont {T.}~\bibnamefont {Khattar}}, \bibinfo {author} {\bibfnamefont {M.}~\bibnamefont {Khezri}}, \bibinfo {author} {\bibfnamefont {M.}~\bibnamefont {Kieferov{\'a}}}, \bibinfo {author} {\bibfnamefont {S.}~\bibnamefont {Kim}}, \bibinfo {author} {\bibfnamefont {A.}~\bibnamefont {Kitaev}}, \bibinfo {author} {\bibfnamefont {A.~R.}\ \bibnamefont {Klots}}, \bibinfo {author} {\bibfnamefont {A.~N.}\ \bibnamefont {Korotkov}}, \bibinfo {author} {\bibfnamefont
  {F.}~\bibnamefont {Kostritsa}}, \bibinfo {author} {\bibfnamefont {J.~M.}\ \bibnamefont {Kreikebaum}}, \bibinfo {author} {\bibfnamefont {D.}~\bibnamefont {Landhuis}}, \bibinfo {author} {\bibfnamefont {P.}~\bibnamefont {Laptev}}, \bibinfo {author} {\bibfnamefont {K.-M.}\ \bibnamefont {Lau}}, \bibinfo {author} {\bibfnamefont {L.}~\bibnamefont {Laws}}, \bibinfo {author} {\bibfnamefont {J.}~\bibnamefont {Lee}}, \bibinfo {author} {\bibfnamefont {K.~W.}\ \bibnamefont {Lee}}, \bibinfo {author} {\bibfnamefont {Y.~D.}\ \bibnamefont {Lensky}}, \bibinfo {author} {\bibfnamefont {B.~J.}\ \bibnamefont {Lester}}, \bibinfo {author} {\bibfnamefont {A.~T.}\ \bibnamefont {Lill}}, \bibinfo {author} {\bibfnamefont {W.}~\bibnamefont {Liu}}, \bibinfo {author} {\bibfnamefont {A.}~\bibnamefont {Locharla}}, \bibinfo {author} {\bibfnamefont {S.}~\bibnamefont {Mandr{\`a}}}, \bibinfo {author} {\bibfnamefont {O.}~\bibnamefont {Martin}}, \bibinfo {author} {\bibfnamefont {S.}~\bibnamefont {Martin}}, \bibinfo {author} {\bibfnamefont
  {J.~R.}\ \bibnamefont {McClean}}, \bibinfo {author} {\bibfnamefont {M.}~\bibnamefont {McEwen}}, \bibinfo {author} {\bibfnamefont {S.}~\bibnamefont {Meeks}}, \bibinfo {author} {\bibfnamefont {K.~C.}\ \bibnamefont {Miao}}, \bibinfo {author} {\bibfnamefont {A.}~\bibnamefont {Mieszala}}, \bibinfo {author} {\bibfnamefont {S.}~\bibnamefont {Montazeri}}, \bibinfo {author} {\bibfnamefont {R.}~\bibnamefont {Movassagh}}, \bibinfo {author} {\bibfnamefont {W.}~\bibnamefont {Mruczkiewicz}}, \bibinfo {author} {\bibfnamefont {A.}~\bibnamefont {Nersisyan}}, \bibinfo {author} {\bibfnamefont {M.}~\bibnamefont {Newman}}, \bibinfo {author} {\bibfnamefont {J.~H.}\ \bibnamefont {Ng}}, \bibinfo {author} {\bibfnamefont {A.}~\bibnamefont {Nguyen}}, \bibinfo {author} {\bibfnamefont {M.}~\bibnamefont {Nguyen}}, \bibinfo {author} {\bibfnamefont {M.~Y.}\ \bibnamefont {Niu}}, \bibinfo {author} {\bibfnamefont {T.~E.}\ \bibnamefont {O'Brien}}, \bibinfo {author} {\bibfnamefont {S.}~\bibnamefont {Omonije}}, \bibinfo {author} {\bibfnamefont
  {A.}~\bibnamefont {Opremcak}}, \bibinfo {author} {\bibfnamefont {R.}~\bibnamefont {Potter}}, \bibinfo {author} {\bibfnamefont {L.~P.}\ \bibnamefont {Pryadko}}, \bibinfo {author} {\bibfnamefont {C.}~\bibnamefont {Quintana}}, \bibinfo {author} {\bibfnamefont {D.~M.}\ \bibnamefont {Rhodes}}, \bibinfo {author} {\bibfnamefont {C.}~\bibnamefont {Rocque}}, \bibinfo {author} {\bibfnamefont {N.~C.}\ \bibnamefont {Rubin}}, \bibinfo {author} {\bibfnamefont {N.}~\bibnamefont {Saei}}, \bibinfo {author} {\bibfnamefont {D.}~\bibnamefont {Sank}}, \bibinfo {author} {\bibfnamefont {K.}~\bibnamefont {Sankaragomathi}}, \bibinfo {author} {\bibfnamefont {K.~J.}\ \bibnamefont {Satzinger}}, \bibinfo {author} {\bibfnamefont {H.~F.}\ \bibnamefont {Schurkus}}, \bibinfo {author} {\bibfnamefont {C.}~\bibnamefont {Schuster}}, \bibinfo {author} {\bibfnamefont {M.~J.}\ \bibnamefont {Shearn}}, \bibinfo {author} {\bibfnamefont {A.}~\bibnamefont {Shorter}}, \bibinfo {author} {\bibfnamefont {N.}~\bibnamefont {Shutty}}, \bibinfo {author}
  {\bibfnamefont {V.}~\bibnamefont {Shvarts}}, \bibinfo {author} {\bibfnamefont {V.}~\bibnamefont {Sivak}}, \bibinfo {author} {\bibfnamefont {J.}~\bibnamefont {Skruzny}}, \bibinfo {author} {\bibfnamefont {W.~Clarke}\ \bibnamefont {Smith}}, \bibinfo {author} {\bibfnamefont {R.~D.}\ \bibnamefont {Somma}}, \bibinfo {author} {\bibfnamefont {G.}~\bibnamefont {Sterling}}, \bibinfo {author} {\bibfnamefont {D.}~\bibnamefont {Strain}}, \bibinfo {author} {\bibfnamefont {M.}~\bibnamefont {Szalay}}, \bibinfo {author} {\bibfnamefont {D.}~\bibnamefont {Thor}}, \bibinfo {author} {\bibfnamefont {A.}~\bibnamefont {Torres}}, \bibinfo {author} {\bibfnamefont {G.}~\bibnamefont {Vidal}}, \bibinfo {author} {\bibfnamefont {B.}~\bibnamefont {Villalonga}}, \bibinfo {author} {\bibfnamefont {C.~Vollgraff}\ \bibnamefont {Heidweiller}}, \bibinfo {author} {\bibfnamefont {T.}~\bibnamefont {White}}, \bibinfo {author} {\bibfnamefont {B.~W.~K.}\ \bibnamefont {Woo}}, \bibinfo {author} {\bibfnamefont {C.}~\bibnamefont {Xing}}, \bibinfo {author}
  {\bibfnamefont {Z.~Jamie}\ \bibnamefont {Yao}}, \bibinfo {author} {\bibfnamefont {P.}~\bibnamefont {Yeh}}, \bibinfo {author} {\bibfnamefont {J.}~\bibnamefont {Yoo}}, \bibinfo {author} {\bibfnamefont {G.}~\bibnamefont {Young}}, \bibinfo {author} {\bibfnamefont {A.}~\bibnamefont {Zalcman}}, \bibinfo {author} {\bibfnamefont {Y.}~\bibnamefont {Zhang}}, \bibinfo {author} {\bibfnamefont {N.}~\bibnamefont {Zhu}}, \bibinfo {author} {\bibfnamefont {N.}~\bibnamefont {Zobrist}}, \bibinfo {author} {\bibfnamefont {H.}~\bibnamefont {Neven}}, \bibinfo {author} {\bibfnamefont {R.}~\bibnamefont {Babbush}}, \bibinfo {author} {\bibfnamefont {D.}~\bibnamefont {Bacon}}, \bibinfo {author} {\bibfnamefont {S.}~\bibnamefont {Boixo}}, \bibinfo {author} {\bibfnamefont {J.}~\bibnamefont {Hilton}}, \bibinfo {author} {\bibfnamefont {E.}~\bibnamefont {Lucero}}, \bibinfo {author} {\bibfnamefont {A.}~\bibnamefont {Megrant}}, \bibinfo {author} {\bibfnamefont {J.}~\bibnamefont {Kelly}}, \bibinfo {author} {\bibfnamefont {Y.}~\bibnamefont
  {Chen}}, \bibinfo {author} {\bibfnamefont {V.}~\bibnamefont {Smelyanskiy}}, \bibinfo {author} {\bibfnamefont {V.}~\bibnamefont {Khemani}}, \bibinfo {author} {\bibfnamefont {S.}~\bibnamefont {Gopalakrishnan}}, \bibinfo {author} {\bibfnamefont {T.}~\bibnamefont {Prosen}}, \ and\ \bibinfo {author} {\bibfnamefont {P.}~\bibnamefont {Roushan}},\ }\bibfield  {title} {\enquote {\bibinfo {title} {Dynamics of magnetization at infinite temperature in a {{Heisenberg}} spin chain},}\ }\href {\doibase 10.1126/science.adi7877} {\bibfield  {journal} {\bibinfo  {journal} {Science}\ }\textbf {\bibinfo {volume} {384}},\ \bibinfo {pages} {48--53} (\bibinfo {year} {2024})}\BibitemShut {NoStop}%
\bibitem [{\citenamefont {Zurek}\ \emph {et~al.}(2005)\citenamefont {Zurek}, \citenamefont {Dorner},\ and\ \citenamefont {Zoller}}]{zurek2005dynamics}%
  \BibitemOpen
  \bibfield  {author} {\bibinfo {author} {\bibfnamefont {Wojciech~H.}\ \bibnamefont {Zurek}}, \bibinfo {author} {\bibfnamefont {Uwe}\ \bibnamefont {Dorner}}, \ and\ \bibinfo {author} {\bibfnamefont {Peter}\ \bibnamefont {Zoller}},\ }\bibfield  {title} {\enquote {\bibinfo {title} {Dynamics of a quantum phase transition},}\ }\href {\doibase 10.1103/PhysRevLett.95.105701} {\bibfield  {journal} {\bibinfo  {journal} {Phys. Rev. Lett.}\ }\textbf {\bibinfo {volume} {95}},\ \bibinfo {pages} {105701} (\bibinfo {year} {2005})}\BibitemShut {NoStop}%
\bibitem [{\citenamefont {Keesling}\ \emph {et~al.}(2019)\citenamefont {Keesling}, \citenamefont {Omran}, \citenamefont {Levine}, \citenamefont {Bernien}, \citenamefont {Pichler}, \citenamefont {Choi}, \citenamefont {Samajdar}, \citenamefont {Schwartz}, \citenamefont {Silvi}, \citenamefont {Sachdev}, \citenamefont {Zoller}, \citenamefont {Endres}, \citenamefont {Greiner}, \citenamefont {Vuleti{\'c}},\ and\ \citenamefont {Lukin}}]{keesling2019quantum}%
  \BibitemOpen
  \bibfield  {author} {\bibinfo {author} {\bibfnamefont {Alexander}\ \bibnamefont {Keesling}}, \bibinfo {author} {\bibfnamefont {Ahmed}\ \bibnamefont {Omran}}, \bibinfo {author} {\bibfnamefont {Harry}\ \bibnamefont {Levine}}, \bibinfo {author} {\bibfnamefont {Hannes}\ \bibnamefont {Bernien}}, \bibinfo {author} {\bibfnamefont {Hannes}\ \bibnamefont {Pichler}}, \bibinfo {author} {\bibfnamefont {Soonwon}\ \bibnamefont {Choi}}, \bibinfo {author} {\bibfnamefont {Rhine}\ \bibnamefont {Samajdar}}, \bibinfo {author} {\bibfnamefont {Sylvain}\ \bibnamefont {Schwartz}}, \bibinfo {author} {\bibfnamefont {Pietro}\ \bibnamefont {Silvi}}, \bibinfo {author} {\bibfnamefont {Subir}\ \bibnamefont {Sachdev}}, \bibinfo {author} {\bibfnamefont {Peter}\ \bibnamefont {Zoller}}, \bibinfo {author} {\bibfnamefont {Manuel}\ \bibnamefont {Endres}}, \bibinfo {author} {\bibfnamefont {Markus}\ \bibnamefont {Greiner}}, \bibinfo {author} {\bibfnamefont {Vladan}\ \bibnamefont {Vuleti{\'c}}}, \ and\ \bibinfo {author} {\bibfnamefont
  {Mikhail~D.}\ \bibnamefont {Lukin}},\ }\bibfield  {title} {\enquote {\bibinfo {title} {Quantum {{Kibble}}--{{Zurek}} mechanism and critical dynamics on a programmable {{Rydberg}} simulator},}\ }\href {\doibase 10.1038/s41586-019-1070-1} {\bibfield  {journal} {\bibinfo  {journal} {Nature}\ }\textbf {\bibinfo {volume} {568}},\ \bibinfo {pages} {207--211} (\bibinfo {year} {2019})}\BibitemShut {NoStop}%
\bibitem [{\citenamefont {Dupont}\ and\ \citenamefont {Moore}(2022)}]{dupont2022quantum}%
  \BibitemOpen
  \bibfield  {author} {\bibinfo {author} {\bibfnamefont {Maxime}\ \bibnamefont {Dupont}}\ and\ \bibinfo {author} {\bibfnamefont {Joel~E.}\ \bibnamefont {Moore}},\ }\bibfield  {title} {\enquote {\bibinfo {title} {Quantum criticality using a superconducting quantum processor},}\ }\href {\doibase 10.1103/PhysRevB.106.L041109} {\bibfield  {journal} {\bibinfo  {journal} {Phys. Rev. B}\ }\textbf {\bibinfo {volume} {106}},\ \bibinfo {pages} {L041109} (\bibinfo {year} {2022})}\BibitemShut {NoStop}%
\bibitem [{\citenamefont {Andersen}\ \emph {et~al.}(2025)\citenamefont {Andersen}, \citenamefont {Astrakhantsev}, \citenamefont {Karamlou}, \citenamefont {Berndtsson}, \citenamefont {Motruk}, \citenamefont {Szasz}, \citenamefont {Gross}, \citenamefont {Schuckert}, \citenamefont {Westerhout}, \citenamefont {Zhang}, \citenamefont {Forati}, \citenamefont {Rossi}, \citenamefont {Kobrin}, \citenamefont {Paolo}, \citenamefont {Klots}, \citenamefont {Drozdov}, \citenamefont {Kurilovich}, \citenamefont {Petukhov}, \citenamefont {Ioffe}, \citenamefont {Elben}, \citenamefont {Rath}, \citenamefont {Vitale}, \citenamefont {Vermersch}, \citenamefont {Acharya}, \citenamefont {Beni}, \citenamefont {Anderson}, \citenamefont {Ansmann}, \citenamefont {Arute}, \citenamefont {Arya}, \citenamefont {Asfaw}, \citenamefont {Atalaya}, \citenamefont {Ballard}, \citenamefont {Bardin}, \citenamefont {Bengtsson}, \citenamefont {Bilmes}, \citenamefont {Bortoli}, \citenamefont {Bourassa}, \citenamefont {Bovaird}, \citenamefont {Brill},
  \citenamefont {Broughton}, \citenamefont {Browne}, \citenamefont {Buchea}, \citenamefont {Buckley}, \citenamefont {Buell}, \citenamefont {Burger}, \citenamefont {Burkett}, \citenamefont {Bushnell}, \citenamefont {Cabrera}, \citenamefont {Campero}, \citenamefont {Chang}, \citenamefont {Chen}, \citenamefont {Chiaro}, \citenamefont {Claes}, \citenamefont {Cleland}, \citenamefont {Cogan}, \citenamefont {Collins}, \citenamefont {Conner}, \citenamefont {Courtney}, \citenamefont {Crook}, \citenamefont {Das}, \citenamefont {Debroy}, \citenamefont {Lorenzo}, \citenamefont {Barba}, \citenamefont {Demura}, \citenamefont {Donohoe}, \citenamefont {Dunsworth}, \citenamefont {Earle}, \citenamefont {Eickbusch}, \citenamefont {Elbag}, \citenamefont {Elzouka}, \citenamefont {Erickson}, \citenamefont {Faoro}, \citenamefont {Fatemi}, \citenamefont {Ferreira}, \citenamefont {Burgos}, \citenamefont {Fowler}, \citenamefont {Foxen}, \citenamefont {Ganjam}, \citenamefont {Gasca}, \citenamefont {Giang}, \citenamefont {Gidney},
  \citenamefont {Gilboa}, \citenamefont {Giustina}, \citenamefont {Gosula}, \citenamefont {Dau}, \citenamefont {Graumann}, \citenamefont {Greene}, \citenamefont {Habegger}, \citenamefont {Hamilton}, \citenamefont {Hansen}, \citenamefont {Harrigan}, \citenamefont {Harrington}, \citenamefont {Heslin}, \citenamefont {Heu}, \citenamefont {Hill}, \citenamefont {Hoffmann}, \citenamefont {Huang}, \citenamefont {Huang}, \citenamefont {Huff}, \citenamefont {Huggins}, \citenamefont {Isakov}, \citenamefont {Jeffrey}, \citenamefont {Jiang}, \citenamefont {Jones}, \citenamefont {Jordan}, \citenamefont {Joshi}, \citenamefont {Juhas}, \citenamefont {Kafri}, \citenamefont {Kang}, \citenamefont {Kechedzhi}, \citenamefont {Khaire}, \citenamefont {Khattar}, \citenamefont {Khezri}, \citenamefont {Kieferov{\'a}}, \citenamefont {Kim}, \citenamefont {Kitaev}, \citenamefont {Klimov}, \citenamefont {Korotkov}, \citenamefont {Kostritsa}, \citenamefont {Kreikebaum}, \citenamefont {Landhuis}, \citenamefont {Langley}, \citenamefont
  {Laptev}, \citenamefont {Lau}, \citenamefont {Guevel}, \citenamefont {Ledford}, \citenamefont {Lee}, \citenamefont {Lee}, \citenamefont {Lensky}, \citenamefont {Lester}, \citenamefont {Li}, \citenamefont {Lill}, \citenamefont {Liu}, \citenamefont {Livingston}, \citenamefont {Locharla}, \citenamefont {Lundahl}, \citenamefont {Lunt}, \citenamefont {Madhuk}, \citenamefont {Maloney}, \citenamefont {Mandr{\`a}}, \citenamefont {Martin}, \citenamefont {Martin}, \citenamefont {Martin}, \citenamefont {Maxfield}, \citenamefont {McClean}, \citenamefont {McEwen}, \citenamefont {Meeks}, \citenamefont {Miao}, \citenamefont {Mieszala}, \citenamefont {Molina}, \citenamefont {Montazeri}, \citenamefont {Morvan}, \citenamefont {Movassagh}, \citenamefont {Neill}, \citenamefont {Nersisyan}, \citenamefont {Newman}, \citenamefont {Nguyen}, \citenamefont {Nguyen}, \citenamefont {Ni}, \citenamefont {Niu}, \citenamefont {Oliver}, \citenamefont {Ottosson}, \citenamefont {Pizzuto}, \citenamefont {Potter}, \citenamefont {Pritchard},
  \citenamefont {Pryadko}, \citenamefont {Quintana}, \citenamefont {Reagor}, \citenamefont {Rhodes}, \citenamefont {Roberts}, \citenamefont {Rocque}, \citenamefont {Rosenberg}, \citenamefont {Rubin}, \citenamefont {Saei}, \citenamefont {Sankaragomathi}, \citenamefont {Satzinger}, \citenamefont {Schurkus}, \citenamefont {Schuster}, \citenamefont {Shearn}, \citenamefont {Shorter}, \citenamefont {Shutty}, \citenamefont {Shvarts}, \citenamefont {Sivak}, \citenamefont {Skruzny}, \citenamefont {Small}, \citenamefont {Smith}, \citenamefont {Springer}, \citenamefont {Sterling}, \citenamefont {Suchard}, \citenamefont {Szalay}, \citenamefont {Sztein}, \citenamefont {Thor}, \citenamefont {Torres}, \citenamefont {Torunbalci}, \citenamefont {Vaishnav}, \citenamefont {Vdovichev}, \citenamefont {Villalonga}, \citenamefont {Heidweiller}, \citenamefont {Waltman}, \citenamefont {Wang}, \citenamefont {White}, \citenamefont {Wong}, \citenamefont {Woo}, \citenamefont {Xing}, \citenamefont {Yao}, \citenamefont {Yeh}, \citenamefont
  {Ying}, \citenamefont {Yoo}, \citenamefont {Yosri}, \citenamefont {Young}, \citenamefont {Zalcman}, \citenamefont {Zhu}, \citenamefont {Zobrist}, \citenamefont {Neven}, \citenamefont {Babbush}, \citenamefont {Boixo}, \citenamefont {Hilton}, \citenamefont {Lucero}, \citenamefont {Megrant}, \citenamefont {Kelly}, \citenamefont {Chen}, \citenamefont {Smelyanskiy}, \citenamefont {Vidal}, \citenamefont {Roushan}, \citenamefont {L{\"a}uchli}, \citenamefont {Abanin},\ and\ \citenamefont {Mi}}]{andersen:2025}%
  \BibitemOpen
  \bibfield  {author} {\bibinfo {author} {\bibfnamefont {T.~I.}\ \bibnamefont {Andersen}}, \bibinfo {author} {\bibfnamefont {N.}~\bibnamefont {Astrakhantsev}}, \bibinfo {author} {\bibfnamefont {A.~H.}\ \bibnamefont {Karamlou}}, \bibinfo {author} {\bibfnamefont {J.}~\bibnamefont {Berndtsson}}, \bibinfo {author} {\bibfnamefont {J.}~\bibnamefont {Motruk}}, \bibinfo {author} {\bibfnamefont {A.}~\bibnamefont {Szasz}}, \bibinfo {author} {\bibfnamefont {J.~A.}\ \bibnamefont {Gross}}, \bibinfo {author} {\bibfnamefont {A.}~\bibnamefont {Schuckert}}, \bibinfo {author} {\bibfnamefont {T.}~\bibnamefont {Westerhout}}, \bibinfo {author} {\bibfnamefont {Y.}~\bibnamefont {Zhang}}, \bibinfo {author} {\bibfnamefont {E.}~\bibnamefont {Forati}}, \bibinfo {author} {\bibfnamefont {D.}~\bibnamefont {Rossi}}, \bibinfo {author} {\bibfnamefont {B.}~\bibnamefont {Kobrin}}, \bibinfo {author} {\bibfnamefont {A.~Di}\ \bibnamefont {Paolo}}, \bibinfo {author} {\bibfnamefont {A.~R.}\ \bibnamefont {Klots}}, \bibinfo {author} {\bibfnamefont
  {I.}~\bibnamefont {Drozdov}}, \bibinfo {author} {\bibfnamefont {V.}~\bibnamefont {Kurilovich}}, \bibinfo {author} {\bibfnamefont {A.}~\bibnamefont {Petukhov}}, \bibinfo {author} {\bibfnamefont {L.~B.}\ \bibnamefont {Ioffe}}, \bibinfo {author} {\bibfnamefont {A.}~\bibnamefont {Elben}}, \bibinfo {author} {\bibfnamefont {A.}~\bibnamefont {Rath}}, \bibinfo {author} {\bibfnamefont {V.}~\bibnamefont {Vitale}}, \bibinfo {author} {\bibfnamefont {B.}~\bibnamefont {Vermersch}}, \bibinfo {author} {\bibfnamefont {R.}~\bibnamefont {Acharya}}, \bibinfo {author} {\bibfnamefont {L.~A.}\ \bibnamefont {Beni}}, \bibinfo {author} {\bibfnamefont {K.}~\bibnamefont {Anderson}}, \bibinfo {author} {\bibfnamefont {M.}~\bibnamefont {Ansmann}}, \bibinfo {author} {\bibfnamefont {F.}~\bibnamefont {Arute}}, \bibinfo {author} {\bibfnamefont {K.}~\bibnamefont {Arya}}, \bibinfo {author} {\bibfnamefont {A.}~\bibnamefont {Asfaw}}, \bibinfo {author} {\bibfnamefont {J.}~\bibnamefont {Atalaya}}, \bibinfo {author} {\bibfnamefont {B.}~\bibnamefont
  {Ballard}}, \bibinfo {author} {\bibfnamefont {J.~C.}\ \bibnamefont {Bardin}}, \bibinfo {author} {\bibfnamefont {A.}~\bibnamefont {Bengtsson}}, \bibinfo {author} {\bibfnamefont {A.}~\bibnamefont {Bilmes}}, \bibinfo {author} {\bibfnamefont {G.}~\bibnamefont {Bortoli}}, \bibinfo {author} {\bibfnamefont {A.}~\bibnamefont {Bourassa}}, \bibinfo {author} {\bibfnamefont {J.}~\bibnamefont {Bovaird}}, \bibinfo {author} {\bibfnamefont {L.}~\bibnamefont {Brill}}, \bibinfo {author} {\bibfnamefont {M.}~\bibnamefont {Broughton}}, \bibinfo {author} {\bibfnamefont {D.~A.}\ \bibnamefont {Browne}}, \bibinfo {author} {\bibfnamefont {B.}~\bibnamefont {Buchea}}, \bibinfo {author} {\bibfnamefont {B.~B.}\ \bibnamefont {Buckley}}, \bibinfo {author} {\bibfnamefont {D.~A.}\ \bibnamefont {Buell}}, \bibinfo {author} {\bibfnamefont {T.}~\bibnamefont {Burger}}, \bibinfo {author} {\bibfnamefont {B.}~\bibnamefont {Burkett}}, \bibinfo {author} {\bibfnamefont {N.}~\bibnamefont {Bushnell}}, \bibinfo {author} {\bibfnamefont {A.}~\bibnamefont
  {Cabrera}}, \bibinfo {author} {\bibfnamefont {J.}~\bibnamefont {Campero}}, \bibinfo {author} {\bibfnamefont {H.-S.}\ \bibnamefont {Chang}}, \bibinfo {author} {\bibfnamefont {Z.}~\bibnamefont {Chen}}, \bibinfo {author} {\bibfnamefont {B.}~\bibnamefont {Chiaro}}, \bibinfo {author} {\bibfnamefont {J.}~\bibnamefont {Claes}}, \bibinfo {author} {\bibfnamefont {A.~Y.}\ \bibnamefont {Cleland}}, \bibinfo {author} {\bibfnamefont {J.}~\bibnamefont {Cogan}}, \bibinfo {author} {\bibfnamefont {R.}~\bibnamefont {Collins}}, \bibinfo {author} {\bibfnamefont {P.}~\bibnamefont {Conner}}, \bibinfo {author} {\bibfnamefont {W.}~\bibnamefont {Courtney}}, \bibinfo {author} {\bibfnamefont {A.~L.}\ \bibnamefont {Crook}}, \bibinfo {author} {\bibfnamefont {S.}~\bibnamefont {Das}}, \bibinfo {author} {\bibfnamefont {D.~M.}\ \bibnamefont {Debroy}}, \bibinfo {author} {\bibfnamefont {L.~De}\ \bibnamefont {Lorenzo}}, \bibinfo {author} {\bibfnamefont {A.~Del~Toro}\ \bibnamefont {Barba}}, \bibinfo {author} {\bibfnamefont {S.}~\bibnamefont
  {Demura}}, \bibinfo {author} {\bibfnamefont {P.}~\bibnamefont {Donohoe}}, \bibinfo {author} {\bibfnamefont {A.}~\bibnamefont {Dunsworth}}, \bibinfo {author} {\bibfnamefont {C.}~\bibnamefont {Earle}}, \bibinfo {author} {\bibfnamefont {A.}~\bibnamefont {Eickbusch}}, \bibinfo {author} {\bibfnamefont {A.~M.}\ \bibnamefont {Elbag}}, \bibinfo {author} {\bibfnamefont {M.}~\bibnamefont {Elzouka}}, \bibinfo {author} {\bibfnamefont {C.}~\bibnamefont {Erickson}}, \bibinfo {author} {\bibfnamefont {L.}~\bibnamefont {Faoro}}, \bibinfo {author} {\bibfnamefont {R.}~\bibnamefont {Fatemi}}, \bibinfo {author} {\bibfnamefont {V.~S.}\ \bibnamefont {Ferreira}}, \bibinfo {author} {\bibfnamefont {L.~Flores}\ \bibnamefont {Burgos}}, \bibinfo {author} {\bibfnamefont {A.~G.}\ \bibnamefont {Fowler}}, \bibinfo {author} {\bibfnamefont {B.}~\bibnamefont {Foxen}}, \bibinfo {author} {\bibfnamefont {S.}~\bibnamefont {Ganjam}}, \bibinfo {author} {\bibfnamefont {R.}~\bibnamefont {Gasca}}, \bibinfo {author} {\bibfnamefont {W.}~\bibnamefont
  {Giang}}, \bibinfo {author} {\bibfnamefont {C.}~\bibnamefont {Gidney}}, \bibinfo {author} {\bibfnamefont {D.}~\bibnamefont {Gilboa}}, \bibinfo {author} {\bibfnamefont {M.}~\bibnamefont {Giustina}}, \bibinfo {author} {\bibfnamefont {R.}~\bibnamefont {Gosula}}, \bibinfo {author} {\bibfnamefont {A.~Grajales}\ \bibnamefont {Dau}}, \bibinfo {author} {\bibfnamefont {D.}~\bibnamefont {Graumann}}, \bibinfo {author} {\bibfnamefont {A.}~\bibnamefont {Greene}}, \bibinfo {author} {\bibfnamefont {S.}~\bibnamefont {Habegger}}, \bibinfo {author} {\bibfnamefont {M.~C.}\ \bibnamefont {Hamilton}}, \bibinfo {author} {\bibfnamefont {M.}~\bibnamefont {Hansen}}, \bibinfo {author} {\bibfnamefont {M.~P.}\ \bibnamefont {Harrigan}}, \bibinfo {author} {\bibfnamefont {S.~D.}\ \bibnamefont {Harrington}}, \bibinfo {author} {\bibfnamefont {S.}~\bibnamefont {Heslin}}, \bibinfo {author} {\bibfnamefont {P.}~\bibnamefont {Heu}}, \bibinfo {author} {\bibfnamefont {G.}~\bibnamefont {Hill}}, \bibinfo {author} {\bibfnamefont {M.~R.}\ \bibnamefont
  {Hoffmann}}, \bibinfo {author} {\bibfnamefont {H.-Y.}\ \bibnamefont {Huang}}, \bibinfo {author} {\bibfnamefont {T.}~\bibnamefont {Huang}}, \bibinfo {author} {\bibfnamefont {A.}~\bibnamefont {Huff}}, \bibinfo {author} {\bibfnamefont {W.~J.}\ \bibnamefont {Huggins}}, \bibinfo {author} {\bibfnamefont {S.~V.}\ \bibnamefont {Isakov}}, \bibinfo {author} {\bibfnamefont {E.}~\bibnamefont {Jeffrey}}, \bibinfo {author} {\bibfnamefont {Z.}~\bibnamefont {Jiang}}, \bibinfo {author} {\bibfnamefont {C.}~\bibnamefont {Jones}}, \bibinfo {author} {\bibfnamefont {S.}~\bibnamefont {Jordan}}, \bibinfo {author} {\bibfnamefont {C.}~\bibnamefont {Joshi}}, \bibinfo {author} {\bibfnamefont {P.}~\bibnamefont {Juhas}}, \bibinfo {author} {\bibfnamefont {D.}~\bibnamefont {Kafri}}, \bibinfo {author} {\bibfnamefont {H.}~\bibnamefont {Kang}}, \bibinfo {author} {\bibfnamefont {K.}~\bibnamefont {Kechedzhi}}, \bibinfo {author} {\bibfnamefont {T.}~\bibnamefont {Khaire}}, \bibinfo {author} {\bibfnamefont {T.}~\bibnamefont {Khattar}}, \bibinfo
  {author} {\bibfnamefont {M.}~\bibnamefont {Khezri}}, \bibinfo {author} {\bibfnamefont {M.}~\bibnamefont {Kieferov{\'a}}}, \bibinfo {author} {\bibfnamefont {S.}~\bibnamefont {Kim}}, \bibinfo {author} {\bibfnamefont {A.}~\bibnamefont {Kitaev}}, \bibinfo {author} {\bibfnamefont {P.}~\bibnamefont {Klimov}}, \bibinfo {author} {\bibfnamefont {A.~N.}\ \bibnamefont {Korotkov}}, \bibinfo {author} {\bibfnamefont {F.}~\bibnamefont {Kostritsa}}, \bibinfo {author} {\bibfnamefont {J.~M.}\ \bibnamefont {Kreikebaum}}, \bibinfo {author} {\bibfnamefont {D.}~\bibnamefont {Landhuis}}, \bibinfo {author} {\bibfnamefont {B.~W.}\ \bibnamefont {Langley}}, \bibinfo {author} {\bibfnamefont {P.}~\bibnamefont {Laptev}}, \bibinfo {author} {\bibfnamefont {K.-M.}\ \bibnamefont {Lau}}, \bibinfo {author} {\bibfnamefont {L.~Le}\ \bibnamefont {Guevel}}, \bibinfo {author} {\bibfnamefont {J.}~\bibnamefont {Ledford}}, \bibinfo {author} {\bibfnamefont {J.}~\bibnamefont {Lee}}, \bibinfo {author} {\bibfnamefont {K.~W.}\ \bibnamefont {Lee}},
  \bibinfo {author} {\bibfnamefont {Y.~D.}\ \bibnamefont {Lensky}}, \bibinfo {author} {\bibfnamefont {B.~J.}\ \bibnamefont {Lester}}, \bibinfo {author} {\bibfnamefont {W.~Y.}\ \bibnamefont {Li}}, \bibinfo {author} {\bibfnamefont {A.~T.}\ \bibnamefont {Lill}}, \bibinfo {author} {\bibfnamefont {W.}~\bibnamefont {Liu}}, \bibinfo {author} {\bibfnamefont {W.~P.}\ \bibnamefont {Livingston}}, \bibinfo {author} {\bibfnamefont {A.}~\bibnamefont {Locharla}}, \bibinfo {author} {\bibfnamefont {D.}~\bibnamefont {Lundahl}}, \bibinfo {author} {\bibfnamefont {A.}~\bibnamefont {Lunt}}, \bibinfo {author} {\bibfnamefont {S.}~\bibnamefont {Madhuk}}, \bibinfo {author} {\bibfnamefont {A.}~\bibnamefont {Maloney}}, \bibinfo {author} {\bibfnamefont {S.}~\bibnamefont {Mandr{\`a}}}, \bibinfo {author} {\bibfnamefont {L.~S.}\ \bibnamefont {Martin}}, \bibinfo {author} {\bibfnamefont {O.}~\bibnamefont {Martin}}, \bibinfo {author} {\bibfnamefont {S.}~\bibnamefont {Martin}}, \bibinfo {author} {\bibfnamefont {C.}~\bibnamefont {Maxfield}},
  \bibinfo {author} {\bibfnamefont {J.~R.}\ \bibnamefont {McClean}}, \bibinfo {author} {\bibfnamefont {M.}~\bibnamefont {McEwen}}, \bibinfo {author} {\bibfnamefont {S.}~\bibnamefont {Meeks}}, \bibinfo {author} {\bibfnamefont {K.~C.}\ \bibnamefont {Miao}}, \bibinfo {author} {\bibfnamefont {A.}~\bibnamefont {Mieszala}}, \bibinfo {author} {\bibfnamefont {S.}~\bibnamefont {Molina}}, \bibinfo {author} {\bibfnamefont {S.}~\bibnamefont {Montazeri}}, \bibinfo {author} {\bibfnamefont {A.}~\bibnamefont {Morvan}}, \bibinfo {author} {\bibfnamefont {R.}~\bibnamefont {Movassagh}}, \bibinfo {author} {\bibfnamefont {C.}~\bibnamefont {Neill}}, \bibinfo {author} {\bibfnamefont {A.}~\bibnamefont {Nersisyan}}, \bibinfo {author} {\bibfnamefont {M.}~\bibnamefont {Newman}}, \bibinfo {author} {\bibfnamefont {A.}~\bibnamefont {Nguyen}}, \bibinfo {author} {\bibfnamefont {M.}~\bibnamefont {Nguyen}}, \bibinfo {author} {\bibfnamefont {C.-H.}\ \bibnamefont {Ni}}, \bibinfo {author} {\bibfnamefont {M.~Y.}\ \bibnamefont {Niu}}, \bibinfo
  {author} {\bibfnamefont {W.~D.}\ \bibnamefont {Oliver}}, \bibinfo {author} {\bibfnamefont {K.}~\bibnamefont {Ottosson}}, \bibinfo {author} {\bibfnamefont {A.}~\bibnamefont {Pizzuto}}, \bibinfo {author} {\bibfnamefont {R.}~\bibnamefont {Potter}}, \bibinfo {author} {\bibfnamefont {O.}~\bibnamefont {Pritchard}}, \bibinfo {author} {\bibfnamefont {L.~P.}\ \bibnamefont {Pryadko}}, \bibinfo {author} {\bibfnamefont {C.}~\bibnamefont {Quintana}}, \bibinfo {author} {\bibfnamefont {M.~J.}\ \bibnamefont {Reagor}}, \bibinfo {author} {\bibfnamefont {D.~M.}\ \bibnamefont {Rhodes}}, \bibinfo {author} {\bibfnamefont {G.}~\bibnamefont {Roberts}}, \bibinfo {author} {\bibfnamefont {C.}~\bibnamefont {Rocque}}, \bibinfo {author} {\bibfnamefont {E.}~\bibnamefont {Rosenberg}}, \bibinfo {author} {\bibfnamefont {N.~C.}\ \bibnamefont {Rubin}}, \bibinfo {author} {\bibfnamefont {N.}~\bibnamefont {Saei}}, \bibinfo {author} {\bibfnamefont {K.}~\bibnamefont {Sankaragomathi}}, \bibinfo {author} {\bibfnamefont {K.~J.}\ \bibnamefont
  {Satzinger}}, \bibinfo {author} {\bibfnamefont {H.~F.}\ \bibnamefont {Schurkus}}, \bibinfo {author} {\bibfnamefont {C.}~\bibnamefont {Schuster}}, \bibinfo {author} {\bibfnamefont {M.~J.}\ \bibnamefont {Shearn}}, \bibinfo {author} {\bibfnamefont {A.}~\bibnamefont {Shorter}}, \bibinfo {author} {\bibfnamefont {N.}~\bibnamefont {Shutty}}, \bibinfo {author} {\bibfnamefont {V.}~\bibnamefont {Shvarts}}, \bibinfo {author} {\bibfnamefont {V.}~\bibnamefont {Sivak}}, \bibinfo {author} {\bibfnamefont {J.}~\bibnamefont {Skruzny}}, \bibinfo {author} {\bibfnamefont {S.}~\bibnamefont {Small}}, \bibinfo {author} {\bibfnamefont {W.~Clarke}\ \bibnamefont {Smith}}, \bibinfo {author} {\bibfnamefont {S.}~\bibnamefont {Springer}}, \bibinfo {author} {\bibfnamefont {G.}~\bibnamefont {Sterling}}, \bibinfo {author} {\bibfnamefont {J.}~\bibnamefont {Suchard}}, \bibinfo {author} {\bibfnamefont {M.}~\bibnamefont {Szalay}}, \bibinfo {author} {\bibfnamefont {A.}~\bibnamefont {Sztein}}, \bibinfo {author} {\bibfnamefont {D.}~\bibnamefont
  {Thor}}, \bibinfo {author} {\bibfnamefont {A.}~\bibnamefont {Torres}}, \bibinfo {author} {\bibfnamefont {M.~M.}\ \bibnamefont {Torunbalci}}, \bibinfo {author} {\bibfnamefont {A.}~\bibnamefont {Vaishnav}}, \bibinfo {author} {\bibfnamefont {S.}~\bibnamefont {Vdovichev}}, \bibinfo {author} {\bibfnamefont {B.}~\bibnamefont {Villalonga}}, \bibinfo {author} {\bibfnamefont {C.~Vollgraff}\ \bibnamefont {Heidweiller}}, \bibinfo {author} {\bibfnamefont {S.}~\bibnamefont {Waltman}}, \bibinfo {author} {\bibfnamefont {S.~X.}\ \bibnamefont {Wang}}, \bibinfo {author} {\bibfnamefont {T.}~\bibnamefont {White}}, \bibinfo {author} {\bibfnamefont {K.}~\bibnamefont {Wong}}, \bibinfo {author} {\bibfnamefont {B.~W.~K.}\ \bibnamefont {Woo}}, \bibinfo {author} {\bibfnamefont {C.}~\bibnamefont {Xing}}, \bibinfo {author} {\bibfnamefont {Z.~Jamie}\ \bibnamefont {Yao}}, \bibinfo {author} {\bibfnamefont {P.}~\bibnamefont {Yeh}}, \bibinfo {author} {\bibfnamefont {B.}~\bibnamefont {Ying}}, \bibinfo {author} {\bibfnamefont
  {J.}~\bibnamefont {Yoo}}, \bibinfo {author} {\bibfnamefont {N.}~\bibnamefont {Yosri}}, \bibinfo {author} {\bibfnamefont {G.}~\bibnamefont {Young}}, \bibinfo {author} {\bibfnamefont {A.}~\bibnamefont {Zalcman}}, \bibinfo {author} {\bibfnamefont {N.}~\bibnamefont {Zhu}}, \bibinfo {author} {\bibfnamefont {N.}~\bibnamefont {Zobrist}}, \bibinfo {author} {\bibfnamefont {H.}~\bibnamefont {Neven}}, \bibinfo {author} {\bibfnamefont {R.}~\bibnamefont {Babbush}}, \bibinfo {author} {\bibfnamefont {S.}~\bibnamefont {Boixo}}, \bibinfo {author} {\bibfnamefont {J.}~\bibnamefont {Hilton}}, \bibinfo {author} {\bibfnamefont {E.}~\bibnamefont {Lucero}}, \bibinfo {author} {\bibfnamefont {A.}~\bibnamefont {Megrant}}, \bibinfo {author} {\bibfnamefont {J.}~\bibnamefont {Kelly}}, \bibinfo {author} {\bibfnamefont {Y.}~\bibnamefont {Chen}}, \bibinfo {author} {\bibfnamefont {V.}~\bibnamefont {Smelyanskiy}}, \bibinfo {author} {\bibfnamefont {G.}~\bibnamefont {Vidal}}, \bibinfo {author} {\bibfnamefont {P.}~\bibnamefont {Roushan}},
  \bibinfo {author} {\bibfnamefont {A.~M.}\ \bibnamefont {L{\"a}uchli}}, \bibinfo {author} {\bibfnamefont {D.~A.}\ \bibnamefont {Abanin}}, \ and\ \bibinfo {author} {\bibfnamefont {X.}~\bibnamefont {Mi}},\ }\bibfield  {title} {\enquote {\bibinfo {title} {Thermalization and criticality on an analogue--digital quantum simulator},}\ }\href {\doibase 10.1038/s41586-024-08460-3} {\bibfield  {journal} {\bibinfo  {journal} {Nature}\ }\textbf {\bibinfo {volume} {638}},\ \bibinfo {pages} {79--85} (\bibinfo {year} {2025})}\BibitemShut {NoStop}%
\bibitem [{\citenamefont {Dborin}\ \emph {et~al.}(2022)\citenamefont {Dborin}, \citenamefont {Wimalaweera}, \citenamefont {Barratt}, \citenamefont {Ostby}, \citenamefont {O'Brien},\ and\ \citenamefont {Green}}]{dborin:2022}%
  \BibitemOpen
  \bibfield  {author} {\bibinfo {author} {\bibfnamefont {James}\ \bibnamefont {Dborin}}, \bibinfo {author} {\bibfnamefont {Vinul}\ \bibnamefont {Wimalaweera}}, \bibinfo {author} {\bibfnamefont {F.}~\bibnamefont {Barratt}}, \bibinfo {author} {\bibfnamefont {Eric}\ \bibnamefont {Ostby}}, \bibinfo {author} {\bibfnamefont {Thomas~E.}\ \bibnamefont {O'Brien}}, \ and\ \bibinfo {author} {\bibfnamefont {A.~G.}\ \bibnamefont {Green}},\ }\bibfield  {title} {\enquote {\bibinfo {title} {Simulating groundstate and dynamical quantum phase transitions on a superconducting quantum computer},}\ }\href {\doibase 10.1038/s41467-022-33737-4} {\bibfield  {journal} {\bibinfo  {journal} {Nature Communications}\ }\textbf {\bibinfo {volume} {13}},\ \bibinfo {pages} {5977} (\bibinfo {year} {2022})}\BibitemShut {NoStop}%
\bibitem [{\citenamefont {Anand}\ \emph {et~al.}(2023)\citenamefont {Anand}, \citenamefont {Hauschild}, \citenamefont {Zhang}, \citenamefont {Potter},\ and\ \citenamefont {Zaletel}}]{anand:2023}%
  \BibitemOpen
  \bibfield  {author} {\bibinfo {author} {\bibfnamefont {Sajant}\ \bibnamefont {Anand}}, \bibinfo {author} {\bibfnamefont {Johannes}\ \bibnamefont {Hauschild}}, \bibinfo {author} {\bibfnamefont {Yuxuan}\ \bibnamefont {Zhang}}, \bibinfo {author} {\bibfnamefont {Andrew~C.}\ \bibnamefont {Potter}}, \ and\ \bibinfo {author} {\bibfnamefont {Michael~P.}\ \bibnamefont {Zaletel}},\ }\bibfield  {title} {\enquote {\bibinfo {title} {Holographic {{Quantum Simulation}} of {{Entanglement Renormalization Circuits}}},}\ }\href {\doibase 10.1103/PRXQuantum.4.030334} {\bibfield  {journal} {\bibinfo  {journal} {PRX Quantum}\ }\textbf {\bibinfo {volume} {4}},\ \bibinfo {pages} {030334} (\bibinfo {year} {2023})}\BibitemShut {NoStop}%
\bibitem [{\citenamefont {Haghshenas}\ \emph {et~al.}(2024)\citenamefont {Haghshenas}, \citenamefont {Chertkov}, \citenamefont {DeCross}, \citenamefont {Gatterman}, \citenamefont {Gerber}, \citenamefont {Gilmore}, \citenamefont {Gresh}, \citenamefont {Hewitt}, \citenamefont {Horst}, \citenamefont {Matheny}, \citenamefont {Mengle}, \citenamefont {Neyenhuis}, \citenamefont {Hayes},\ and\ \citenamefont {{Foss-Feig}}}]{haghshenas:2024}%
  \BibitemOpen
  \bibfield  {author} {\bibinfo {author} {\bibfnamefont {Reza}\ \bibnamefont {Haghshenas}}, \bibinfo {author} {\bibfnamefont {Eli}\ \bibnamefont {Chertkov}}, \bibinfo {author} {\bibfnamefont {Matthew}\ \bibnamefont {DeCross}}, \bibinfo {author} {\bibfnamefont {Thomas~M.}\ \bibnamefont {Gatterman}}, \bibinfo {author} {\bibfnamefont {Justin~A.}\ \bibnamefont {Gerber}}, \bibinfo {author} {\bibfnamefont {Kevin}\ \bibnamefont {Gilmore}}, \bibinfo {author} {\bibfnamefont {Dan}\ \bibnamefont {Gresh}}, \bibinfo {author} {\bibfnamefont {Nathan}\ \bibnamefont {Hewitt}}, \bibinfo {author} {\bibfnamefont {Chandler~V.}\ \bibnamefont {Horst}}, \bibinfo {author} {\bibfnamefont {Mitchell}\ \bibnamefont {Matheny}}, \bibinfo {author} {\bibfnamefont {Tanner}\ \bibnamefont {Mengle}}, \bibinfo {author} {\bibfnamefont {Brian}\ \bibnamefont {Neyenhuis}}, \bibinfo {author} {\bibfnamefont {David}\ \bibnamefont {Hayes}}, \ and\ \bibinfo {author} {\bibfnamefont {Michael}\ \bibnamefont {{Foss-Feig}}},\ }\bibfield  {title} {\enquote
  {\bibinfo {title} {Probing {{Critical States}} of {{Matter}} on a {{Digital Quantum Computer}}},}\ }\href {\doibase 10.1103/PhysRevLett.133.266502} {\bibfield  {journal} {\bibinfo  {journal} {Physical Review Letters}\ }\textbf {\bibinfo {volume} {133}},\ \bibinfo {pages} {266502} (\bibinfo {year} {2024})}\BibitemShut {NoStop}%
\bibitem [{\citenamefont {Fang}\ \emph {et~al.}(2024)\citenamefont {Fang}, \citenamefont {Wang}, \citenamefont {Liu}, \citenamefont {Wang}, \citenamefont {Cimmino}, \citenamefont {Wei}, \citenamefont {Bintz}, \citenamefont {Parr}, \citenamefont {Kemp}, \citenamefont {Ni},\ and\ \citenamefont {Yao}}]{fang2024probing}%
  \BibitemOpen
  \bibfield  {author} {\bibinfo {author} {\bibfnamefont {Fang}\ \bibnamefont {Fang}}, \bibinfo {author} {\bibfnamefont {Kenneth}\ \bibnamefont {Wang}}, \bibinfo {author} {\bibfnamefont {Vincent~S.}\ \bibnamefont {Liu}}, \bibinfo {author} {\bibfnamefont {Yu}~\bibnamefont {Wang}}, \bibinfo {author} {\bibfnamefont {Ryan}\ \bibnamefont {Cimmino}}, \bibinfo {author} {\bibfnamefont {Julia}\ \bibnamefont {Wei}}, \bibinfo {author} {\bibfnamefont {Marcus}\ \bibnamefont {Bintz}}, \bibinfo {author} {\bibfnamefont {Avery}\ \bibnamefont {Parr}}, \bibinfo {author} {\bibfnamefont {Jack}\ \bibnamefont {Kemp}}, \bibinfo {author} {\bibfnamefont {Kang-Kuen}\ \bibnamefont {Ni}}, \ and\ \bibinfo {author} {\bibfnamefont {Norman~Y.}\ \bibnamefont {Yao}},\ }\bibfield  {title} {\enquote {\bibinfo {title} {Probing critical phenomena in open quantum systems using atom arrays},}\ }\href {\doibase 10.48550/arXiv.2402.15376} {\  (\bibinfo {year} {2024}),\ 10.48550/arXiv.2402.15376}\BibitemShut {NoStop}%
\bibitem [{\citenamefont {Heyl}\ \emph {et~al.}(2013)\citenamefont {Heyl}, \citenamefont {Polkovnikov},\ and\ \citenamefont {Kehrein}}]{heyl:2013}%
  \BibitemOpen
  \bibfield  {author} {\bibinfo {author} {\bibfnamefont {Markus}\ \bibnamefont {Heyl}}, \bibinfo {author} {\bibfnamefont {Anatoli}\ \bibnamefont {Polkovnikov}}, \ and\ \bibinfo {author} {\bibfnamefont {Stefan}\ \bibnamefont {Kehrein}},\ }\bibfield  {title} {\enquote {\bibinfo {title} {Dynamical {{Quantum Phase Transitions}} in the {{Transverse Field Ising Model}}},}\ }\href {\doibase 10.1103/PhysRevLett.110.135704} {\bibfield  {journal} {\bibinfo  {journal} {Physical Review Letters}\ }\textbf {\bibinfo {volume} {110}},\ \bibinfo {pages} {135704} (\bibinfo {year} {2013})}\BibitemShut {NoStop}%
\bibitem [{\citenamefont {Heyl}(2014)}]{heyl:2014}%
  \BibitemOpen
  \bibfield  {author} {\bibinfo {author} {\bibfnamefont {Markus}\ \bibnamefont {Heyl}},\ }\bibfield  {title} {\enquote {\bibinfo {title} {Dynamical quantum phase transitions in systems with broken-symmetry phases},}\ }\href {\doibase 10.1103/PhysRevLett.113.205701} {\bibfield  {journal} {\bibinfo  {journal} {Physical Review Letters}\ } (\bibinfo {year} {2014}),\ 10.1103/PhysRevLett.113.205701}\BibitemShut {NoStop}%
\bibitem [{\citenamefont {Heyl}(2015)}]{heyl:2015}%
  \BibitemOpen
  \bibfield  {author} {\bibinfo {author} {\bibfnamefont {Markus}\ \bibnamefont {Heyl}},\ }\bibfield  {title} {\enquote {\bibinfo {title} {Scaling and {{Universality}} at {{Dynamical Quantum Phase Transitions}}},}\ }\href {\doibase 10.1103/PhysRevLett.115.140602} {\bibfield  {journal} {\bibinfo  {journal} {Physical Review Letters}\ }\textbf {\bibinfo {volume} {115}},\ \bibinfo {pages} {140602} (\bibinfo {year} {2015})}\BibitemShut {NoStop}%
\bibitem [{\citenamefont {Heyl}(2016)}]{heyl:2016}%
  \BibitemOpen
  \bibfield  {author} {\bibinfo {author} {\bibfnamefont {Markus}\ \bibnamefont {Heyl}},\ }\bibfield  {title} {\enquote {\bibinfo {title} {Quenching a {{Quantum Critical State}} by the {{Order Parameter}}: {{Dynamical Quantum Phase Transitions}} and {{Quantum Speed Limits}}},}\ }\href {\doibase 10.1103/PhysRevB.95.060504} {\bibfield  {journal} {\bibinfo  {journal} {Physical Review B}\ } (\bibinfo {year} {2016}),\ 10.1103/PhysRevB.95.060504}\BibitemShut {NoStop}%
\bibitem [{\citenamefont {Zunkovic}\ \emph {et~al.}(2016)\citenamefont {Zunkovic}, \citenamefont {Heyl}, \citenamefont {Knap},\ and\ \citenamefont {Silva}}]{zunkovic:2016}%
  \BibitemOpen
  \bibfield  {author} {\bibinfo {author} {\bibfnamefont {Bojan}\ \bibnamefont {Zunkovic}}, \bibinfo {author} {\bibfnamefont {Markus}\ \bibnamefont {Heyl}}, \bibinfo {author} {\bibfnamefont {Michael}\ \bibnamefont {Knap}}, \ and\ \bibinfo {author} {\bibfnamefont {Alessandro}\ \bibnamefont {Silva}},\ }\bibfield  {title} {\enquote {\bibinfo {title} {Dynamical {{Quantum Phase Transitions}} in {{Spin Chains}} with {{Long-Range Interactions}}: {{Merging}} different concepts of non-equilibrium criticality},}\ }\href {\doibase 10.1103/PhysRevLett.120.130601} {\bibfield  {journal} {\bibinfo  {journal} {Physical Review Letters}\ } (\bibinfo {year} {2016}),\ 10.1103/PhysRevLett.120.130601}\BibitemShut {NoStop}%
\bibitem [{\citenamefont {Karrasch}\ and\ \citenamefont {Schuricht}(2017)}]{karrasch:2017}%
  \BibitemOpen
  \bibfield  {author} {\bibinfo {author} {\bibfnamefont {C.}~\bibnamefont {Karrasch}}\ and\ \bibinfo {author} {\bibfnamefont {D.}~\bibnamefont {Schuricht}},\ }\bibfield  {title} {\enquote {\bibinfo {title} {Dynamical quantum phase transitions in the quantum {{Potts}} chain},}\ }\href {\doibase 10.1103/PhysRevB.95.075143} {\bibfield  {journal} {\bibinfo  {journal} {Physical Review B}\ }\textbf {\bibinfo {volume} {95}},\ \bibinfo {pages} {075143} (\bibinfo {year} {2017})}\BibitemShut {NoStop}%
\bibitem [{\citenamefont {Titum}\ \emph {et~al.}(2019)\citenamefont {Titum}, \citenamefont {Iosue}, \citenamefont {Garrison}, \citenamefont {Gorshkov},\ and\ \citenamefont {Gong}}]{titum:2019}%
  \BibitemOpen
  \bibfield  {author} {\bibinfo {author} {\bibfnamefont {Paraj}\ \bibnamefont {Titum}}, \bibinfo {author} {\bibfnamefont {Joseph~T.}\ \bibnamefont {Iosue}}, \bibinfo {author} {\bibfnamefont {James~R.}\ \bibnamefont {Garrison}}, \bibinfo {author} {\bibfnamefont {Alexey~V.}\ \bibnamefont {Gorshkov}}, \ and\ \bibinfo {author} {\bibfnamefont {Zhe-Xuan}\ \bibnamefont {Gong}},\ }\bibfield  {title} {\enquote {\bibinfo {title} {Probing ground-state phase transitions through quench dynamics},}\ }\href {\doibase 10.1103/PhysRevLett.123.115701} {\bibfield  {journal} {\bibinfo  {journal} {Physical Review Letters}\ }\textbf {\bibinfo {volume} {123}},\ \bibinfo {pages} {115701} (\bibinfo {year} {2019})}\BibitemShut {NoStop}%
\bibitem [{\citenamefont {Titum}\ and\ \citenamefont {Maghrebi}(2020)}]{titum:2020}%
  \BibitemOpen
  \bibfield  {author} {\bibinfo {author} {\bibfnamefont {Paraj}\ \bibnamefont {Titum}}\ and\ \bibinfo {author} {\bibfnamefont {Mohammad~F.}\ \bibnamefont {Maghrebi}},\ }\bibfield  {title} {\enquote {\bibinfo {title} {Nonequilibrium {{Criticality}} in {{Quench Dynamics}} of {{Long-Range Spin Models}}},}\ }\href {\doibase 10.1103/PhysRevLett.125.040602} {\bibfield  {journal} {\bibinfo  {journal} {Physical Review Letters}\ }\textbf {\bibinfo {volume} {125}},\ \bibinfo {pages} {040602} (\bibinfo {year} {2020})}\BibitemShut {NoStop}%
\bibitem [{\citenamefont {Halimeh}\ \emph {et~al.}(2020)\citenamefont {Halimeh}, \citenamefont {Trapin}, \citenamefont {Van~Damme},\ and\ \citenamefont {Heyl}}]{halimeh:2020}%
  \BibitemOpen
  \bibfield  {author} {\bibinfo {author} {\bibfnamefont {Jad~C.}\ \bibnamefont {Halimeh}}, \bibinfo {author} {\bibfnamefont {Daniele}\ \bibnamefont {Trapin}}, \bibinfo {author} {\bibfnamefont {Maarten}\ \bibnamefont {Van~Damme}}, \ and\ \bibinfo {author} {\bibfnamefont {Markus}\ \bibnamefont {Heyl}},\ }\bibfield  {title} {\enquote {\bibinfo {title} {Local measures of dynamical quantum phase transitions},}\ }\href {\doibase 10.1103/PhysRevB.104.075130} {\bibfield  {journal} {\bibinfo  {journal} {Physical Review B}\ } (\bibinfo {year} {2020}),\ 10.1103/PhysRevB.104.075130}\BibitemShut {NoStop}%
\bibitem [{\citenamefont {Rossini}\ and\ \citenamefont {Vicari}(2020)}]{rossini:2020}%
  \BibitemOpen
  \bibfield  {author} {\bibinfo {author} {\bibfnamefont {Davide}\ \bibnamefont {Rossini}}\ and\ \bibinfo {author} {\bibfnamefont {Ettore}\ \bibnamefont {Vicari}},\ }\bibfield  {title} {\enquote {\bibinfo {title} {Dynamics after quenches in one-dimensional quantum {{Ising-like}} systems},}\ }\href {\doibase 10.1103/PhysRevB.102.054444} {\bibfield  {journal} {\bibinfo  {journal} {Physical Review B}\ }\textbf {\bibinfo {volume} {102}},\ \bibinfo {pages} {054444} (\bibinfo {year} {2020})}\BibitemShut {NoStop}%
\bibitem [{\citenamefont {Da{\u g}}\ \emph {et~al.}(2023{\natexlab{a}})\citenamefont {Da{\u g}}, \citenamefont {Wang}, \citenamefont {Uhrich}, \citenamefont {Na},\ and\ \citenamefont {Halimeh}}]{dag:2023}%
  \BibitemOpen
  \bibfield  {author} {\bibinfo {author} {\bibfnamefont {Ceren~B.}\ \bibnamefont {Da{\u g}}}, \bibinfo {author} {\bibfnamefont {Yidan}\ \bibnamefont {Wang}}, \bibinfo {author} {\bibfnamefont {Philipp}\ \bibnamefont {Uhrich}}, \bibinfo {author} {\bibfnamefont {Xuesen}\ \bibnamefont {Na}}, \ and\ \bibinfo {author} {\bibfnamefont {Jad~C.}\ \bibnamefont {Halimeh}},\ }\bibfield  {title} {\enquote {\bibinfo {title} {Critical slowing down in sudden quench dynamics},}\ }\href {\doibase 10.1103/PhysRevB.107.L121113} {\bibfield  {journal} {\bibinfo  {journal} {Physical Review B}\ }\textbf {\bibinfo {volume} {107}},\ \bibinfo {pages} {L121113} (\bibinfo {year} {2023}{\natexlab{a}})}\BibitemShut {NoStop}%
\bibitem [{\citenamefont {Da{\u g}}\ \emph {et~al.}(2023{\natexlab{b}})\citenamefont {Da{\u g}}, \citenamefont {Uhrich}, \citenamefont {Wang}, \citenamefont {McCulloch},\ and\ \citenamefont {Halimeh}}]{dag:2023a}%
  \BibitemOpen
  \bibfield  {author} {\bibinfo {author} {\bibfnamefont {Ceren~B.}\ \bibnamefont {Da{\u g}}}, \bibinfo {author} {\bibfnamefont {Philipp}\ \bibnamefont {Uhrich}}, \bibinfo {author} {\bibfnamefont {Yidan}\ \bibnamefont {Wang}}, \bibinfo {author} {\bibfnamefont {Ian~P.}\ \bibnamefont {McCulloch}}, \ and\ \bibinfo {author} {\bibfnamefont {Jad~C.}\ \bibnamefont {Halimeh}},\ }\bibfield  {title} {\enquote {\bibinfo {title} {Detecting quantum phase transitions in the quasistationary regime of {{Ising}} chains},}\ }\href {\doibase 10.1103/PhysRevB.107.094432} {\bibfield  {journal} {\bibinfo  {journal} {Physical Review B}\ }\textbf {\bibinfo {volume} {107}},\ \bibinfo {pages} {094432} (\bibinfo {year} {2023}{\natexlab{b}})}\BibitemShut {NoStop}%
\bibitem [{\citenamefont {Robertson}\ \emph {et~al.}(2023)\citenamefont {Robertson}, \citenamefont {Senese},\ and\ \citenamefont {Essler}}]{robertson:2023}%
  \BibitemOpen
  \bibfield  {author} {\bibinfo {author} {\bibfnamefont {Jacob}\ \bibnamefont {Robertson}}, \bibinfo {author} {\bibfnamefont {Riccardo}\ \bibnamefont {Senese}}, \ and\ \bibinfo {author} {\bibfnamefont {Fabian}\ \bibnamefont {Essler}},\ }\bibfield  {title} {\enquote {\bibinfo {title} {A simple theory for quantum quenches in the {{ANNNI}} model},}\ }\href {\doibase 10.21468/SciPostPhys.15.1.032} {\bibfield  {journal} {\bibinfo  {journal} {SciPost Physics}\ }\textbf {\bibinfo {volume} {15}},\ \bibinfo {pages} {032} (\bibinfo {year} {2023})}\BibitemShut {NoStop}%
\bibitem [{\citenamefont {Bandyopadhyay}\ \emph {et~al.}(2023)\citenamefont {Bandyopadhyay}, \citenamefont {Polkovnikov},\ and\ \citenamefont {Dutta}}]{bandyopadhyay:2023}%
  \BibitemOpen
  \bibfield  {author} {\bibinfo {author} {\bibfnamefont {Souvik}\ \bibnamefont {Bandyopadhyay}}, \bibinfo {author} {\bibfnamefont {Anatoli}\ \bibnamefont {Polkovnikov}}, \ and\ \bibinfo {author} {\bibfnamefont {Amit}\ \bibnamefont {Dutta}},\ }\bibfield  {title} {\enquote {\bibinfo {title} {Late-time critical behavior of local string-like observables under quantum quenches},}\ }\href {\doibase 10.1103/PhysRevB.107.064105} {\bibfield  {journal} {\bibinfo  {journal} {Physical Review B}\ }\textbf {\bibinfo {volume} {107}},\ \bibinfo {pages} {064105} (\bibinfo {year} {2023})},\ \Eprint {http://arxiv.org/abs/2205.00201} {arXiv:2205.00201 [cond-mat]} \BibitemShut {NoStop}%
\bibitem [{\citenamefont {Calabrese}\ and\ \citenamefont {Cardy}(2016)}]{calabrese2016quantum}%
  \BibitemOpen
  \bibfield  {author} {\bibinfo {author} {\bibfnamefont {Pasquale}\ \bibnamefont {Calabrese}}\ and\ \bibinfo {author} {\bibfnamefont {John}\ \bibnamefont {Cardy}},\ }\bibfield  {title} {\enquote {\bibinfo {title} {Quantum quenches in 1 + 1 dimensional conformal field theories},}\ }\href {\doibase 10.1088/1742-5468/2016/06/064003} {\bibfield  {journal} {\bibinfo  {journal} {Journal of Statistical Mechanics: Theory and Experiment}\ }\textbf {\bibinfo {volume} {2016}},\ \bibinfo {pages} {064003} (\bibinfo {year} {2016})}\BibitemShut {NoStop}%
\bibitem [{SM()}]{SM}%
  \BibitemOpen
  \href@noop {} {}\bibinfo {note} {Please see the Supplementary Materials for more discussion on timescales, convergence of numerical methods, and details of the conformal perturbation theory analysis.}\BibitemShut {Stop}%
\bibitem [{\citenamefont {Calabrese}\ and\ \citenamefont {Cardy}(2005)}]{calabrese2005evolution}%
  \BibitemOpen
  \bibfield  {author} {\bibinfo {author} {\bibfnamefont {Pasquale}\ \bibnamefont {Calabrese}}\ and\ \bibinfo {author} {\bibfnamefont {John}\ \bibnamefont {Cardy}},\ }\bibfield  {title} {\enquote {\bibinfo {title} {Evolution of entanglement entropy in one-dimensional systems},}\ }\href {\doibase 10.1088/1742-5468/2005/04/P04010} {\bibfield  {journal} {\bibinfo  {journal} {Journal of Statistical Mechanics: Theory and Experiment}\ }\textbf {\bibinfo {volume} {2005}},\ \bibinfo {pages} {P04010} (\bibinfo {year} {2005})}\BibitemShut {NoStop}%
\bibitem [{\citenamefont {Calabrese}\ and\ \citenamefont {Cardy}(2006)}]{calabrese2006time}%
  \BibitemOpen
  \bibfield  {author} {\bibinfo {author} {\bibfnamefont {Pasquale}\ \bibnamefont {Calabrese}}\ and\ \bibinfo {author} {\bibfnamefont {John}\ \bibnamefont {Cardy}},\ }\bibfield  {title} {\enquote {\bibinfo {title} {Time {{Dependence}} of {{Correlation Functions Following}} a {{Quantum Quench}}},}\ }\href {\doibase 10.1103/PhysRevLett.96.136801} {\bibfield  {journal} {\bibinfo  {journal} {Physical Review Letters}\ }\textbf {\bibinfo {volume} {96}},\ \bibinfo {pages} {136801} (\bibinfo {year} {2006})}\BibitemShut {NoStop}%
\bibitem [{\citenamefont {Calabrese}\ and\ \citenamefont {Cardy}(2007)}]{Calabrese_Cardy_2007}%
  \BibitemOpen
  \bibfield  {author} {\bibinfo {author} {\bibfnamefont {Pasquale}\ \bibnamefont {Calabrese}}\ and\ \bibinfo {author} {\bibfnamefont {John}\ \bibnamefont {Cardy}},\ }\bibfield  {title} {\enquote {\bibinfo {title} {Quantum quenches in extended systems},}\ }\href {\doibase 10.1088/1742-5468/2007/06/P06008} {\bibfield  {journal} {\bibinfo  {journal} {Journal of Statistical Mechanics: Theory and Experiment}\ }\textbf {\bibinfo {volume} {2007}},\ \bibinfo {pages} {P06008–P06008} (\bibinfo {year} {2007})}\BibitemShut {NoStop}%
\bibitem [{\citenamefont {Cardy}(2016)}]{cardy2016furtherresults}%
  \BibitemOpen
  \bibfield  {author} {\bibinfo {author} {\bibfnamefont {John}\ \bibnamefont {Cardy}},\ }\bibfield  {title} {\enquote {\bibinfo {title} {Quantum quenches to a critical point in one dimension: Some further results},}\ }\href {\doibase 10.1088/1742-5468/2016/02/023103} {\bibfield  {journal} {\bibinfo  {journal} {Journal of Statistical Mechanics: Theory and Experiment}\ }\textbf {\bibinfo {volume} {2016}},\ \bibinfo {pages} {023103} (\bibinfo {year} {2016})}\BibitemShut {NoStop}%
\bibitem [{\citenamefont {Kim}\ and\ \citenamefont {Huse}(2013)}]{kim:2013}%
  \BibitemOpen
  \bibfield  {author} {\bibinfo {author} {\bibfnamefont {Hyungwon}\ \bibnamefont {Kim}}\ and\ \bibinfo {author} {\bibfnamefont {David~A.}\ \bibnamefont {Huse}},\ }\bibfield  {title} {\enquote {\bibinfo {title} {Ballistic {{Spreading}} of {{Entanglement}} in a {{Diffusive Nonintegrable System}}},}\ }\href {\doibase 10.1103/PhysRevLett.111.127205} {\bibfield  {journal} {\bibinfo  {journal} {Physical Review Letters}\ }\textbf {\bibinfo {volume} {111}},\ \bibinfo {pages} {127205} (\bibinfo {year} {2013})}\BibitemShut {NoStop}%
\bibitem [{\citenamefont {Bonnes}\ \emph {et~al.}(2014)\citenamefont {Bonnes}, \citenamefont {Essler},\ and\ \citenamefont {L{\"a}uchli}}]{bonnes:2014}%
  \BibitemOpen
  \bibfield  {author} {\bibinfo {author} {\bibfnamefont {Lars}\ \bibnamefont {Bonnes}}, \bibinfo {author} {\bibfnamefont {Fabian H.~L.}\ \bibnamefont {Essler}}, \ and\ \bibinfo {author} {\bibfnamefont {Andreas~M.}\ \bibnamefont {L{\"a}uchli}},\ }\bibfield  {title} {\enquote {\bibinfo {title} {``{{Light-Cone}}'' {{Dynamics After Quantum Quenches}} in {{Spin Chains}}},}\ }\href {\doibase 10.1103/PhysRevLett.113.187203} {\bibfield  {journal} {\bibinfo  {journal} {Physical Review Letters}\ }\textbf {\bibinfo {volume} {113}},\ \bibinfo {pages} {187203} (\bibinfo {year} {2014})}\BibitemShut {NoStop}%
\bibitem [{\citenamefont {Cotler}\ \emph {et~al.}(2016)\citenamefont {Cotler}, \citenamefont {Hertzberg}, \citenamefont {Mezei},\ and\ \citenamefont {Mueller}}]{cotler:2016}%
  \BibitemOpen
  \bibfield  {author} {\bibinfo {author} {\bibfnamefont {Jordan~S.}\ \bibnamefont {Cotler}}, \bibinfo {author} {\bibfnamefont {Mark~P.}\ \bibnamefont {Hertzberg}}, \bibinfo {author} {\bibfnamefont {M{\'a}rk}\ \bibnamefont {Mezei}}, \ and\ \bibinfo {author} {\bibfnamefont {Mark~T.}\ \bibnamefont {Mueller}},\ }\bibfield  {title} {\enquote {\bibinfo {title} {Entanglement {{Growth}} after a {{Global Quench}} in {{Free Scalar Field Theory}}},}\ }\href {\doibase 10.1007/JHEP11(2016)166} {\bibfield  {journal} {\bibinfo  {journal} {Journal of High Energy Physics}\ }\textbf {\bibinfo {volume} {2016}},\ \bibinfo {pages} {166} (\bibinfo {year} {2016})}\BibitemShut {NoStop}%
\bibitem [{\citenamefont {Alba}\ and\ \citenamefont {Calabrese}(2018)}]{alba:2018}%
  \BibitemOpen
  \bibfield  {author} {\bibinfo {author} {\bibfnamefont {Vincenzo}\ \bibnamefont {Alba}}\ and\ \bibinfo {author} {\bibfnamefont {Pasquale}\ \bibnamefont {Calabrese}},\ }\bibfield  {title} {\enquote {\bibinfo {title} {Entanglement dynamics after quantum quenches in generic integrable systems},}\ }\href {\doibase 10.21468/SciPostPhys.4.3.017} {\bibfield  {journal} {\bibinfo  {journal} {SciPost Physics}\ }\textbf {\bibinfo {volume} {4}},\ \bibinfo {pages} {017} (\bibinfo {year} {2018})}\BibitemShut {NoStop}%
\bibitem [{\citenamefont {Bastianello}\ and\ \citenamefont {Calabrese}(2018)}]{bastianello:2018}%
  \BibitemOpen
  \bibfield  {author} {\bibinfo {author} {\bibfnamefont {Alvise}\ \bibnamefont {Bastianello}}\ and\ \bibinfo {author} {\bibfnamefont {Pasquale}\ \bibnamefont {Calabrese}},\ }\bibfield  {title} {\enquote {\bibinfo {title} {Spreading of entanglement and correlations after a quench with intertwined quasiparticles},}\ }\href {\doibase 10.21468/SciPostPhys.5.4.033} {\bibfield  {journal} {\bibinfo  {journal} {SciPost Physics}\ }\textbf {\bibinfo {volume} {5}},\ \bibinfo {pages} {033} (\bibinfo {year} {2018})}\BibitemShut {NoStop}%
\bibitem [{\citenamefont {Bertini}\ \emph {et~al.}(2018)\citenamefont {Bertini}, \citenamefont {Fagotti}, \citenamefont {Piroli},\ and\ \citenamefont {Calabrese}}]{bertini:2018}%
  \BibitemOpen
  \bibfield  {author} {\bibinfo {author} {\bibfnamefont {Bruno}\ \bibnamefont {Bertini}}, \bibinfo {author} {\bibfnamefont {Maurizio}\ \bibnamefont {Fagotti}}, \bibinfo {author} {\bibfnamefont {Lorenzo}\ \bibnamefont {Piroli}}, \ and\ \bibinfo {author} {\bibfnamefont {Pasquale}\ \bibnamefont {Calabrese}},\ }\bibfield  {title} {\enquote {\bibinfo {title} {Entanglement evolution and generalised hydrodynamics: Noninteracting systems},}\ }\href {\doibase 10.1088/1751-8121/aad82e} {\bibfield  {journal} {\bibinfo  {journal} {Journal of Physics A: Mathematical and Theoretical}\ }\textbf {\bibinfo {volume} {51}},\ \bibinfo {pages} {39LT01} (\bibinfo {year} {2018})}\BibitemShut {NoStop}%
\bibitem [{\citenamefont {Cevolani}\ \emph {et~al.}(2018)\citenamefont {Cevolani}, \citenamefont {Despres}, \citenamefont {Carleo}, \citenamefont {Tagliacozzo},\ and\ \citenamefont {{Sanchez-Palencia}}}]{cevolani:2018}%
  \BibitemOpen
  \bibfield  {author} {\bibinfo {author} {\bibfnamefont {Lorenzo}\ \bibnamefont {Cevolani}}, \bibinfo {author} {\bibfnamefont {Julien}\ \bibnamefont {Despres}}, \bibinfo {author} {\bibfnamefont {Giuseppe}\ \bibnamefont {Carleo}}, \bibinfo {author} {\bibfnamefont {Luca}\ \bibnamefont {Tagliacozzo}}, \ and\ \bibinfo {author} {\bibfnamefont {Laurent}\ \bibnamefont {{Sanchez-Palencia}}},\ }\bibfield  {title} {\enquote {\bibinfo {title} {Universal scaling laws for correlation spreading in quantum systems with short- and long-range interactions},}\ }\href {\doibase 10.1103/PhysRevB.98.024302} {\bibfield  {journal} {\bibinfo  {journal} {Physical Review B}\ }\textbf {\bibinfo {volume} {98}},\ \bibinfo {pages} {024302} (\bibinfo {year} {2018})}\BibitemShut {NoStop}%
\bibitem [{\citenamefont {{von Keyserlingk}}\ \emph {et~al.}(2018)\citenamefont {{von Keyserlingk}}, \citenamefont {Rakovszky}, \citenamefont {Pollmann},\ and\ \citenamefont {Sondhi}}]{vonkeyserlingk:2018a}%
  \BibitemOpen
  \bibfield  {author} {\bibinfo {author} {\bibfnamefont {C.~W.}\ \bibnamefont {{von Keyserlingk}}}, \bibinfo {author} {\bibfnamefont {Tibor}\ \bibnamefont {Rakovszky}}, \bibinfo {author} {\bibfnamefont {Frank}\ \bibnamefont {Pollmann}}, \ and\ \bibinfo {author} {\bibfnamefont {S.~L.}\ \bibnamefont {Sondhi}},\ }\bibfield  {title} {\enquote {\bibinfo {title} {Operator {{Hydrodynamics}}, {{OTOCs}}, and {{Entanglement Growth}} in {{Systems}} without {{Conservation Laws}}},}\ }\href {\doibase 10.1103/PhysRevX.8.021013} {\bibfield  {journal} {\bibinfo  {journal} {Physical Review X}\ }\textbf {\bibinfo {volume} {8}},\ \bibinfo {pages} {021013} (\bibinfo {year} {2018})}\BibitemShut {NoStop}%
\bibitem [{\citenamefont {Najafi}\ \emph {et~al.}(2018)\citenamefont {Najafi}, \citenamefont {Rajabpour},\ and\ \citenamefont {Viti}}]{najafi:2018}%
  \BibitemOpen
  \bibfield  {author} {\bibinfo {author} {\bibfnamefont {K.}~\bibnamefont {Najafi}}, \bibinfo {author} {\bibfnamefont {M.~A.}\ \bibnamefont {Rajabpour}}, \ and\ \bibinfo {author} {\bibfnamefont {J.}~\bibnamefont {Viti}},\ }\bibfield  {title} {\enquote {\bibinfo {title} {Light-cone velocities after a global quench in a noninteracting model},}\ }\href {\doibase 10.1103/PhysRevB.97.205103} {\bibfield  {journal} {\bibinfo  {journal} {Physical Review B}\ }\textbf {\bibinfo {volume} {97}},\ \bibinfo {pages} {205103} (\bibinfo {year} {2018})}\BibitemShut {NoStop}%
\bibitem [{\citenamefont {Bertini}\ \emph {et~al.}(2019)\citenamefont {Bertini}, \citenamefont {Kos},\ and\ \citenamefont {Prosen}}]{bertini:2019}%
  \BibitemOpen
  \bibfield  {author} {\bibinfo {author} {\bibfnamefont {Bruno}\ \bibnamefont {Bertini}}, \bibinfo {author} {\bibfnamefont {Pavel}\ \bibnamefont {Kos}}, \ and\ \bibinfo {author} {\bibfnamefont {Toma{\v z}}\ \bibnamefont {Prosen}},\ }\bibfield  {title} {\enquote {\bibinfo {title} {Entanglement {{Spreading}} in a {{Minimal Model}} of {{Maximal Many-Body Quantum Chaos}}},}\ }\href {\doibase 10.1103/PhysRevX.9.021033} {\bibfield  {journal} {\bibinfo  {journal} {Physical Review X}\ }\textbf {\bibinfo {volume} {9}},\ \bibinfo {pages} {021033} (\bibinfo {year} {2019})}\BibitemShut {NoStop}%
\bibitem [{\citenamefont {Cao}\ \emph {et~al.}(2019)\citenamefont {Cao}, \citenamefont {Tilloy},\ and\ \citenamefont {De~Luca}}]{cao:2019}%
  \BibitemOpen
  \bibfield  {author} {\bibinfo {author} {\bibfnamefont {Xiangyu}\ \bibnamefont {Cao}}, \bibinfo {author} {\bibfnamefont {Antoine}\ \bibnamefont {Tilloy}}, \ and\ \bibinfo {author} {\bibfnamefont {Andrea}\ \bibnamefont {De~Luca}},\ }\bibfield  {title} {\enquote {\bibinfo {title} {Entanglement in a fermion chain under continuous monitoring},}\ }\href {\doibase 10.21468/SciPostPhys.7.2.024} {\bibfield  {journal} {\bibinfo  {journal} {SciPost Physics}\ }\textbf {\bibinfo {volume} {7}},\ \bibinfo {pages} {024} (\bibinfo {year} {2019})}\BibitemShut {NoStop}%
\bibitem [{\citenamefont {Calabrese}(2020)}]{calabrese:2020}%
  \BibitemOpen
  \bibfield  {author} {\bibinfo {author} {\bibfnamefont {Pasquale}\ \bibnamefont {Calabrese}},\ }\bibfield  {title} {\enquote {\bibinfo {title} {Entanglement spreading in non-equilibrium integrable systems},}\ }\href {\doibase 10.21468/SciPostPhysLectNotes.20} {\bibfield  {journal} {\bibinfo  {journal} {SciPost Physics Lecture Notes}\ ,\ \bibinfo {pages} {020}} (\bibinfo {year} {2020})}\BibitemShut {NoStop}%
\bibitem [{\citenamefont {Klobas}\ and\ \citenamefont {Bertini}(2021)}]{klobas:2021}%
  \BibitemOpen
  \bibfield  {author} {\bibinfo {author} {\bibfnamefont {Katja}\ \bibnamefont {Klobas}}\ and\ \bibinfo {author} {\bibfnamefont {Bruno}\ \bibnamefont {Bertini}},\ }\bibfield  {title} {\enquote {\bibinfo {title} {Entanglement dynamics in {{Rule}} 54: {{Exact}} results and quasiparticle picture},}\ }\href {\doibase 10.21468/SciPostPhys.11.6.107} {\bibfield  {journal} {\bibinfo  {journal} {SciPost Physics}\ }\textbf {\bibinfo {volume} {11}},\ \bibinfo {pages} {107} (\bibinfo {year} {2021})}\BibitemShut {NoStop}%
\bibitem [{\citenamefont {Schneider}\ \emph {et~al.}(2021)\citenamefont {Schneider}, \citenamefont {Despres}, \citenamefont {Thomson}, \citenamefont {Tagliacozzo},\ and\ \citenamefont {{Sanchez-Palencia}}}]{schneider:2021}%
  \BibitemOpen
  \bibfield  {author} {\bibinfo {author} {\bibfnamefont {J.~T.}\ \bibnamefont {Schneider}}, \bibinfo {author} {\bibfnamefont {J.}~\bibnamefont {Despres}}, \bibinfo {author} {\bibfnamefont {S.~J.}\ \bibnamefont {Thomson}}, \bibinfo {author} {\bibfnamefont {L.}~\bibnamefont {Tagliacozzo}}, \ and\ \bibinfo {author} {\bibfnamefont {L.}~\bibnamefont {{Sanchez-Palencia}}},\ }\bibfield  {title} {\enquote {\bibinfo {title} {Spreading of correlations and entanglement in the long-range transverse {{Ising}} chain},}\ }\href {\doibase 10.1103/PhysRevResearch.3.L012022} {\bibfield  {journal} {\bibinfo  {journal} {Physical Review Research}\ }\textbf {\bibinfo {volume} {3}},\ \bibinfo {pages} {L012022} (\bibinfo {year} {2021})}\BibitemShut {NoStop}%
\bibitem [{\citenamefont {Rottoli}\ \emph {et~al.}(2025)\citenamefont {Rottoli}, \citenamefont {Rylands},\ and\ \citenamefont {Calabrese}}]{rottoli:2025}%
  \BibitemOpen
  \bibfield  {author} {\bibinfo {author} {\bibfnamefont {Federico}\ \bibnamefont {Rottoli}}, \bibinfo {author} {\bibfnamefont {Colin}\ \bibnamefont {Rylands}}, \ and\ \bibinfo {author} {\bibfnamefont {Pasquale}\ \bibnamefont {Calabrese}},\ }\bibfield  {title} {\enquote {\bibinfo {title} {Entanglement {{Hamiltonians}} and the quasiparticle picture},}\ }\href {\doibase 10.1103/PhysRevB.111.L140302} {\bibfield  {journal} {\bibinfo  {journal} {Physical Review B}\ }\textbf {\bibinfo {volume} {111}},\ \bibinfo {pages} {L140302} (\bibinfo {year} {2025})}\BibitemShut {NoStop}%
\bibitem [{\citenamefont {Cheneau}\ \emph {et~al.}(2012)\citenamefont {Cheneau}, \citenamefont {Barmettler}, \citenamefont {Poletti}, \citenamefont {Endres}, \citenamefont {Schau{\ss}}, \citenamefont {Fukuhara}, \citenamefont {Gross}, \citenamefont {Bloch}, \citenamefont {Kollath},\ and\ \citenamefont {Kuhr}}]{cheneau:2012}%
  \BibitemOpen
  \bibfield  {author} {\bibinfo {author} {\bibfnamefont {Marc}\ \bibnamefont {Cheneau}}, \bibinfo {author} {\bibfnamefont {Peter}\ \bibnamefont {Barmettler}}, \bibinfo {author} {\bibfnamefont {Dario}\ \bibnamefont {Poletti}}, \bibinfo {author} {\bibfnamefont {Manuel}\ \bibnamefont {Endres}}, \bibinfo {author} {\bibfnamefont {Peter}\ \bibnamefont {Schau{\ss}}}, \bibinfo {author} {\bibfnamefont {Takeshi}\ \bibnamefont {Fukuhara}}, \bibinfo {author} {\bibfnamefont {Christian}\ \bibnamefont {Gross}}, \bibinfo {author} {\bibfnamefont {Immanuel}\ \bibnamefont {Bloch}}, \bibinfo {author} {\bibfnamefont {Corinna}\ \bibnamefont {Kollath}}, \ and\ \bibinfo {author} {\bibfnamefont {Stefan}\ \bibnamefont {Kuhr}},\ }\bibfield  {title} {\enquote {\bibinfo {title} {Light-cone-like spreading of correlations in a quantum many-body system},}\ }\href {\doibase 10.1038/nature10748} {\bibfield  {journal} {\bibinfo  {journal} {Nature}\ }\textbf {\bibinfo {volume} {481}},\ \bibinfo {pages} {484--487} (\bibinfo {year}
  {2012})}\BibitemShut {NoStop}%
\bibitem [{\citenamefont {Jurcevic}\ \emph {et~al.}(2014)\citenamefont {Jurcevic}, \citenamefont {Lanyon}, \citenamefont {Hauke}, \citenamefont {Hempel}, \citenamefont {Zoller}, \citenamefont {Blatt},\ and\ \citenamefont {Roos}}]{jurcevic:2014}%
  \BibitemOpen
  \bibfield  {author} {\bibinfo {author} {\bibfnamefont {P.}~\bibnamefont {Jurcevic}}, \bibinfo {author} {\bibfnamefont {B.~P.}\ \bibnamefont {Lanyon}}, \bibinfo {author} {\bibfnamefont {P.}~\bibnamefont {Hauke}}, \bibinfo {author} {\bibfnamefont {C.}~\bibnamefont {Hempel}}, \bibinfo {author} {\bibfnamefont {P.}~\bibnamefont {Zoller}}, \bibinfo {author} {\bibfnamefont {R.}~\bibnamefont {Blatt}}, \ and\ \bibinfo {author} {\bibfnamefont {C.~F.}\ \bibnamefont {Roos}},\ }\bibfield  {title} {\enquote {\bibinfo {title} {Quasiparticle engineering and entanglement propagation in a quantum many-body system},}\ }\href {\doibase 10.1038/nature13461} {\bibfield  {journal} {\bibinfo  {journal} {Nature}\ }\textbf {\bibinfo {volume} {511}},\ \bibinfo {pages} {202--205} (\bibinfo {year} {2014})}\BibitemShut {NoStop}%
\bibitem [{\citenamefont {Kaufman}\ \emph {et~al.}(2016)\citenamefont {Kaufman}, \citenamefont {Tai}, \citenamefont {Lukin}, \citenamefont {Rispoli}, \citenamefont {Schittko}, \citenamefont {Preiss},\ and\ \citenamefont {Greiner}}]{kaufman:2016}%
  \BibitemOpen
  \bibfield  {author} {\bibinfo {author} {\bibfnamefont {Adam~M.}\ \bibnamefont {Kaufman}}, \bibinfo {author} {\bibfnamefont {M.~Eric}\ \bibnamefont {Tai}}, \bibinfo {author} {\bibfnamefont {Alexander}\ \bibnamefont {Lukin}}, \bibinfo {author} {\bibfnamefont {Matthew}\ \bibnamefont {Rispoli}}, \bibinfo {author} {\bibfnamefont {Robert}\ \bibnamefont {Schittko}}, \bibinfo {author} {\bibfnamefont {Philipp~M.}\ \bibnamefont {Preiss}}, \ and\ \bibinfo {author} {\bibfnamefont {Markus}\ \bibnamefont {Greiner}},\ }\bibfield  {title} {\enquote {\bibinfo {title} {Quantum thermalization through entanglement in an isolated many-body system},}\ }\href {\doibase 10.1126/science.aaf6725} {\bibfield  {journal} {\bibinfo  {journal} {Science}\ }\textbf {\bibinfo {volume} {353}},\ \bibinfo {pages} {794--800} (\bibinfo {year} {2016})}\BibitemShut {NoStop}%
\bibitem [{\citenamefont {Kormos}\ \emph {et~al.}(2017)\citenamefont {Kormos}, \citenamefont {Collura}, \citenamefont {Tak{\'a}cs},\ and\ \citenamefont {Calabrese}}]{kormos:2017}%
  \BibitemOpen
  \bibfield  {author} {\bibinfo {author} {\bibfnamefont {Marton}\ \bibnamefont {Kormos}}, \bibinfo {author} {\bibfnamefont {Mario}\ \bibnamefont {Collura}}, \bibinfo {author} {\bibfnamefont {Gabor}\ \bibnamefont {Tak{\'a}cs}}, \ and\ \bibinfo {author} {\bibfnamefont {Pasquale}\ \bibnamefont {Calabrese}},\ }\bibfield  {title} {\enquote {\bibinfo {title} {Real-time confinement following a quantum quench to a non-integrable model},}\ }\href {\doibase 10.1038/nphys3934} {\bibfield  {journal} {\bibinfo  {journal} {Nature Physics}\ }\textbf {\bibinfo {volume} {13}},\ \bibinfo {pages} {246--249} (\bibinfo {year} {2017})}\BibitemShut {NoStop}%
\bibitem [{\citenamefont {Vovrosh}\ and\ \citenamefont {Knolle}(2021)}]{vovrosh:2021}%
  \BibitemOpen
  \bibfield  {author} {\bibinfo {author} {\bibfnamefont {Joseph}\ \bibnamefont {Vovrosh}}\ and\ \bibinfo {author} {\bibfnamefont {Johannes}\ \bibnamefont {Knolle}},\ }\bibfield  {title} {\enquote {\bibinfo {title} {Confinement and entanglement dynamics on a digital quantum computer},}\ }\href {\doibase 10.1038/s41598-021-90849-5} {\bibfield  {journal} {\bibinfo  {journal} {Scientific Reports}\ }\textbf {\bibinfo {volume} {11}},\ \bibinfo {pages} {11577} (\bibinfo {year} {2021})}\BibitemShut {NoStop}%
\bibitem [{nim()}]{nimag}%
  \BibitemOpen
  \href@noop {} {}\bibinfo {note} {Implementing this requires post-selection on exponentially small outcomes \cite{mcardle2019variational,motta2020determining}.}\BibitemShut {Stop}%
\bibitem [{\citenamefont {Hauschild}\ \emph {et~al.}(2024)\citenamefont {Hauschild}, \citenamefont {Unfried}, \citenamefont {Anand}, \citenamefont {Andrews}, \citenamefont {Bintz}, \citenamefont {Borla}, \citenamefont {Divic}, \citenamefont {Drescher}, \citenamefont {Geiger}, \citenamefont {Hefel}, \citenamefont {H{\'e}mery}, \citenamefont {Kadow}, \citenamefont {Kemp}, \citenamefont {Kirchner}, \citenamefont {Liu}, \citenamefont {Moller}, \citenamefont {Parker}, \citenamefont {Rader}, \citenamefont {Romen}, \citenamefont {Scalet}, \citenamefont {Schoonderwoerd}, \citenamefont {Schulz}, \citenamefont {Soejima}, \citenamefont {Thoma}, \citenamefont {Wu}, \citenamefont {Zechmann}, \citenamefont {Zweng}, \citenamefont {Mong}, \citenamefont {Zaletel},\ and\ \citenamefont {Pollmann}}]{hauschild:2024}%
  \BibitemOpen
  \bibfield  {author} {\bibinfo {author} {\bibfnamefont {Johannes}\ \bibnamefont {Hauschild}}, \bibinfo {author} {\bibfnamefont {Jakob}\ \bibnamefont {Unfried}}, \bibinfo {author} {\bibfnamefont {Sajant}\ \bibnamefont {Anand}}, \bibinfo {author} {\bibfnamefont {Bartholomew}\ \bibnamefont {Andrews}}, \bibinfo {author} {\bibfnamefont {Marcus}\ \bibnamefont {Bintz}}, \bibinfo {author} {\bibfnamefont {Umberto}\ \bibnamefont {Borla}}, \bibinfo {author} {\bibfnamefont {Stefan}\ \bibnamefont {Divic}}, \bibinfo {author} {\bibfnamefont {Markus}\ \bibnamefont {Drescher}}, \bibinfo {author} {\bibfnamefont {Jan}\ \bibnamefont {Geiger}}, \bibinfo {author} {\bibfnamefont {Martin}\ \bibnamefont {Hefel}}, \bibinfo {author} {\bibfnamefont {K{\'e}vin}\ \bibnamefont {H{\'e}mery}}, \bibinfo {author} {\bibfnamefont {Wilhelm}\ \bibnamefont {Kadow}}, \bibinfo {author} {\bibfnamefont {Jack}\ \bibnamefont {Kemp}}, \bibinfo {author} {\bibfnamefont {Nico}\ \bibnamefont {Kirchner}}, \bibinfo {author} {\bibfnamefont {Vincent~S.}\
  \bibnamefont {Liu}}, \bibinfo {author} {\bibfnamefont {Gunnar}\ \bibnamefont {Moller}}, \bibinfo {author} {\bibfnamefont {Daniel}\ \bibnamefont {Parker}}, \bibinfo {author} {\bibfnamefont {Michael}\ \bibnamefont {Rader}}, \bibinfo {author} {\bibfnamefont {Anton}\ \bibnamefont {Romen}}, \bibinfo {author} {\bibfnamefont {Samuel}\ \bibnamefont {Scalet}}, \bibinfo {author} {\bibfnamefont {Leon}\ \bibnamefont {Schoonderwoerd}}, \bibinfo {author} {\bibfnamefont {Maximilian}\ \bibnamefont {Schulz}}, \bibinfo {author} {\bibfnamefont {Tomohiro}\ \bibnamefont {Soejima}}, \bibinfo {author} {\bibfnamefont {Philipp}\ \bibnamefont {Thoma}}, \bibinfo {author} {\bibfnamefont {Yantao}\ \bibnamefont {Wu}}, \bibinfo {author} {\bibfnamefont {Philip}\ \bibnamefont {Zechmann}}, \bibinfo {author} {\bibfnamefont {Ludwig}\ \bibnamefont {Zweng}}, \bibinfo {author} {\bibfnamefont {Roger}\ \bibnamefont {Mong}}, \bibinfo {author} {\bibfnamefont {Michael~P.}\ \bibnamefont {Zaletel}}, \ and\ \bibinfo {author} {\bibfnamefont {Frank}\
  \bibnamefont {Pollmann}},\ }\bibfield  {title} {\enquote {\bibinfo {title} {Tensor network {{Python}} ({{TeNPy}}) version 1},}\ }\href {\doibase 10.21468/SciPostPhysCodeb.41} {\bibfield  {journal} {\bibinfo  {journal} {SciPost Physics Codebases}\ ,\ \bibinfo {pages} {041}} (\bibinfo {year} {2024})}\BibitemShut {NoStop}%
\bibitem [{\citenamefont {Barthel}\ and\ \citenamefont {Zhang}(2020)}]{barthelOptimizedLieTrotter2020}%
  \BibitemOpen
  \bibfield  {author} {\bibinfo {author} {\bibfnamefont {Thomas}\ \bibnamefont {Barthel}}\ and\ \bibinfo {author} {\bibfnamefont {Yikang}\ \bibnamefont {Zhang}},\ }\bibfield  {title} {\enquote {\bibinfo {title} {Optimized {{Lie}}--{{Trotter}}--{{Suzuki}} decompositions for two and three non-commuting terms},}\ }\href {\doibase 10.1016/j.aop.2020.168165} {\bibfield  {journal} {\bibinfo  {journal} {Annals of Physics}\ }\textbf {\bibinfo {volume} {418}},\ \bibinfo {pages} {168165} (\bibinfo {year} {2020})}\BibitemShut {NoStop}%
\bibitem [{\citenamefont {Schollw{\"o}ck}(2011)}]{SCHOLLWOCK201196}%
  \BibitemOpen
  \bibfield  {author} {\bibinfo {author} {\bibfnamefont {Ulrich}\ \bibnamefont {Schollw{\"o}ck}},\ }\bibfield  {title} {\enquote {\bibinfo {title} {The density-matrix renormalization group in the age of matrix product states},}\ }\href {\doibase 10.1016/j.aop.2010.09.012} {\bibfield  {journal} {\bibinfo  {journal} {Annals of Physics}\ }\textbf {\bibinfo {volume} {326}},\ \bibinfo {pages} {96--192} (\bibinfo {year} {2011})}\BibitemShut {NoStop}%
\bibitem [{\citenamefont {Karrasch}\ \emph {et~al.}(2013)\citenamefont {Karrasch}, \citenamefont {Bardarson},\ and\ \citenamefont {Moore}}]{karraschReducingNumericalEffort2013}%
  \BibitemOpen
  \bibfield  {author} {\bibinfo {author} {\bibfnamefont {C.}~\bibnamefont {Karrasch}}, \bibinfo {author} {\bibfnamefont {J.~H.}\ \bibnamefont {Bardarson}}, \ and\ \bibinfo {author} {\bibfnamefont {J.~E.}\ \bibnamefont {Moore}},\ }\bibfield  {title} {\enquote {\bibinfo {title} {Reducing the numerical effort of finite-temperature density matrix renormalization group calculations},}\ }\href {\doibase 10.1088/1367-2630/15/8/083031} {\bibfield  {journal} {\bibinfo  {journal} {New Journal of Physics}\ }\textbf {\bibinfo {volume} {15}},\ \bibinfo {pages} {083031} (\bibinfo {year} {2013})}\BibitemShut {NoStop}%
\bibitem [{nin()}]{nintegr}%
  \BibitemOpen
  \href@noop {} {}\bibinfo {note} {At large length- and time-scales the dynamics should generally be insensitive to the details of the short-range correlations. However, the integrability of 1+1D rational conformal field theories causes this expectation to break down~\cite{calabrese2012quantum2,calabrese2016quantum, cardy2016furtherresults}.}\BibitemShut {Stop}%
\bibitem [{\citenamefont {Zou}\ \emph {et~al.}(2020)\citenamefont {Zou}, \citenamefont {Milsted},\ and\ \citenamefont {Vidal}}]{zou2020conformal}%
  \BibitemOpen
  \bibfield  {author} {\bibinfo {author} {\bibfnamefont {Yijian}\ \bibnamefont {Zou}}, \bibinfo {author} {\bibfnamefont {Ashley}\ \bibnamefont {Milsted}}, \ and\ \bibinfo {author} {\bibfnamefont {Guifre}\ \bibnamefont {Vidal}},\ }\bibfield  {title} {\enquote {\bibinfo {title} {Conformal {{Fields}} and {{Operator Product Expansion}} in {{Critical Quantum Spin Chains}}},}\ }\href {\doibase 10.1103/PhysRevLett.124.040604} {\bibfield  {journal} {\bibinfo  {journal} {Physical Review Letters}\ }\textbf {\bibinfo {volume} {124}},\ \bibinfo {pages} {040604} (\bibinfo {year} {2020})}\BibitemShut {NoStop}%
\bibitem [{\citenamefont {Mong}\ \emph {et~al.}(2014)\citenamefont {Mong}, \citenamefont {Clarke}, \citenamefont {Alicea}, \citenamefont {Lindner},\ and\ \citenamefont {Fendley}}]{mong2014parafermionic}%
  \BibitemOpen
  \bibfield  {author} {\bibinfo {author} {\bibfnamefont {Roger S~K}\ \bibnamefont {Mong}}, \bibinfo {author} {\bibfnamefont {David~J}\ \bibnamefont {Clarke}}, \bibinfo {author} {\bibfnamefont {Jason}\ \bibnamefont {Alicea}}, \bibinfo {author} {\bibfnamefont {Netanel~H}\ \bibnamefont {Lindner}}, \ and\ \bibinfo {author} {\bibfnamefont {Paul}\ \bibnamefont {Fendley}},\ }\bibfield  {title} {\enquote {\bibinfo {title} {Parafermionic conformal field theory on the lattice},}\ }\href {\doibase 10.1088/1751-8113/47/45/452001} {\bibfield  {journal} {\bibinfo  {journal} {Journal of Physics A: Mathematical and Theoretical}\ }\textbf {\bibinfo {volume} {47}},\ \bibinfo {pages} {452001} (\bibinfo {year} {2014})}\BibitemShut {NoStop}%
\bibitem [{\citenamefont {Barouch}\ \emph {et~al.}(1970)\citenamefont {Barouch}, \citenamefont {McCoy},\ and\ \citenamefont {Dresden}}]{barouch1970statistical}%
  \BibitemOpen
  \bibfield  {author} {\bibinfo {author} {\bibfnamefont {Eytan}\ \bibnamefont {Barouch}}, \bibinfo {author} {\bibfnamefont {Barry~M.}\ \bibnamefont {McCoy}}, \ and\ \bibinfo {author} {\bibfnamefont {Max}\ \bibnamefont {Dresden}},\ }\bibfield  {title} {\enquote {\bibinfo {title} {Statistical mechanics of the $\mathrm{XY}$ model. i},}\ }\href {\doibase 10.1103/PhysRevA.2.1075} {\bibfield  {journal} {\bibinfo  {journal} {Phys. Rev. A}\ }\textbf {\bibinfo {volume} {2}},\ \bibinfo {pages} {1075--1092} (\bibinfo {year} {1970})}\BibitemShut {NoStop}%
\bibitem [{\citenamefont {Li}\ \emph {et~al.}(2009)\citenamefont {Li}, \citenamefont {Huo},\ and\ \citenamefont {Song}}]{li:2009}%
  \BibitemOpen
  \bibfield  {author} {\bibinfo {author} {\bibfnamefont {Ying}\ \bibnamefont {Li}}, \bibinfo {author} {\bibfnamefont {MingXia}\ \bibnamefont {Huo}}, \ and\ \bibinfo {author} {\bibfnamefont {Zhi}\ \bibnamefont {Song}},\ }\bibfield  {title} {\enquote {\bibinfo {title} {Exact results for the criticality of quench dynamics in quantum {{Ising}} models},}\ }\href {\doibase 10.1103/PhysRevB.80.054404} {\bibfield  {journal} {\bibinfo  {journal} {Physical Review B}\ }\textbf {\bibinfo {volume} {80}},\ \bibinfo {pages} {054404} (\bibinfo {year} {2009})}\BibitemShut {NoStop}%
\bibitem [{\citenamefont {De~Grandi}\ \emph {et~al.}(2010)\citenamefont {De~Grandi}, \citenamefont {Gritsev},\ and\ \citenamefont {Polkovnikov}}]{degrandi:2010}%
  \BibitemOpen
  \bibfield  {author} {\bibinfo {author} {\bibfnamefont {C.}~\bibnamefont {De~Grandi}}, \bibinfo {author} {\bibfnamefont {V.}~\bibnamefont {Gritsev}}, \ and\ \bibinfo {author} {\bibfnamefont {A.}~\bibnamefont {Polkovnikov}},\ }\bibfield  {title} {\enquote {\bibinfo {title} {Quench dynamics near a quantum critical point: {{Application}} to the sine-{{Gordon}} model},}\ }\href {\doibase 10.1103/PhysRevB.81.224301} {\bibfield  {journal} {\bibinfo  {journal} {Physical Review B}\ }\textbf {\bibinfo {volume} {81}},\ \bibinfo {pages} {224301} (\bibinfo {year} {2010})}\BibitemShut {NoStop}%
\bibitem [{\citenamefont {Calabrese}\ \emph {et~al.}(2011)\citenamefont {Calabrese}, \citenamefont {Essler},\ and\ \citenamefont {Fagotti}}]{Calabrese_Essler_Fagotti_2011}%
  \BibitemOpen
  \bibfield  {author} {\bibinfo {author} {\bibfnamefont {Pasquale}\ \bibnamefont {Calabrese}}, \bibinfo {author} {\bibfnamefont {Fabian H.~L.}\ \bibnamefont {Essler}}, \ and\ \bibinfo {author} {\bibfnamefont {Maurizio}\ \bibnamefont {Fagotti}},\ }\bibfield  {title} {\enquote {\bibinfo {title} {Quantum quench in the transverse field ising chain},}\ }\href {\doibase 10.1103/PhysRevLett.106.227203} {\bibfield  {journal} {\bibinfo  {journal} {Physical Review Letters}\ }\textbf {\bibinfo {volume} {106}},\ \bibinfo {pages} {227203} (\bibinfo {year} {2011})}\BibitemShut {NoStop}%
\bibitem [{\citenamefont {Calabrese}\ \emph {et~al.}(2012{\natexlab{a}})\citenamefont {Calabrese}, \citenamefont {Essler},\ and\ \citenamefont {Fagotti}}]{calabrese2012quantum}%
  \BibitemOpen
  \bibfield  {author} {\bibinfo {author} {\bibfnamefont {Pasquale}\ \bibnamefont {Calabrese}}, \bibinfo {author} {\bibfnamefont {Fabian H.~L.}\ \bibnamefont {Essler}}, \ and\ \bibinfo {author} {\bibfnamefont {Maurizio}\ \bibnamefont {Fagotti}},\ }\bibfield  {title} {\enquote {\bibinfo {title} {Quantum quench in the transverse field {{Ising}} chain: {{I}}. {{Time}} evolution of order parameter correlators},}\ }\href {\doibase 10.1088/1742-5468/2012/07/P07016} {\bibfield  {journal} {\bibinfo  {journal} {Journal of Statistical Mechanics: Theory and Experiment}\ }\textbf {\bibinfo {volume} {2012}},\ \bibinfo {pages} {P07016} (\bibinfo {year} {2012}{\natexlab{a}})}\BibitemShut {NoStop}%
\bibitem [{\citenamefont {Calabrese}\ \emph {et~al.}(2012{\natexlab{b}})\citenamefont {Calabrese}, \citenamefont {Essler},\ and\ \citenamefont {Fagotti}}]{calabrese2012quantum2}%
  \BibitemOpen
  \bibfield  {author} {\bibinfo {author} {\bibfnamefont {Pasquale}\ \bibnamefont {Calabrese}}, \bibinfo {author} {\bibfnamefont {Fabian H.~L.}\ \bibnamefont {Essler}}, \ and\ \bibinfo {author} {\bibfnamefont {Maurizio}\ \bibnamefont {Fagotti}},\ }\bibfield  {title} {\enquote {\bibinfo {title} {Quantum quenches in the transverse field {{Ising}} chain: {{II}}. {{Stationary}} state properties},}\ }\href {\doibase 10.1088/1742-5468/2012/07/P07022} {\bibfield  {journal} {\bibinfo  {journal} {Journal of Statistical Mechanics: Theory and Experiment}\ }\textbf {\bibinfo {volume} {2012}},\ \bibinfo {pages} {P07022} (\bibinfo {year} {2012}{\natexlab{b}})}\BibitemShut {NoStop}%
\bibitem [{\citenamefont {Delfino}(2014)}]{delfino:2014}%
  \BibitemOpen
  \bibfield  {author} {\bibinfo {author} {\bibfnamefont {Gesualdo}\ \bibnamefont {Delfino}},\ }\bibfield  {title} {\enquote {\bibinfo {title} {Quantum quenches with integrable pre-quench dynamics},}\ }\href {\doibase 10.1088/1751-8113/47/40/402001} {\bibfield  {journal} {\bibinfo  {journal} {Journal of Physics A: Mathematical and Theoretical}\ }\textbf {\bibinfo {volume} {47}},\ \bibinfo {pages} {402001} (\bibinfo {year} {2014})}\BibitemShut {NoStop}%
\bibitem [{\citenamefont {Essler}\ and\ \citenamefont {Fagotti}(2016)}]{Essler_Fagotti_2016}%
  \BibitemOpen
  \bibfield  {author} {\bibinfo {author} {\bibfnamefont {Fabian H.~L.}\ \bibnamefont {Essler}}\ and\ \bibinfo {author} {\bibfnamefont {Maurizio}\ \bibnamefont {Fagotti}},\ }\bibfield  {title} {\enquote {\bibinfo {title} {Quench dynamics and relaxation in isolated integrable quantum spin chains},}\ }\href {\doibase 10.1088/1742-5468/2016/06/064002} {\bibfield  {journal} {\bibinfo  {journal} {Journal of Statistical Mechanics: Theory and Experiment}\ }\textbf {\bibinfo {volume} {2016}},\ \bibinfo {pages} {064002} (\bibinfo {year} {2016})}\BibitemShut {NoStop}%
\bibitem [{\citenamefont {Delfino}\ and\ \citenamefont {Viti}(2017)}]{delfino:2017}%
  \BibitemOpen
  \bibfield  {author} {\bibinfo {author} {\bibfnamefont {Gesualdo}\ \bibnamefont {Delfino}}\ and\ \bibinfo {author} {\bibfnamefont {Jacopo}\ \bibnamefont {Viti}},\ }\bibfield  {title} {\enquote {\bibinfo {title} {On the theory of quantum quenches in near-critical systems},}\ }\href {\doibase 10.1088/1751-8121/aa5660} {\bibfield  {journal} {\bibinfo  {journal} {Journal of Physics A: Mathematical and Theoretical}\ }\textbf {\bibinfo {volume} {50}},\ \bibinfo {pages} {084004} (\bibinfo {year} {2017})}\BibitemShut {NoStop}%
\bibitem [{\citenamefont {Das}\ \emph {et~al.}(2017)\citenamefont {Das}, \citenamefont {Das}, \citenamefont {Galante}, \citenamefont {Myers},\ and\ \citenamefont {Sengupta}}]{das:2017}%
  \BibitemOpen
  \bibfield  {author} {\bibinfo {author} {\bibfnamefont {Diptarka}\ \bibnamefont {Das}}, \bibinfo {author} {\bibfnamefont {Sumit~R.}\ \bibnamefont {Das}}, \bibinfo {author} {\bibfnamefont {Dami{\'a}n~A.}\ \bibnamefont {Galante}}, \bibinfo {author} {\bibfnamefont {Robert~C.}\ \bibnamefont {Myers}}, \ and\ \bibinfo {author} {\bibfnamefont {Krishnendu}\ \bibnamefont {Sengupta}},\ }\bibfield  {title} {\enquote {\bibinfo {title} {An exactly solvable quench protocol for integrable spin models},}\ }\href {\doibase 10.1007/JHEP11(2017)157} {\bibfield  {journal} {\bibinfo  {journal} {Journal of High Energy Physics}\ }\textbf {\bibinfo {volume} {2017}},\ \bibinfo {pages} {157} (\bibinfo {year} {2017})}\BibitemShut {NoStop}%
\bibitem [{\citenamefont {H{\'o}ds{\'a}gi}\ \emph {et~al.}(2018)\citenamefont {H{\'o}ds{\'a}gi}, \citenamefont {Kormos},\ and\ \citenamefont {Tak{\'a}cs}}]{hodsagi:2018}%
  \BibitemOpen
  \bibfield  {author} {\bibinfo {author} {\bibfnamefont {Krist{\'o}f}\ \bibnamefont {H{\'o}ds{\'a}gi}}, \bibinfo {author} {\bibfnamefont {M{\'a}rton}\ \bibnamefont {Kormos}}, \ and\ \bibinfo {author} {\bibfnamefont {G{\'a}bor}\ \bibnamefont {Tak{\'a}cs}},\ }\bibfield  {title} {\enquote {\bibinfo {title} {Quench dynamics of the {{Ising}} field theory in a magnetic field},}\ }\href {\doibase 10.21468/SciPostPhys.5.3.027} {\bibfield  {journal} {\bibinfo  {journal} {SciPost Physics}\ }\textbf {\bibinfo {volume} {5}},\ \bibinfo {pages} {027} (\bibinfo {year} {2018})}\BibitemShut {NoStop}%
\bibitem [{\citenamefont {Granet}\ \emph {et~al.}(2020)\citenamefont {Granet}, \citenamefont {Fagotti},\ and\ \citenamefont {Essler}}]{Granet_Fagotti_Essler_2020}%
  \BibitemOpen
  \bibfield  {author} {\bibinfo {author} {\bibfnamefont {Etienne}\ \bibnamefont {Granet}}, \bibinfo {author} {\bibfnamefont {Maurizio}\ \bibnamefont {Fagotti}}, \ and\ \bibinfo {author} {\bibfnamefont {Fabian H.~L.}\ \bibnamefont {Essler}},\ }\bibfield  {title} {\enquote {\bibinfo {title} {Finite temperature and quench dynamics in the transverse field ising model from form factor expansions},}\ }\href {\doibase 10.21468/SciPostPhys.9.3.033} {\bibfield  {journal} {\bibinfo  {journal} {SciPost Physics}\ }\textbf {\bibinfo {volume} {9}},\ \bibinfo {pages} {033} (\bibinfo {year} {2020})}\BibitemShut {NoStop}%
\bibitem [{\citenamefont {Granet}\ \emph {et~al.}(2022)\citenamefont {Granet}, \citenamefont {Dreyer},\ and\ \citenamefont {Essler}}]{granet:2022}%
  \BibitemOpen
  \bibfield  {author} {\bibinfo {author} {\bibfnamefont {Etienne}\ \bibnamefont {Granet}}, \bibinfo {author} {\bibfnamefont {Henrik}\ \bibnamefont {Dreyer}}, \ and\ \bibinfo {author} {\bibfnamefont {Fabian}\ \bibnamefont {Essler}},\ }\bibfield  {title} {\enquote {\bibinfo {title} {Out-of-equilibrium dynamics of the {{XY}} spin chain from form factor expansion},}\ }\href {\doibase 10.21468/SciPostPhys.12.1.019} {\bibfield  {journal} {\bibinfo  {journal} {SciPost Physics}\ }\textbf {\bibinfo {volume} {12}},\ \bibinfo {pages} {019} (\bibinfo {year} {2022})}\BibitemShut {NoStop}%
\bibitem [{sur()}]{surace}%
  \BibitemOpen
  \href@noop {} {}\bibinfo {note} {Crucially, the model is weakly integrability-breaking, meaning that there is a prethermal regime where the dynamics are still governed by the underlying integrable CFT\cite{Mori_Ikeda_Kaminishi_Ueda_2018,Surace_Motrunich_2023}. We work within these timescales by choosing $\gamma$ accordingly.}\BibitemShut {Stop}%
\bibitem [{\citenamefont {Rahmani}\ \emph {et~al.}(2015)\citenamefont {Rahmani}, \citenamefont {Zhu}, \citenamefont {Franz},\ and\ \citenamefont {Affleck}}]{rahmani2015mzm}%
  \BibitemOpen
  \bibfield  {author} {\bibinfo {author} {\bibfnamefont {Armin}\ \bibnamefont {Rahmani}}, \bibinfo {author} {\bibfnamefont {Xiaoyu}\ \bibnamefont {Zhu}}, \bibinfo {author} {\bibfnamefont {Marcel}\ \bibnamefont {Franz}}, \ and\ \bibinfo {author} {\bibfnamefont {Ian}\ \bibnamefont {Affleck}},\ }\bibfield  {title} {\enquote {\bibinfo {title} {Emergent supersymmetry from strongly interacting majorana zero modes},}\ }\href {\doibase 10.1103/PhysRevLett.115.166401} {\bibfield  {journal} {\bibinfo  {journal} {Physical Review Letters}\ }\textbf {\bibinfo {volume} {115}},\ \bibinfo {pages} {166401} (\bibinfo {year} {2015})}\BibitemShut {NoStop}%
\bibitem [{\citenamefont {Milsted}\ \emph {et~al.}(2015)\citenamefont {Milsted}, \citenamefont {Seabra}, \citenamefont {Fulga}, \citenamefont {Beenakker},\ and\ \citenamefont {Cobanera}}]{milsted2015annni}%
  \BibitemOpen
  \bibfield  {author} {\bibinfo {author} {\bibfnamefont {A.}~\bibnamefont {Milsted}}, \bibinfo {author} {\bibfnamefont {L.}~\bibnamefont {Seabra}}, \bibinfo {author} {\bibfnamefont {I.~C.}\ \bibnamefont {Fulga}}, \bibinfo {author} {\bibfnamefont {C.~W.~J.}\ \bibnamefont {Beenakker}}, \ and\ \bibinfo {author} {\bibfnamefont {E.}~\bibnamefont {Cobanera}},\ }\bibfield  {title} {\enquote {\bibinfo {title} {Statistical translation invariance protects a topological insulator from interactions},}\ }\href {\doibase 10.1103/PhysRevB.92.085139} {\bibfield  {journal} {\bibinfo  {journal} {Phys. Rev. B}\ }\textbf {\bibinfo {volume} {92}},\ \bibinfo {pages} {085139} (\bibinfo {year} {2015})}\BibitemShut {NoStop}%
\bibitem [{\citenamefont {Milsted}\ and\ \citenamefont {Vidal}(2017)}]{milsted2017extraction}%
  \BibitemOpen
  \bibfield  {author} {\bibinfo {author} {\bibfnamefont {Ashley}\ \bibnamefont {Milsted}}\ and\ \bibinfo {author} {\bibfnamefont {Guifre}\ \bibnamefont {Vidal}},\ }\bibfield  {title} {\enquote {\bibinfo {title} {Extraction of conformal data in critical quantum spin chains using the koo-saleur formula},}\ }\href {\doibase 10.1103/PhysRevB.96.245105} {\bibfield  {journal} {\bibinfo  {journal} {Phys. Rev. B}\ }\textbf {\bibinfo {volume} {96}},\ \bibinfo {pages} {245105} (\bibinfo {year} {2017})}\BibitemShut {NoStop}%
\bibitem [{\citenamefont {Surace}\ \emph {et~al.}(2020)\citenamefont {Surace}, \citenamefont {Tagliacozzo},\ and\ \citenamefont {Tonni}}]{surace:2020}%
  \BibitemOpen
  \bibfield  {author} {\bibinfo {author} {\bibfnamefont {Jacopo}\ \bibnamefont {Surace}}, \bibinfo {author} {\bibfnamefont {Luca}\ \bibnamefont {Tagliacozzo}}, \ and\ \bibinfo {author} {\bibfnamefont {Erik}\ \bibnamefont {Tonni}},\ }\bibfield  {title} {\enquote {\bibinfo {title} {Operator content of entanglement spectra in the transverse field {{Ising}} chain after global quenches},}\ }\href {\doibase 10.1103/PhysRevB.101.241107} {\bibfield  {journal} {\bibinfo  {journal} {Physical Review B}\ }\textbf {\bibinfo {volume} {101}},\ \bibinfo {pages} {241107} (\bibinfo {year} {2020})}\BibitemShut {NoStop}%
\bibitem [{\citenamefont {Robertson}\ \emph {et~al.}(2022)\citenamefont {Robertson}, \citenamefont {Surace},\ and\ \citenamefont {Tagliacozzo}}]{robertson:2022}%
  \BibitemOpen
  \bibfield  {author} {\bibinfo {author} {\bibfnamefont {Niall~F.}\ \bibnamefont {Robertson}}, \bibinfo {author} {\bibfnamefont {Jacopo}\ \bibnamefont {Surace}}, \ and\ \bibinfo {author} {\bibfnamefont {Luca}\ \bibnamefont {Tagliacozzo}},\ }\bibfield  {title} {\enquote {\bibinfo {title} {Quenches to the critical point of the three-state {{Potts}} model: {{Matrix}} product state simulations and conformal field theory},}\ }\href {\doibase 10.1103/PhysRevB.105.195103} {\bibfield  {journal} {\bibinfo  {journal} {Physical Review B}\ }\textbf {\bibinfo {volume} {105}},\ \bibinfo {pages} {195103} (\bibinfo {year} {2022})}\BibitemShut {NoStop}%
\bibitem [{\citenamefont {Ginsparg}(1988)}]{ginsparg1988applied}%
  \BibitemOpen
  \bibfield  {author} {\bibinfo {author} {\bibfnamefont {Paul}\ \bibnamefont {Ginsparg}},\ }\bibfield  {title} {\enquote {\bibinfo {title} {Applied {{Conformal Field Theory}}},}\ }\href {\doibase 10.48550/arXiv.hep-th/9108028} {\  (\bibinfo {year} {1988}),\ 10.48550/arXiv.hep-th/9108028}\BibitemShut {NoStop}%
\bibitem [{\citenamefont {Sachdev}(2011)}]{sachdevQuantumPhaseTransitions2011}%
  \BibitemOpen
  \bibfield  {author} {\bibinfo {author} {\bibfnamefont {Subir}\ \bibnamefont {Sachdev}},\ }\href {\doibase 10.1017/CBO9780511973765} {\emph {\bibinfo {title} {Quantum {{Phase Transitions}}}}},\ \bibinfo {edition} {2nd}\ ed.\ (\bibinfo  {publisher} {Cambridge University Press},\ \bibinfo {address} {Cambridge},\ \bibinfo {year} {2011})\BibitemShut {NoStop}%
\bibitem [{nOP()}]{nOPE}%
  \BibitemOpen
  \href@noop {} {}\bibinfo {note} {The $H_\Psi$ perturbations correspond to the primary fields up to an additive constant that leaves the dynamics unchanged. In addition, we note that any lattice perturbation can be written in terms of primary fields and their descendants by way of the lattice-to-field mapping and the operator product expansion~\cite{cardy1984conformal,Cardy_1986,sachdevQuantumPhaseTransitions2011,Di_Francesco_Mathieu_Sénéchal_1997,SM}. Since correlation functions of descendants can be calculated in principle using Ward identities, this does not constitute a particular restriction.}\BibitemShut {Stop}%
\bibitem [{nsi()}]{nsigmaquench}%
  \BibitemOpen
  \href@noop {} {}\bibinfo {note} {Notice we do not consider such a perturbation in the ground state as it results in a first order phase transition.}\BibitemShut {Stop}%
\bibitem [{nta()}]{ntau0}%
  \BibitemOpen
  \href@noop {} {}\bibinfo {note} {This amounts to the identification $\beta = 4\tau_0$ \cite{calabrese2016quantum}, where $\tau_0$ is the extrapolation length of the boundary-CFT quench (Fig.~\ref{fig:groundstate}).}\BibitemShut {Stop}%
\bibitem [{\citenamefont {Zamolodchikov}(1991)}]{Zamolodchikov_1991}%
  \BibitemOpen
  \bibfield  {author} {\bibinfo {author} {\bibfnamefont {A.~B.}\ \bibnamefont {Zamolodchikov}},\ }\bibfield  {title} {\enquote {\bibinfo {title} {Two-point correlation function in scaling lee-yang model},}\ }\href {\doibase 10.1016/0550-3213(91)90207-E} {\bibfield  {journal} {\bibinfo  {journal} {Nuclear Physics B}\ }\textbf {\bibinfo {volume} {348}},\ \bibinfo {pages} {619–641} (\bibinfo {year} {1991})}\BibitemShut {NoStop}%
\bibitem [{\citenamefont {Mussardo}(2020)}]{Mussardo_2010}%
  \BibitemOpen
  \bibfield  {author} {\bibinfo {author} {\bibfnamefont {Giuseppe}\ \bibnamefont {Mussardo}},\ }\href@noop {} {\emph {\bibinfo {title} {Statistical {{Field Theory}}: {{An Introduction}} to {{Exactly Solved Models}} in {{Statistical Physics}}}}},\ \bibinfo {edition} {2nd}\ ed.,\ Oxford {{Graduate Texts}}\ (\bibinfo  {publisher} {Oxford University Press},\ \bibinfo {address} {Oxford, New York},\ \bibinfo {year} {2020})\BibitemShut {NoStop}%
\bibitem [{ndg()}]{ndga}%
  \BibitemOpen
  \href@noop {} {}\bibinfo {note} {This is the digamma function, which is defined as $\psi^{\prime}(x)=\Gamma'(x)/\Gamma(x)$.}\BibitemShut {Stop}%
\bibitem [{ndi()}]{ndims}%
  \BibitemOpen
  \href@noop {} {}\bibinfo {note} {An attentive reader may observe that the dimensions on either side of the equation do not explicitly match; omitted from the main text are dimensionful pre-factors introduced by the model-specific lattice-to-field mapping that ultimately ensure the correct overall scaling dimension of the observable to all orders $\mathcal{O}(g^n)$--- see \cite{SM}.}\BibitemShut {Stop}%
\bibitem [{\citenamefont {Deutsch}(1991)}]{Deutsch_1991}%
  \BibitemOpen
  \bibfield  {author} {\bibinfo {author} {\bibfnamefont {J.~M.}\ \bibnamefont {Deutsch}},\ }\bibfield  {title} {\enquote {\bibinfo {title} {Quantum statistical mechanics in a closed system},}\ }\href {\doibase 10.1103/PhysRevA.43.2046} {\bibfield  {journal} {\bibinfo  {journal} {Physical Review A}\ }\textbf {\bibinfo {volume} {43}},\ \bibinfo {pages} {2046–2049} (\bibinfo {year} {1991})}\BibitemShut {NoStop}%
\bibitem [{\citenamefont {Srednicki}(1994)}]{Srednicki_1994}%
  \BibitemOpen
  \bibfield  {author} {\bibinfo {author} {\bibfnamefont {Mark}\ \bibnamefont {Srednicki}},\ }\bibfield  {title} {\enquote {\bibinfo {title} {Chaos and quantum thermalization},}\ }\href {\doibase 10.1103/PhysRevE.50.888} {\bibfield  {journal} {\bibinfo  {journal} {Physical Review E}\ }\textbf {\bibinfo {volume} {50}},\ \bibinfo {pages} {888–901} (\bibinfo {year} {1994})}\BibitemShut {NoStop}%
\bibitem [{\citenamefont {Rigol}\ \emph {et~al.}(2008)\citenamefont {Rigol}, \citenamefont {Dunjko},\ and\ \citenamefont {Olshanii}}]{Rigol_Dunjko_Olshanii_2008}%
  \BibitemOpen
  \bibfield  {author} {\bibinfo {author} {\bibfnamefont {Marcos}\ \bibnamefont {Rigol}}, \bibinfo {author} {\bibfnamefont {Vanja}\ \bibnamefont {Dunjko}}, \ and\ \bibinfo {author} {\bibfnamefont {Maxim}\ \bibnamefont {Olshanii}},\ }\bibfield  {title} {\enquote {\bibinfo {title} {Thermalization and its mechanism for generic isolated quantum systems},}\ }\href {\doibase 10.1038/nature06838} {\bibfield  {journal} {\bibinfo  {journal} {Nature}\ }\textbf {\bibinfo {volume} {452}},\ \bibinfo {pages} {854–858} (\bibinfo {year} {2008})}\BibitemShut {NoStop}%
\bibitem [{nlt()}]{nltc}%
  \BibitemOpen
  \href@noop {} {}\bibinfo {note} {We note that if the full lattice-to-field mapping is known, then the scaling forms Eqn.~\eqref{eqn:g^1} and Eqn.~\eqref{eqn:g^2} allow the extraction of arbitrary scaling dimensions. However, even when the lattice-to-field mapping is not fully known, experimentally, one can still simply measure the ratios of the scaling dimensions of the most relevant primary fields distinguished by symmetry. In fact, this offers an experimental strategy for extracting the lattice-to-field mapping.}\BibitemShut {Stop}%
\bibitem [{\citenamefont {Cardy}(1984{\natexlab{a}})}]{cardy1984conformal}%
  \BibitemOpen
  \bibfield  {author} {\bibinfo {author} {\bibfnamefont {John~L}\ \bibnamefont {Cardy}},\ }\bibfield  {title} {\enquote {\bibinfo {title} {Conformal invariance and universality in finite-size scaling},}\ }\href@noop {} {\bibfield  {journal} {\bibinfo  {journal} {Journal of Physics A: Mathematical and General}\ }\textbf {\bibinfo {volume} {17}},\ \bibinfo {pages} {L385} (\bibinfo {year} {1984}{\natexlab{a}})}\BibitemShut {NoStop}%
\bibitem [{lit()}]{littlejohn}%
  \BibitemOpen
  \href@noop {} {\enquote {\bibinfo {title} {Quantum mechanics},}\ }\bibinfo {howpublished} {\url{http://bohr.physics.berkeley.edu/classes/221/2122/221.html}},\ \bibinfo {note} {accessed: 2022-5-18}\BibitemShut {NoStop}%
\bibitem [{\citenamefont {Schuricht}\ and\ \citenamefont {Essler}(2012)}]{schuricht2012dynamics}%
  \BibitemOpen
  \bibfield  {author} {\bibinfo {author} {\bibfnamefont {Dirk}\ \bibnamefont {Schuricht}}\ and\ \bibinfo {author} {\bibfnamefont {Fabian~HL}\ \bibnamefont {Essler}},\ }\bibfield  {title} {\enquote {\bibinfo {title} {Dynamics in the ising field theory after a quantum quench},}\ }\href@noop {} {\bibfield  {journal} {\bibinfo  {journal} {Journal of Statistical Mechanics: Theory and Experiment}\ }\textbf {\bibinfo {volume} {2012}},\ \bibinfo {pages} {P04017} (\bibinfo {year} {2012})}\BibitemShut {NoStop}%
\bibitem [{\citenamefont {Dupont}\ \emph {et~al.}(2021)\citenamefont {Dupont}, \citenamefont {Sherman},\ and\ \citenamefont {Moore}}]{dupont2021spatiotemporal}%
  \BibitemOpen
  \bibfield  {author} {\bibinfo {author} {\bibfnamefont {Maxime}\ \bibnamefont {Dupont}}, \bibinfo {author} {\bibfnamefont {Nicholas~E}\ \bibnamefont {Sherman}}, \ and\ \bibinfo {author} {\bibfnamefont {Joel~E}\ \bibnamefont {Moore}},\ }\bibfield  {title} {\enquote {\bibinfo {title} {Spatiotemporal crossover between low-and high-temperature dynamical regimes in the quantum heisenberg magnet},}\ }\href@noop {} {\bibfield  {journal} {\bibinfo  {journal} {Physical review letters}\ }\textbf {\bibinfo {volume} {127}},\ \bibinfo {pages} {107201} (\bibinfo {year} {2021})}\BibitemShut {NoStop}%
\bibitem [{\citenamefont {Ye}\ \emph {et~al.}(2022)\citenamefont {Ye}, \citenamefont {Machado}, \citenamefont {Kemp}, \citenamefont {Hutson},\ and\ \citenamefont {Yao}}]{ye2022universal}%
  \BibitemOpen
  \bibfield  {author} {\bibinfo {author} {\bibfnamefont {Bingtian}\ \bibnamefont {Ye}}, \bibinfo {author} {\bibfnamefont {Francisco}\ \bibnamefont {Machado}}, \bibinfo {author} {\bibfnamefont {Jack}\ \bibnamefont {Kemp}}, \bibinfo {author} {\bibfnamefont {Ross~B}\ \bibnamefont {Hutson}}, \ and\ \bibinfo {author} {\bibfnamefont {Norman~Y}\ \bibnamefont {Yao}},\ }\bibfield  {title} {\enquote {\bibinfo {title} {Universal kardar-parisi-zhang dynamics in integrable quantum systems},}\ }\href@noop {} {\bibfield  {journal} {\bibinfo  {journal} {Physical review letters}\ }\textbf {\bibinfo {volume} {129}},\ \bibinfo {pages} {230602} (\bibinfo {year} {2022})}\BibitemShut {NoStop}%
\bibitem [{\citenamefont {Ye}\ \emph {et~al.}(2020)\citenamefont {Ye}, \citenamefont {Machado}, \citenamefont {White}, \citenamefont {Mong},\ and\ \citenamefont {Yao}}]{ye2020emergent}%
  \BibitemOpen
  \bibfield  {author} {\bibinfo {author} {\bibfnamefont {Bingtian}\ \bibnamefont {Ye}}, \bibinfo {author} {\bibfnamefont {Francisco}\ \bibnamefont {Machado}}, \bibinfo {author} {\bibfnamefont {Christopher~David}\ \bibnamefont {White}}, \bibinfo {author} {\bibfnamefont {Roger~SK}\ \bibnamefont {Mong}}, \ and\ \bibinfo {author} {\bibfnamefont {Norman~Y}\ \bibnamefont {Yao}},\ }\bibfield  {title} {\enquote {\bibinfo {title} {Emergent hydrodynamics in nonequilibrium quantum systems},}\ }\href@noop {} {\bibfield  {journal} {\bibinfo  {journal} {Physical Review Letters}\ }\textbf {\bibinfo {volume} {125}},\ \bibinfo {pages} {030601} (\bibinfo {year} {2020})}\BibitemShut {NoStop}%
\bibitem [{\citenamefont {Kumar}(1965)}]{kumar1965expanding}%
  \BibitemOpen
  \bibfield  {author} {\bibinfo {author} {\bibfnamefont {Kailash}\ \bibnamefont {Kumar}},\ }\bibfield  {title} {\enquote {\bibinfo {title} {On expanding the exponential},}\ }\href@noop {} {\bibfield  {journal} {\bibinfo  {journal} {Journal of Mathematical Physics}\ }\textbf {\bibinfo {volume} {6}},\ \bibinfo {pages} {1928--1934} (\bibinfo {year} {1965})}\BibitemShut {NoStop}%
\bibitem [{\citenamefont {McArdle}\ \emph {et~al.}(2019)\citenamefont {McArdle}, \citenamefont {Jones}, \citenamefont {Endo}, \citenamefont {Li}, \citenamefont {Benjamin},\ and\ \citenamefont {Yuan}}]{mcardle2019variational}%
  \BibitemOpen
  \bibfield  {author} {\bibinfo {author} {\bibfnamefont {Sam}\ \bibnamefont {McArdle}}, \bibinfo {author} {\bibfnamefont {Tyson}\ \bibnamefont {Jones}}, \bibinfo {author} {\bibfnamefont {Suguru}\ \bibnamefont {Endo}}, \bibinfo {author} {\bibfnamefont {Ying}\ \bibnamefont {Li}}, \bibinfo {author} {\bibfnamefont {Simon~C.}\ \bibnamefont {Benjamin}}, \ and\ \bibinfo {author} {\bibfnamefont {Xiao}\ \bibnamefont {Yuan}},\ }\bibfield  {title} {\enquote {\bibinfo {title} {Variational ansatz-based quantum simulation of imaginary time evolution},}\ }\href {\doibase 10.1038/s41534-019-0187-2} {\bibfield  {journal} {\bibinfo  {journal} {npj Quantum Information}\ }\textbf {\bibinfo {volume} {5}},\ \bibinfo {pages} {75} (\bibinfo {year} {2019})}\BibitemShut {NoStop}%
\bibitem [{\citenamefont {Motta}\ \emph {et~al.}(2020)\citenamefont {Motta}, \citenamefont {Sun}, \citenamefont {Tan}, \citenamefont {O'Rourke}, \citenamefont {Ye}, \citenamefont {Minnich}, \citenamefont {Brand{\~a}o},\ and\ \citenamefont {Chan}}]{motta2020determining}%
  \BibitemOpen
  \bibfield  {author} {\bibinfo {author} {\bibfnamefont {Mario}\ \bibnamefont {Motta}}, \bibinfo {author} {\bibfnamefont {Chong}\ \bibnamefont {Sun}}, \bibinfo {author} {\bibfnamefont {Adrian T.~K.}\ \bibnamefont {Tan}}, \bibinfo {author} {\bibfnamefont {Matthew~J.}\ \bibnamefont {O'Rourke}}, \bibinfo {author} {\bibfnamefont {Erika}\ \bibnamefont {Ye}}, \bibinfo {author} {\bibfnamefont {Austin~J.}\ \bibnamefont {Minnich}}, \bibinfo {author} {\bibfnamefont {Fernando G. S.~L.}\ \bibnamefont {Brand{\~a}o}}, \ and\ \bibinfo {author} {\bibfnamefont {Garnet Kin-Lic}\ \bibnamefont {Chan}},\ }\bibfield  {title} {\enquote {\bibinfo {title} {Determining eigenstates and thermal states on a quantum computer using quantum imaginary time evolution},}\ }\href {\doibase 10.1038/s41567-019-0704-4} {\bibfield  {journal} {\bibinfo  {journal} {Nature Physics}\ }\textbf {\bibinfo {volume} {16}},\ \bibinfo {pages} {205--210} (\bibinfo {year} {2020})}\BibitemShut {NoStop}%
\bibitem [{\citenamefont {Di~Francesco}\ \emph {et~al.}(1997)\citenamefont {Di~Francesco}, \citenamefont {Mathieu},\ and\ \citenamefont {Sénéchal}}]{Di_Francesco_Mathieu_Sénéchal_1997}%
  \BibitemOpen
  \bibfield  {author} {\bibinfo {author} {\bibfnamefont {Philippe}\ \bibnamefont {Di~Francesco}}, \bibinfo {author} {\bibfnamefont {Pierre}\ \bibnamefont {Mathieu}}, \ and\ \bibinfo {author} {\bibfnamefont {David}\ \bibnamefont {Sénéchal}},\ }\href {\doibase 10.1007/978-1-4612-2256-9} {\emph {\bibinfo {title} {Conformal Field Theory}}},\ Graduate Texts in Contemporary Physics\ (\bibinfo  {publisher} {Springer},\ \bibinfo {address} {New York, NY},\ \bibinfo {year} {1997})\BibitemShut {NoStop}%
\bibitem [{\citenamefont {Surace}\ and\ \citenamefont {Motrunich}(2023)}]{Surace_Motrunich_2023}%
  \BibitemOpen
  \bibfield  {author} {\bibinfo {author} {\bibfnamefont {Federica~Maria}\ \bibnamefont {Surace}}\ and\ \bibinfo {author} {\bibfnamefont {Olexei}\ \bibnamefont {Motrunich}},\ }\bibfield  {title} {\enquote {\bibinfo {title} {Weak integrability breaking perturbations of integrable models},}\ }\href {\doibase 10.1103/PhysRevResearch.5.043019} {\bibfield  {journal} {\bibinfo  {journal} {Physical Review Research}\ }\textbf {\bibinfo {volume} {5}},\ \bibinfo {pages} {043019} (\bibinfo {year} {2023})}\BibitemShut {NoStop}%
\bibitem [{\citenamefont {Canovi}\ \emph {et~al.}(2011)\citenamefont {Canovi}, \citenamefont {Rossini}, \citenamefont {Fazio}, \citenamefont {Santoro},\ and\ \citenamefont {Silva}}]{Canovi_Rossini_Fazio_Santoro_Silva_2011}%
  \BibitemOpen
  \bibfield  {author} {\bibinfo {author} {\bibfnamefont {Elena}\ \bibnamefont {Canovi}}, \bibinfo {author} {\bibfnamefont {Davide}\ \bibnamefont {Rossini}}, \bibinfo {author} {\bibfnamefont {Rosario}\ \bibnamefont {Fazio}}, \bibinfo {author} {\bibfnamefont {Giuseppe~E.}\ \bibnamefont {Santoro}}, \ and\ \bibinfo {author} {\bibfnamefont {Alessandro}\ \bibnamefont {Silva}},\ }\bibfield  {title} {\enquote {\bibinfo {title} {Quantum quenches, thermalization and many-body localization},}\ }\href {\doibase 10.1103/PhysRevB.83.094431} {\bibfield  {journal} {\bibinfo  {journal} {Physical Review B}\ }\textbf {\bibinfo {volume} {83}},\ \bibinfo {pages} {094431} (\bibinfo {year} {2011})},\ \bibinfo {note} {arXiv:1006.1634 [cond-mat]}\BibitemShut {NoStop}%
\bibitem [{\citenamefont {Michailidis}\ \emph {et~al.}(2020)\citenamefont {Michailidis}, \citenamefont {Turner}, \citenamefont {Papić}, \citenamefont {Abanin},\ and\ \citenamefont {Serbyn}}]{Michailidis_Turner_Papić_Abanin_Serbyn_2020}%
  \BibitemOpen
  \bibfield  {author} {\bibinfo {author} {\bibfnamefont {A.~A.}\ \bibnamefont {Michailidis}}, \bibinfo {author} {\bibfnamefont {C.~J.}\ \bibnamefont {Turner}}, \bibinfo {author} {\bibfnamefont {Z.}~\bibnamefont {Papić}}, \bibinfo {author} {\bibfnamefont {D.~A.}\ \bibnamefont {Abanin}}, \ and\ \bibinfo {author} {\bibfnamefont {M.}~\bibnamefont {Serbyn}},\ }\bibfield  {title} {\enquote {\bibinfo {title} {Slow quantum thermalization and many-body revivals from mixed phase space},}\ }\href {\doibase 10.1103/PhysRevX.10.011055} {\bibfield  {journal} {\bibinfo  {journal} {Physical Review X}\ }\textbf {\bibinfo {volume} {10}},\ \bibinfo {pages} {011055} (\bibinfo {year} {2020})},\ \bibinfo {note} {arXiv:1905.08564 [quant-ph]}\BibitemShut {NoStop}%
\bibitem [{\citenamefont {Mori}\ \emph {et~al.}(2018)\citenamefont {Mori}, \citenamefont {Ikeda}, \citenamefont {Kaminishi},\ and\ \citenamefont {Ueda}}]{Mori_Ikeda_Kaminishi_Ueda_2018}%
  \BibitemOpen
  \bibfield  {author} {\bibinfo {author} {\bibfnamefont {Takashi}\ \bibnamefont {Mori}}, \bibinfo {author} {\bibfnamefont {Tatsuhiko~N.}\ \bibnamefont {Ikeda}}, \bibinfo {author} {\bibfnamefont {Eriko}\ \bibnamefont {Kaminishi}}, \ and\ \bibinfo {author} {\bibfnamefont {Masahito}\ \bibnamefont {Ueda}},\ }\bibfield  {title} {\enquote {\bibinfo {title} {Thermalization and prethermalization in isolated quantum systems: a theoretical overview},}\ }\href {\doibase 10.1088/1361-6455/aabcdf} {\bibfield  {journal} {\bibinfo  {journal} {Journal of Physics B: Atomic, Molecular and Optical Physics}\ }\textbf {\bibinfo {volume} {51}},\ \bibinfo {pages} {112001} (\bibinfo {year} {2018})}\BibitemShut {NoStop}%
\bibitem [{\citenamefont {Abanin}\ \emph {et~al.}(2019)\citenamefont {Abanin}, \citenamefont {Altman}, \citenamefont {Bloch},\ and\ \citenamefont {Serbyn}}]{Abanin_Altman_Bloch_Serbyn_2019}%
  \BibitemOpen
  \bibfield  {author} {\bibinfo {author} {\bibfnamefont {Dmitry~A.}\ \bibnamefont {Abanin}}, \bibinfo {author} {\bibfnamefont {Ehud}\ \bibnamefont {Altman}}, \bibinfo {author} {\bibfnamefont {Immanuel}\ \bibnamefont {Bloch}}, \ and\ \bibinfo {author} {\bibfnamefont {Maksym}\ \bibnamefont {Serbyn}},\ }\bibfield  {title} {{\selectlanguage {english}\enquote {\bibinfo {title} {Colloquium: Many-body localization, thermalization, and entanglement},}\ }}\href {\doibase 10.1103/RevModPhys.91.021001} {\bibfield  {journal} {\bibinfo  {journal} {Reviews of Modern Physics}\ }\textbf {\bibinfo {volume} {91}},\ \bibinfo {pages} {021001} (\bibinfo {year} {2019})}\BibitemShut {NoStop}%
\bibitem [{\citenamefont {Cardy}(1984{\natexlab{b}})}]{Cardy_1984}%
  \BibitemOpen
  \bibfield  {author} {\bibinfo {author} {\bibfnamefont {John~L.}\ \bibnamefont {Cardy}},\ }\bibfield  {title} {\enquote {\bibinfo {title} {Conformal invariance and surface critical behavior},}\ }\href {\doibase 10.1016/0550-3213(84)90241-4} {\bibfield  {journal} {\bibinfo  {journal} {Nuclear Physics B}\ }\textbf {\bibinfo {volume} {240}},\ \bibinfo {pages} {514–532} (\bibinfo {year} {1984}{\natexlab{b}})}\BibitemShut {NoStop}%
\bibitem [{\citenamefont {Cardy}(1986)}]{Cardy_1986}%
  \BibitemOpen
  \bibfield  {author} {\bibinfo {author} {\bibfnamefont {John~L.}\ \bibnamefont {Cardy}},\ }\bibfield  {title} {\enquote {\bibinfo {title} {Operator content of two-dimensional conformally invariant theories},}\ }\href {\doibase 10.1016/0550-3213(86)90552-3} {\bibfield  {journal} {\bibinfo  {journal} {Nuclear Physics B}\ }\textbf {\bibinfo {volume} {270}},\ \bibinfo {pages} {186–204} (\bibinfo {year} {1986})}\BibitemShut {NoStop}%
\bibitem [{\citenamefont {Belavin}\ \emph {et~al.}(1984)\citenamefont {Belavin}, \citenamefont {Polyakov},\ and\ \citenamefont {Zamolodchikov}}]{Belavin_Polyakov_Zamolodchikov_1984}%
  \BibitemOpen
  \bibfield  {author} {\bibinfo {author} {\bibfnamefont {A~A}\ \bibnamefont {Belavin}}, \bibinfo {author} {\bibfnamefont {A~M}\ \bibnamefont {Polyakov}}, \ and\ \bibinfo {author} {\bibfnamefont {A~B}\ \bibnamefont {Zamolodchikov}},\ }\bibfield  {title} {{\selectlanguage {english}\enquote {\bibinfo {title} {Infinite conformal symmetry in two-dimensional quantum field theory},}\ }}\href@noop {} {\bibfield  {journal} {\bibinfo  {journal} {Nuclear Physics B}\ }\textbf {\bibinfo {volume} {241}} (\bibinfo {year} {1984})}\BibitemShut {NoStop}%
\bibitem [{\citenamefont {Zamolodchikov}(1987)}]{Zamolodchikov_1987}%
  \BibitemOpen
  \bibfield  {author} {\bibinfo {author} {\bibfnamefont {A.~B.}\ \bibnamefont {Zamolodchikov}},\ }\bibfield  {title} {\enquote {\bibinfo {title} {Renormalization group and perturbation theory near fixed points in two-dimensional field theory},}\ }\href@noop {} {\bibfield  {journal} {\bibinfo  {journal} {Sov. J. Nucl. Phys.}\ }\textbf {\bibinfo {volume} {46}},\ \bibinfo {pages} {1090} (\bibinfo {year} {1987})}\BibitemShut {NoStop}%
\bibitem [{\citenamefont {Zamolodchikov}(1989)}]{Zamolodchikov_1989}%
  \BibitemOpen
  \bibfield  {author} {\bibinfo {author} {\bibfnamefont {A.~B.}\ \bibnamefont {Zamolodchikov}},\ }\enquote {\bibinfo {title} {Integrable field theory from conformal field theory},}\ in\ \href {\doibase 10.2969/aspm/01910641} {\emph {\bibinfo {booktitle} {Integrable Systems in Quantum Field Theory and Statistical Mechanics}}},\ Vol.~\bibinfo {volume} {19}\ (\bibinfo  {publisher} {Mathematical Society of Japan},\ \bibinfo {year} {1989})\ p.\ \bibinfo {pages} {641–675}\BibitemShut {NoStop}%
\bibitem [{\citenamefont {Maldacena}(1999)}]{Maldacena_1999}%
  \BibitemOpen
  \bibfield  {author} {\bibinfo {author} {\bibfnamefont {Juan~M.}\ \bibnamefont {Maldacena}},\ }\bibfield  {title} {\enquote {\bibinfo {title} {The large n limit of superconformal field theories and supergravity},}\ }\href@noop {} {\bibfield  {journal} {\bibinfo  {journal} {International Journal of Theoretical Physics}\ }\textbf {\bibinfo {volume} {38}},\ \bibinfo {pages} {1113–1133} (\bibinfo {year} {1999})}\BibitemShut {NoStop}%
\bibitem [{\citenamefont {Da{\u g}}\ and\ \citenamefont {Sun}(2021)}]{dag:2021}%
  \BibitemOpen
  \bibfield  {author} {\bibinfo {author} {\bibfnamefont {Ceren~B.}\ \bibnamefont {Da{\u g}}}\ and\ \bibinfo {author} {\bibfnamefont {Kai}\ \bibnamefont {Sun}},\ }\bibfield  {title} {\enquote {\bibinfo {title} {Dynamical crossover in the transient quench dynamics of short-range transverse field {{Ising}} models},}\ }\href {\doibase 10.1103/PhysRevB.103.214402} {\bibfield  {journal} {\bibinfo  {journal} {Physical Review B}\ }\textbf {\bibinfo {volume} {103}},\ \bibinfo {pages} {214402} (\bibinfo {year} {2021})},\ \Eprint {http://arxiv.org/abs/2004.12287} {arXiv:2004.12287 [cond-mat, physics:quant-ph]} \BibitemShut {NoStop}%
\bibitem [{\citenamefont {Berenstein}\ and\ \citenamefont {Miller}(2014)}]{Berenstein_Miller_2014}%
  \BibitemOpen
  \bibfield  {author} {\bibinfo {author} {\bibfnamefont {David}\ \bibnamefont {Berenstein}}\ and\ \bibinfo {author} {\bibfnamefont {Alexandra}\ \bibnamefont {Miller}},\ }\bibfield  {title} {\enquote {\bibinfo {title} {Conformal perturbation theory, dimensional regularization and ads/cft},}\ }\href {\doibase 10.1103/PhysRevD.90.086011} {\bibfield  {journal} {\bibinfo  {journal} {Physical Review D}\ }\textbf {\bibinfo {volume} {90}},\ \bibinfo {pages} {086011} (\bibinfo {year} {2014})}\BibitemShut {NoStop}%
\bibitem [{\citenamefont {Haravifard}\ \emph {et~al.}(2015)\citenamefont {Haravifard}, \citenamefont {Yamani},\ and\ \citenamefont {Gaulin}}]{Haravifard_Yamani_Gaulin_2015}%
  \BibitemOpen
  \bibfield  {author} {\bibinfo {author} {\bibfnamefont {Sara}\ \bibnamefont {Haravifard}}, \bibinfo {author} {\bibfnamefont {Zahra}\ \bibnamefont {Yamani}}, \ and\ \bibinfo {author} {\bibfnamefont {Bruce~D.}\ \bibnamefont {Gaulin}},\ }\enquote {\bibinfo {title} {Chapter 2 - quantum phase transitions},}\ in\ \href {\doibase 10.1016/B978-0-12-802049-4.00002-6} {\emph {\bibinfo {booktitle} {Experimental Methods in the Physical Sciences}}},\ \bibinfo {series} {Neutron Scattering - Magnetic and Quantum Phenomena}, Vol.~\bibinfo {volume} {48},\ \bibinfo {editor} {edited by\ \bibinfo {editor} {\bibfnamefont {Felix}\ \bibnamefont {Fernandez-Alonso}}\ and\ \bibinfo {editor} {\bibfnamefont {David~L.}\ \bibnamefont {Price}}}\ (\bibinfo  {publisher} {Academic Press},\ \bibinfo {year} {2015})\ p.\ \bibinfo {pages} {43–144}\BibitemShut {NoStop}%
\bibitem [{\citenamefont {Mallayya}\ \emph {et~al.}(2019)\citenamefont {Mallayya}, \citenamefont {Rigol},\ and\ \citenamefont {Roeck}}]{Mallayya_Rigol_Roeck_2019}%
  \BibitemOpen
  \bibfield  {author} {\bibinfo {author} {\bibfnamefont {Krishnanand}\ \bibnamefont {Mallayya}}, \bibinfo {author} {\bibfnamefont {Marcos}\ \bibnamefont {Rigol}}, \ and\ \bibinfo {author} {\bibfnamefont {Wojciech~De}\ \bibnamefont {Roeck}},\ }\bibfield  {title} {\enquote {\bibinfo {title} {Prethermalization and thermalization in isolated quantum systems},}\ }\href {\doibase 10.1103/PhysRevX.9.021027} {\bibfield  {journal} {\bibinfo  {journal} {Physical Review X}\ }\textbf {\bibinfo {volume} {9}} (\bibinfo {year} {2019}),\ 10.1103/PhysRevX.9.021027}\BibitemShut {NoStop}%
\bibitem [{\citenamefont {Stark}\ and\ \citenamefont {Kollar}(2013)}]{Stark_Kollar_2013}%
  \BibitemOpen
  \bibfield  {author} {\bibinfo {author} {\bibfnamefont {Michael}\ \bibnamefont {Stark}}\ and\ \bibinfo {author} {\bibfnamefont {Marcus}\ \bibnamefont {Kollar}},\ }\bibfield  {title} {\enquote {\bibinfo {title} {Kinetic description of thermalization dynamics in weakly interacting quantum systems},}\ }\href {\doibase 10.48550/arXiv.1308.1610} {\  (\bibinfo {year} {2013}),\ 10.48550/arXiv.1308.1610},\ \bibinfo {note} {arXiv:1308.1610 [cond-mat]}\BibitemShut {NoStop}%
\end{thebibliography}%


\begin{thebibliography}{25}%
\makeatletter
\providecommand \@ifxundefined [1]{%
 \@ifx{#1\undefined}
}%
\providecommand \@ifnum [1]{%
 \ifnum #1\expandafter \@firstoftwo
 \else \expandafter \@secondoftwo
 \fi
}%
\providecommand \@ifx [1]{%
 \ifx #1\expandafter \@firstoftwo
 \else \expandafter \@secondoftwo
 \fi
}%
\providecommand \natexlab [1]{#1}%
\providecommand \enquote  [1]{``#1''}%
\providecommand \bibnamefont  [1]{#1}%
\providecommand \bibfnamefont [1]{#1}%
\providecommand \citenamefont [1]{#1}%
\providecommand \href@noop [0]{\@secondoftwo}%
\providecommand \href [0]{\begingroup \@sanitize@url \@href}%
\providecommand \@href[1]{\@@startlink{#1}\@@href}%
\providecommand \@@href[1]{\endgroup#1\@@endlink}%
\providecommand \@sanitize@url [0]{\catcode `\\12\catcode `\$12\catcode `\&12\catcode `\#12\catcode `\^12\catcode `\_12\catcode `\%12\relax}%
\providecommand \@@startlink[1]{}%
\providecommand \@@endlink[0]{}%
\providecommand \url  [0]{\begingroup\@sanitize@url \@url }%
\providecommand \@url [1]{\endgroup\@href {#1}{\urlprefix }}%
\providecommand \urlprefix  [0]{URL }%
\providecommand \Eprint [0]{\href }%
\providecommand \doibase [0]{http://dx.doi.org/}%
\providecommand \selectlanguage [0]{\@gobble}%
\providecommand \bibinfo  [0]{\@secondoftwo}%
\providecommand \bibfield  [0]{\@secondoftwo}%
\providecommand \translation [1]{[#1]}%
\providecommand \BibitemOpen [0]{}%
\providecommand \bibitemStop [0]{}%
\providecommand \bibitemNoStop [0]{.\EOS\space}%
\providecommand \EOS [0]{\spacefactor3000\relax}%
\providecommand \BibitemShut  [1]{\csname bibitem#1\endcsname}%
\let\auto@bib@innerbib\@empty
\bibitem [{\citenamefont {Mong}\ \emph {et~al.}(2014)\citenamefont {Mong}, \citenamefont {Clarke}, \citenamefont {Alicea}, \citenamefont {Lindner},\ and\ \citenamefont {Fendley}}]{S_mong2014parafermionic}%
  \BibitemOpen
  \bibfield  {author} {\bibinfo {author} {\bibfnamefont {R.~S.~K.}\ \bibnamefont {Mong}}, \bibinfo {author} {\bibfnamefont {D.~J.}\ \bibnamefont {Clarke}}, \bibinfo {author} {\bibfnamefont {J.}~\bibnamefont {Alicea}}, \bibinfo {author} {\bibfnamefont {N.~H.}\ \bibnamefont {Lindner}}, \ and\ \bibinfo {author} {\bibfnamefont {P.}~\bibnamefont {Fendley}},\ }\href {\doibase 10.1088/1751-8113/47/45/452001} {\bibfield  {journal} {\bibinfo  {journal} {Journal of Physics A: Mathematical and Theoretical}\ }\textbf {\bibinfo {volume} {47}},\ \bibinfo {pages} {452001} (\bibinfo {year} {2014})}\BibitemShut {NoStop}%
\bibitem [{\citenamefont {Hauschild}\ \emph {et~al.}(2024)\citenamefont {Hauschild}, \citenamefont {Unfried}, \citenamefont {Anand}, \citenamefont {Andrews}, \citenamefont {Bintz}, \citenamefont {Borla}, \citenamefont {Divic}, \citenamefont {Drescher}, \citenamefont {Geiger}, \citenamefont {Hefel}, \citenamefont {H{\'e}mery}, \citenamefont {Kadow}, \citenamefont {Kemp}, \citenamefont {Kirchner}, \citenamefont {Liu}, \citenamefont {Moller}, \citenamefont {Parker}, \citenamefont {Rader}, \citenamefont {Romen}, \citenamefont {Scalet}, \citenamefont {Schoonderwoerd}, \citenamefont {Schulz}, \citenamefont {Soejima}, \citenamefont {Thoma}, \citenamefont {Wu}, \citenamefont {Zechmann}, \citenamefont {Zweng}, \citenamefont {Mong}, \citenamefont {Zaletel},\ and\ \citenamefont {Pollmann}}]{S_hauschild:2024}%
  \BibitemOpen
  \bibfield  {author} {\bibinfo {author} {\bibfnamefont {J.}~\bibnamefont {Hauschild}}, \bibinfo {author} {\bibfnamefont {J.}~\bibnamefont {Unfried}}, \bibinfo {author} {\bibfnamefont {S.}~\bibnamefont {Anand}}, \bibinfo {author} {\bibfnamefont {B.}~\bibnamefont {Andrews}}, \bibinfo {author} {\bibfnamefont {M.}~\bibnamefont {Bintz}}, \bibinfo {author} {\bibfnamefont {U.}~\bibnamefont {Borla}}, \bibinfo {author} {\bibfnamefont {S.}~\bibnamefont {Divic}}, \bibinfo {author} {\bibfnamefont {M.}~\bibnamefont {Drescher}}, \bibinfo {author} {\bibfnamefont {J.}~\bibnamefont {Geiger}}, \bibinfo {author} {\bibfnamefont {M.}~\bibnamefont {Hefel}}, \bibinfo {author} {\bibfnamefont {K.}~\bibnamefont {H{\'e}mery}}, \bibinfo {author} {\bibfnamefont {W.}~\bibnamefont {Kadow}}, \bibinfo {author} {\bibfnamefont {J.}~\bibnamefont {Kemp}}, \bibinfo {author} {\bibfnamefont {N.}~\bibnamefont {Kirchner}}, \bibinfo {author} {\bibfnamefont {V.~S.}\ \bibnamefont {Liu}}, \bibinfo {author} {\bibfnamefont {G.}~\bibnamefont {Moller}},
  \bibinfo {author} {\bibfnamefont {D.}~\bibnamefont {Parker}}, \bibinfo {author} {\bibfnamefont {M.}~\bibnamefont {Rader}}, \bibinfo {author} {\bibfnamefont {A.}~\bibnamefont {Romen}}, \bibinfo {author} {\bibfnamefont {S.}~\bibnamefont {Scalet}}, \bibinfo {author} {\bibfnamefont {L.}~\bibnamefont {Schoonderwoerd}}, \bibinfo {author} {\bibfnamefont {M.}~\bibnamefont {Schulz}}, \bibinfo {author} {\bibfnamefont {T.}~\bibnamefont {Soejima}}, \bibinfo {author} {\bibfnamefont {P.}~\bibnamefont {Thoma}}, \bibinfo {author} {\bibfnamefont {Y.}~\bibnamefont {Wu}}, \bibinfo {author} {\bibfnamefont {P.}~\bibnamefont {Zechmann}}, \bibinfo {author} {\bibfnamefont {L.}~\bibnamefont {Zweng}}, \bibinfo {author} {\bibfnamefont {R.}~\bibnamefont {Mong}}, \bibinfo {author} {\bibfnamefont {M.~P.}\ \bibnamefont {Zaletel}}, \ and\ \bibinfo {author} {\bibfnamefont {F.}~\bibnamefont {Pollmann}},\ }\href {\doibase 10.21468/SciPostPhysCodeb.41} {\bibfield  {journal} {\bibinfo  {journal} {SciPost Physics Codebases}\ ,\ \bibinfo {pages}
  {041}} (\bibinfo {year} {2024})}\BibitemShut {NoStop}%
\bibitem [{\citenamefont {Barthel}\ and\ \citenamefont {Zhang}(2020)}]{S_barthelOptimizedLieTrotter2020}%
  \BibitemOpen
  \bibfield  {author} {\bibinfo {author} {\bibfnamefont {T.}~\bibnamefont {Barthel}}\ and\ \bibinfo {author} {\bibfnamefont {Y.}~\bibnamefont {Zhang}},\ }\href {\doibase 10.1016/j.aop.2020.168165} {\bibfield  {journal} {\bibinfo  {journal} {Annals of Physics}\ }\textbf {\bibinfo {volume} {418}},\ \bibinfo {pages} {168165} (\bibinfo {year} {2020})}\BibitemShut {NoStop}%
\bibitem [{\citenamefont {Schollw{\"o}ck}(2011)}]{S_SCHOLLWOCK201196}%
  \BibitemOpen
  \bibfield  {author} {\bibinfo {author} {\bibfnamefont {U.}~\bibnamefont {Schollw{\"o}ck}},\ }\href {\doibase 10.1016/j.aop.2010.09.012} {\bibfield  {journal} {\bibinfo  {journal} {Annals of Physics}\ }\textbf {\bibinfo {volume} {326}},\ \bibinfo {pages} {96} (\bibinfo {year} {2011})}\BibitemShut {NoStop}%
\bibitem [{\citenamefont {Sachdev}(2011)}]{S_sachdevQuantumPhaseTransitions2011}%
  \BibitemOpen
  \bibfield  {author} {\bibinfo {author} {\bibfnamefont {S.}~\bibnamefont {Sachdev}},\ }\href {\doibase 10.1017/CBO9780511973765} {\emph {\bibinfo {title} {Quantum {{Phase Transitions}}}}},\ \bibinfo {edition} {2nd}\ ed.\ (\bibinfo  {publisher} {Cambridge University Press},\ \bibinfo {address} {Cambridge},\ \bibinfo {year} {2011})\BibitemShut {NoStop}%
\bibitem [{\citenamefont {Karrasch}\ \emph {et~al.}(2013)\citenamefont {Karrasch}, \citenamefont {Bardarson},\ and\ \citenamefont {Moore}}]{S_karraschReducingNumericalEffort2013}%
  \BibitemOpen
  \bibfield  {author} {\bibinfo {author} {\bibfnamefont {C.}~\bibnamefont {Karrasch}}, \bibinfo {author} {\bibfnamefont {J.~H.}\ \bibnamefont {Bardarson}}, \ and\ \bibinfo {author} {\bibfnamefont {J.~E.}\ \bibnamefont {Moore}},\ }\href {\doibase 10.1088/1367-2630/15/8/083031} {\bibfield  {journal} {\bibinfo  {journal} {New Journal of Physics}\ }\textbf {\bibinfo {volume} {15}},\ \bibinfo {pages} {083031} (\bibinfo {year} {2013})}\BibitemShut {NoStop}%
\bibitem [{\citenamefont {Calabrese}\ \emph {et~al.}(2011)\citenamefont {Calabrese}, \citenamefont {Essler},\ and\ \citenamefont {Fagotti}}]{S_Calabrese_Essler_Fagotti_2011}%
  \BibitemOpen
  \bibfield  {author} {\bibinfo {author} {\bibfnamefont {P.}~\bibnamefont {Calabrese}}, \bibinfo {author} {\bibfnamefont {F.~H.~L.}\ \bibnamefont {Essler}}, \ and\ \bibinfo {author} {\bibfnamefont {M.}~\bibnamefont {Fagotti}},\ }\href {\doibase 10.1103/PhysRevLett.106.227203} {\bibfield  {journal} {\bibinfo  {journal} {Physical Review Letters}\ }\textbf {\bibinfo {volume} {106}},\ \bibinfo {pages} {227203} (\bibinfo {year} {2011})}\BibitemShut {NoStop}%
\bibitem [{\citenamefont {Barouch}\ \emph {et~al.}(1970)\citenamefont {Barouch}, \citenamefont {McCoy},\ and\ \citenamefont {Dresden}}]{S_barouch1970statistical}%
  \BibitemOpen
  \bibfield  {author} {\bibinfo {author} {\bibfnamefont {E.}~\bibnamefont {Barouch}}, \bibinfo {author} {\bibfnamefont {B.~M.}\ \bibnamefont {McCoy}}, \ and\ \bibinfo {author} {\bibfnamefont {M.}~\bibnamefont {Dresden}},\ }\href {\doibase 10.1103/PhysRevA.2.1075} {\bibfield  {journal} {\bibinfo  {journal} {Phys. Rev. A}\ }\textbf {\bibinfo {volume} {2}},\ \bibinfo {pages} {1075} (\bibinfo {year} {1970})}\BibitemShut {NoStop}%
\bibitem [{\citenamefont {Di~Francesco}\ \emph {et~al.}(1997)\citenamefont {Di~Francesco}, \citenamefont {Mathieu},\ and\ \citenamefont {Sénéchal}}]{S_Di_Francesco_Mathieu_Sénéchal_1997}%
  \BibitemOpen
  \bibfield  {author} {\bibinfo {author} {\bibfnamefont {P.}~\bibnamefont {Di~Francesco}}, \bibinfo {author} {\bibfnamefont {P.}~\bibnamefont {Mathieu}}, \ and\ \bibinfo {author} {\bibfnamefont {D.}~\bibnamefont {Sénéchal}},\ }\href {\doibase 10.1007/978-1-4612-2256-9} {\emph {\bibinfo {title} {Conformal Field Theory}}},\ Graduate Texts in Contemporary Physics\ (\bibinfo  {publisher} {Springer},\ \bibinfo {address} {New York, NY},\ \bibinfo {year} {1997})\BibitemShut {NoStop}%
\bibitem [{Note1()}]{Note1}%
  \BibitemOpen
  \bibinfo {note} {This notation for the scaling dimension deviates from the main text's convention, $x_\Phi $, in order to avoid confusion with the conformal space-time co-ordinates $\protect \mathbf {x}$.}\BibitemShut {Stop}%
\bibitem [{\citenamefont {Milsted}\ and\ \citenamefont {Vidal}(2017)}]{S_milsted2017extraction}%
  \BibitemOpen
  \bibfield  {author} {\bibinfo {author} {\bibfnamefont {A.}~\bibnamefont {Milsted}}\ and\ \bibinfo {author} {\bibfnamefont {G.}~\bibnamefont {Vidal}},\ }\href {\doibase 10.1103/PhysRevB.96.245105} {\bibfield  {journal} {\bibinfo  {journal} {Phys. Rev. B}\ }\textbf {\bibinfo {volume} {96}},\ \bibinfo {pages} {245105} (\bibinfo {year} {2017})}\BibitemShut {NoStop}%
\bibitem [{\citenamefont {Zamolodchikov}(1991)}]{S_Zamolodchikov_1991}%
  \BibitemOpen
  \bibfield  {author} {\bibinfo {author} {\bibfnamefont {A.~B.}\ \bibnamefont {Zamolodchikov}},\ }\href {\doibase 10.1016/0550-3213(91)90207-E} {\bibfield  {journal} {\bibinfo  {journal} {Nuclear Physics B}\ }\textbf {\bibinfo {volume} {348}},\ \bibinfo {pages} {619–641} (\bibinfo {year} {1991})}\BibitemShut {NoStop}%
\bibitem [{\citenamefont {Mussardo}(2020)}]{S_Mussardo_2010}%
  \BibitemOpen
  \bibfield  {author} {\bibinfo {author} {\bibfnamefont {G.}~\bibnamefont {Mussardo}},\ }\href@noop {} {\emph {\bibinfo {title} {Statistical {{Field Theory}}: {{An Introduction}} to {{Exactly Solved Models}} in {{Statistical Physics}}}}},\ \bibinfo {edition} {2nd}\ ed.,\ Oxford {{Graduate Texts}}\ (\bibinfo  {publisher} {Oxford University Press},\ \bibinfo {address} {Oxford, New York},\ \bibinfo {year} {2020})\BibitemShut {NoStop}%
\bibitem [{\citenamefont {Berenstein}\ and\ \citenamefont {Miller}(2014)}]{S_Berenstein_Miller_2014}%
  \BibitemOpen
  \bibfield  {author} {\bibinfo {author} {\bibfnamefont {D.}~\bibnamefont {Berenstein}}\ and\ \bibinfo {author} {\bibfnamefont {A.}~\bibnamefont {Miller}},\ }\href {\doibase 10.1103/PhysRevD.90.086011} {\bibfield  {journal} {\bibinfo  {journal} {Physical Review D}\ }\textbf {\bibinfo {volume} {90}},\ \bibinfo {pages} {086011} (\bibinfo {year} {2014})}\BibitemShut {NoStop}%
\bibitem [{Note2()}]{S_Note2}%
  \BibitemOpen
  \bibinfo {note} {$\Lambda $ should be distinguished from $a$, which is the UV cutoff corresponding to the lattice spacing}\BibitemShut {NoStop}%
\bibitem [{Note3()}]{S_Note3}%
  \BibitemOpen
  \bibinfo {note} {We point out that an equivalent formulation up until the Wick rotation was obtained by Ref.~\cite {Berenstein_Miller_2014}, but the following calculation is distinct.}\BibitemShut {Stop}%
\bibitem [{\citenamefont {Calabrese}\ and\ \citenamefont {Cardy}(2006)}]{S_calabrese2006time}%
  \BibitemOpen
  \bibfield  {author} {\bibinfo {author} {\bibfnamefont {P.}~\bibnamefont {Calabrese}}\ and\ \bibinfo {author} {\bibfnamefont {J.}~\bibnamefont {Cardy}},\ }\href {\doibase 10.1103/PhysRevLett.96.136801} {\bibfield  {journal} {\bibinfo  {journal} {Physical Review Letters}\ }\textbf {\bibinfo {volume} {96}},\ \bibinfo {pages} {136801} (\bibinfo {year} {2006})}\BibitemShut {NoStop}%
\bibitem [{\citenamefont {Calabrese}\ and\ \citenamefont {Cardy}(2005)}]{S_calabrese2005evolution}%
  \BibitemOpen
  \bibfield  {author} {\bibinfo {author} {\bibfnamefont {P.}~\bibnamefont {Calabrese}}\ and\ \bibinfo {author} {\bibfnamefont {J.}~\bibnamefont {Cardy}},\ }\href {\doibase 10.1088/1742-5468/2005/04/P04010} {\bibfield  {journal} {\bibinfo  {journal} {Journal of Statistical Mechanics: Theory and Experiment}\ }\textbf {\bibinfo {volume} {2005}},\ \bibinfo {pages} {P04010} (\bibinfo {year} {2005})}\BibitemShut {NoStop}%
\bibitem [{\citenamefont {Cardy}(2016)}]{S_cardy2016furtherresults}%
  \BibitemOpen
  \bibfield  {author} {\bibinfo {author} {\bibfnamefont {J.}~\bibnamefont {Cardy}},\ }\href {\doibase 10.1088/1742-5468/2016/02/023103} {\bibfield  {journal} {\bibinfo  {journal} {Journal of Statistical Mechanics: Theory and Experiment}\ }\textbf {\bibinfo {volume} {2016}},\ \bibinfo {pages} {023103} (\bibinfo {year} {2016})}\BibitemShut {NoStop}%
\bibitem [{\citenamefont {Calabrese}\ and\ \citenamefont {Cardy}(2007)}]{S_Calabrese_Cardy_2007}%
  \BibitemOpen
  \bibfield  {author} {\bibinfo {author} {\bibfnamefont {P.}~\bibnamefont {Calabrese}}\ and\ \bibinfo {author} {\bibfnamefont {J.}~\bibnamefont {Cardy}},\ }\href {\doibase 10.1088/1742-5468/2007/06/P06008} {\bibfield  {journal} {\bibinfo  {journal} {Journal of Statistical Mechanics: Theory and Experiment}\ }\textbf {\bibinfo {volume} {2007}},\ \bibinfo {pages} {P06008–P06008} (\bibinfo {year} {2007})}\BibitemShut {NoStop}%
\bibitem [{\citenamefont {Essler}\ and\ \citenamefont {Fagotti}(2016)}]{S_Essler_Fagotti_2016}%
  \BibitemOpen
  \bibfield  {author} {\bibinfo {author} {\bibfnamefont {F.~H.~L.}\ \bibnamefont {Essler}}\ and\ \bibinfo {author} {\bibfnamefont {M.}~\bibnamefont {Fagotti}},\ }\href {\doibase 10.1088/1742-5468/2016/06/064002} {\bibfield  {journal} {\bibinfo  {journal} {Journal of Statistical Mechanics: Theory and Experiment}\ }\textbf {\bibinfo {volume} {2016}},\ \bibinfo {pages} {064002} (\bibinfo {year} {2016})}\BibitemShut {NoStop}%
\bibitem [{\citenamefont {Calabrese}\ \emph {et~al.}(2012)\citenamefont {Calabrese}, \citenamefont {Essler},\ and\ \citenamefont {Fagotti}}]{S_calabrese2012quantum}%
  \BibitemOpen
  \bibfield  {author} {\bibinfo {author} {\bibfnamefont {P.}~\bibnamefont {Calabrese}}, \bibinfo {author} {\bibfnamefont {F.~H.~L.}\ \bibnamefont {Essler}}, \ and\ \bibinfo {author} {\bibfnamefont {M.}~\bibnamefont {Fagotti}},\ }\href {\doibase 10.1088/1742-5468/2012/07/P07016} {\bibfield  {journal} {\bibinfo  {journal} {Journal of Statistical Mechanics: Theory and Experiment}\ }\textbf {\bibinfo {volume} {2012}},\ \bibinfo {pages} {P07016} (\bibinfo {year} {2012})}\BibitemShut {NoStop}%
\bibitem [{\citenamefont {Granet}\ \emph {et~al.}(2020)\citenamefont {Granet}, \citenamefont {Fagotti},\ and\ \citenamefont {Essler}}]{S_Granet_Fagotti_Essler_2020}%
  \BibitemOpen
  \bibfield  {author} {\bibinfo {author} {\bibfnamefont {E.}~\bibnamefont {Granet}}, \bibinfo {author} {\bibfnamefont {M.}~\bibnamefont {Fagotti}}, \ and\ \bibinfo {author} {\bibfnamefont {F.~H.~L.}\ \bibnamefont {Essler}},\ }\href {\doibase 10.21468/SciPostPhys.9.3.033} {\bibfield  {journal} {\bibinfo  {journal} {SciPost Physics}\ }\textbf {\bibinfo {volume} {9}},\ \bibinfo {pages} {033} (\bibinfo {year} {2020})}\BibitemShut {NoStop}%
\bibitem [{\citenamefont {Mori}\ \emph {et~al.}(2018)\citenamefont {Mori}, \citenamefont {Ikeda}, \citenamefont {Kaminishi},\ and\ \citenamefont {Ueda}}]{S_Mori_Ikeda_Kaminishi_Ueda_2018}%
  \BibitemOpen
  \bibfield  {author} {\bibinfo {author} {\bibfnamefont {T.}~\bibnamefont {Mori}}, \bibinfo {author} {\bibfnamefont {T.~N.}\ \bibnamefont {Ikeda}}, \bibinfo {author} {\bibfnamefont {E.}~\bibnamefont {Kaminishi}}, \ and\ \bibinfo {author} {\bibfnamefont {M.}~\bibnamefont {Ueda}},\ }\href {\doibase 10.1088/1361-6455/aabcdf} {\bibfield  {journal} {\bibinfo  {journal} {Journal of Physics B: Atomic, Molecular and Optical Physics}\ }\textbf {\bibinfo {volume} {51}},\ \bibinfo {pages} {112001} (\bibinfo {year} {2018})}\BibitemShut {NoStop}%
\bibitem [{\citenamefont {Surace}\ and\ \citenamefont {Motrunich}(2023)}]{S_Surace_Motrunich_2023}%
  \BibitemOpen
  \bibfield  {author} {\bibinfo {author} {\bibfnamefont {F.~M.}\ \bibnamefont {Surace}}\ and\ \bibinfo {author} {\bibfnamefont {O.}~\bibnamefont {Motrunich}},\ }\href {\doibase 10.1103/PhysRevResearch.5.043019} {\bibfield  {journal} {\bibinfo  {journal} {Physical Review Research}\ }\textbf {\bibinfo {volume} {5}},\ \bibinfo {pages} {043019} (\bibinfo {year} {2023})}\BibitemShut {NoStop}%
\end{thebibliography}

\onecolumngrid
\section*{End Matter}
\twocolumngrid

\appendix
\renewcommand{\theequation}{A\arabic{equation}}
\setcounter{equation}{0}

\begin{figure*}
\centering
\includegraphics[width=1.0\textwidth]{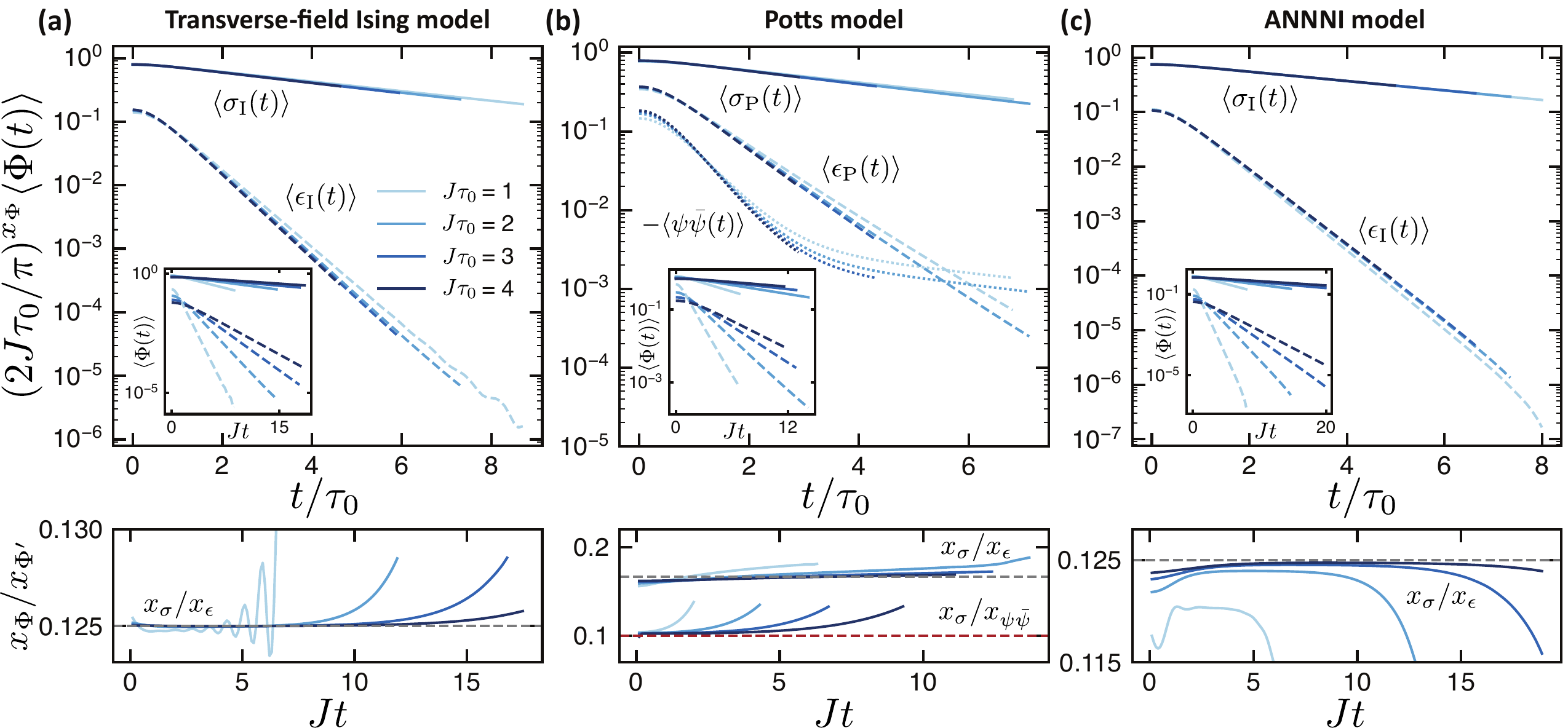}
\caption{Quench dynamics starting in the boundary-CFT initial state for (a) the TFI ($L=200$), (b) Potts ($L=200$), and (c) ANNNI ($L=400$) spin chains. The $x$-axis is rescaled by the initial state's extrapolation length, $\tau_0$. TEBD simulations at bond dimension $\chi = 512$ are shown. The dynamics of different primary fields are fitted to an exponential decay curve using a sliding window of $Jt = 1$, in order to extract field-dependent decay constants. The ratios of these decay constants are shown in the lower plots, where the horizontal dotted lines indicate universal ratios of the fields' scaling dimensions. The time-scale of the lower plot is truncated as the ratio diverges from boundary-CFT expectations, illustrating the onset of the nonuniversal regime. \label{fig:cc}}
\end{figure*}

\emph{Appendix A: Boundary CFT Quenches}---\label{app:bcft}In this appendix, we present supporting numerical data for the quench dynamics starting in a boundary-CFT initial state, for all three lattice models considered in the main text. 
In particular, Calabrese and Cardy determined the quench dynamics for a conformal field theory starting from a special initial state, $|\psi_0\rangle$~\cite{calabrese2005evolution,calabrese2016quantum}:
\begin{equation}
|\psi_0\rangle \propto e^{-\tau_0 H_{\textrm{CFT}}}|\psi_0^* \rangle,
\label{eqn:cardystate}
\end{equation}
where $|\psi_0^*\rangle$ is a conformally invariant boundary state, and the extrapolation length, $\tau_0$, determines the correlation length in the initial state. For both the Ising and Potts universality classes, these conformally invariant boundary states have been calculated exactly. For example,  for the TFIM, they correspond to the two (symmetry-broken) ferromagnetic product states, or the $|{+}X\rangle$ (symmetric) paramagnetic state. 

After real-time evolution under the critical Hamiltonian, Calabrese and Cardy showed that this initial state thermalizes locally to a Gibbs ensemble at temperature $\beta = 4 \tau_0$. Furthermore, for $t \gg \tau_0$, the primary fields decay exponentially:
\begin{align}
\langle \psi_0| \Phi(t) |\psi_0\rangle &\simeq A_b^{\Phi}\left(\frac{\pi}{2 \tau_0}\right)^{x_\Phi} e^{-x_\Phi \pi t / 2 \tau_0}
\label{eqn:cardydecay}
\end{align}
where $A_b^{\Phi}$ is a constant depending on the boundary conditions, $b$. This may vanish: for instance, if we quench from $|{+}X\rangle$, $A_X^\sigma$ is zero due to the $\mathbb{Z}_2$ symmetry. 

Crucially, because $\tau_0$ only depends on the initial state, the ratio of the decay times of two primary fields is universal: it is simply the ratio of their scaling dimensions. We confirm this prediction carries over to shallow quenches on the lattice by simulating open spin chains for the TFI, Potts, and ANNNI models. For all three models, we choose a symmetry-broken product state (corresponding to $\sigma > 0$) and simulate imaginary-time evolution for various $\tau_0$ values. We observe scaling collapse in good agreement with the asymptotic expectations. The scaling collapse improves once $J\tau_0 > 1$, so that the initial correlation length is much larger than the lattice spacing and the continuum limit may be applied. Up to $t \propto$ $2\tau_0 - 4\tau_0$ for the models, there is quantitatively good agreement with the expected ratios, shown in the lower plots of Fig.~\ref{fig:cc}. At longer times, there are deviations from the CFT prediction, particularly $\langle \psi \bar \psi(t)\rangle$ in the Potts chain. This indicates a crossover into the nonuniversal regime and the breakdown of the low-energy effective CFT.

\renewcommand{\theequation}{B\arabic{equation}}
\setcounter{equation}{0}
\emph{Appendix B: Finite-Temperature Conformal Perturbation Theory}---\label{app:FT_CFT}In this appendix, we present details of  some essential aspects of the analytical technique~\cite{Zamolodchikov_1991,Mussardo_2010, Berenstein_Miller_2014} used for calculating the relaxation dynamics of primary fields after a finite-temperature critical quench, and qualitatively sketch how the quasiparticle light-cone picture manifests~\cite{calabrese2005evolution}.
Corrections to the dynamics of a local observable, $\langle\Phi(t)\rangle$, after a quench from the finite-temperature Gibbs ensemble of
\begin{align}
    H = H_{\mathrm{CFT}} + \tilde{g}  \int dr \Psi\left( r \right)
\end{align}
can be calculated perturbatively by integrating over $n$-point finite-temperature correlation functions:
\begin{equation}
\begin{aligned}
    &\left\langle \Phi\left(t\right)\right\rangle 
    =  \tilde{g} \int_\beta  d r d\tau  \left\langle \Psi(r,\tau) \Phi\left(t\right)\right\rangle \quad\quad\\
    &\quad+
    \frac{1}{2}\tilde{g}^2\int_\beta  d r d\tau d r' d\tau' \left\langle \Psi(r,\tau)\Psi(r',\tau') \Phi\left(t\right)\right\rangle + \ldots,
    \label{eq:perteexpn}
\end{aligned}
\end{equation}
where the notation $\int_\beta$ indicates that the integrals are evaluated over infinite real space, $r$, and periodic imaginary time, $\tau$, with circumference give by the inverse temperature, $\beta$ The correlators, $\langle \ldots \rangle$, can be evaluated using the operator product expansion~\cite{Di_Francesco_Mathieu_Sénéchal_1997}; for simplicity, both $\Phi$ and $\Psi$ are (real components of) primary fields. Even the first-order correction to the dynamics illustrates the light-cone picture. Here, the integrand is the two-point finite-temperature correlation function, which is given by (for $2\pi t\gg  \beta$)
\begin{equation}
    \left\langle \Psi(0,t)\Phi\left( r,\tau\right)\right\rangle  
    \simeq \frac{\delta_{x_\Psi x_\Phi}}{\beta^{2 x_\Phi}\left( \cosh\frac{2\pi r}{v\beta}-e^{\frac{2\pi}{\beta}(t-i\tau)}\right)^{x_\Phi}},
    \label{eq:twopointcorrs}
\end{equation}
where $v$ is the propagation speed of the quasiparticles in the CFT~\cite{ cardy2016furtherresults}. 
We see from Eqn.~\ref{eq:twopointcorrs}, that for large spatial separations, $|r|\gg vt$, the spatial decay term, $\cosh\frac{2\pi r}{v\beta} \sim \frac{1}{2} e^{2\pi |r|/\beta}$, dominates, meaning that the two-point correlation function decays exponentially with distance $r$. On the other hand, for small spatial separations, $|r|\ll vt$, the exponential term $e^{2\pi(t-i\tau)/\beta}$ dominates, meaning that the two-point correlation function is constant over distance $r$. These distinct regimes of $r$, which are causally separated by the ``horizon" of the light-cone, $|r| = vt$, leads to two qualitatively different contributions to the first-order corrections of the relaxation dynamics, as in Eqn. \ref{eqn:g^1}. The horizon effect due to the causal structure of the CFT naturally extends to understanding second-
and higher-order corrections $\mathcal{O}(g^n)$, where the interplay of $n$
light-cones leads to more complex scaling forms~\cite{SM}.

\onecolumngrid
\clearpage

\widetext

\setcounter{equation}{0}
\setcounter{figure}{0}
\setcounter{table}{0}
\setcounter{page}{1}
\makeatletter
\renewcommand{\theequation}{S\arabic{equation}}
\renewcommand{\thefigure}{S\arabic{figure}}
\renewcommand{\bibnumfmt}[1]{[S#1]}
\renewcommand{\citenumfont}[1]{S#1}
\renewcommand{\theHfigure}{Supplement.\thefigure}



\begin{center}
\textbf{\large Supplementary Information: \\ Universality of Shallow Global Quenches in Critical Spin Chains}

\vspace{10pt}
\thispagestyle{plain}
Julia Wei,\textsuperscript{1,*}, Méabh Allen,\textsuperscript{2,*} Jack Kemp\textsuperscript{1,3,*}, Chenbing Wang,\textsuperscript{1} Zixia Wei,\textsuperscript{1} Joel E. Moore,\textsuperscript{2,4} and Norman Y. Yao\textsuperscript{1}

\textsuperscript{1}\textit{\small Department of Physics, Harvard University, Cambridge, MA 02138 USA} \\
\textsuperscript{2}\textit{\small Department of Physics, University of California, Berkeley, California 94720 USA} \\
\textsuperscript{3}\textit{\small TCM Group, University of Cambridge, JJ Thomson Avenue, Cambridge, CB3 0US UK} \\
\textsuperscript{4}\textit{\small Materials Science Division, Lawrence Berkeley National Laboratory, Berkeley, CA 94720, USA} \\

\end{center}

\section{Operators in the three-state Potts model}
In the three-state Potts model, the $\mathbb{Z}_3$ clock and shift operators are defined as
\begin{align}
    s_i &= \begin{pmatrix}
                    1 & 0 & 0\\
                    0 & \omega & 0\\
                    0 & 0 & \omega^2
                    \end{pmatrix} \ \rm{ and} \\
    \tau_i &=  \begin{pmatrix}
                    0 & 0 & 1\\
                    1 & 0 & 0\\
                    0 & 1 & 0
                \end{pmatrix},
\end{align}

where $\omega = e^{\frac{2\pi i}{3}}$~\cite{S_mong2014parafermionic}.

\section{Simulation parameters and convergence}

All matrix product state (MPS) simulations were performed using the \texttt{tenpy} library~\cite{S_hauschild:2024}. We use the time-evolving block decimation algorithm (TEBD) with a fourth-order Trotter decomposition, where the time step is fixed to $\delta t = 0.1$~\cite{S_barthelOptimizedLieTrotter2020}. TFI and Potts are both nearest-neighbor models, so we can immediately apply TEBD; for the ANNNI model, we first group pairs of nearest-neighbor sites together. Measurements are performed on the central sites.  

For ground state quenches, we first use the density matrix renormalization group (DMRG) method to find near-critical ground states of the Potts and ANNNI models~\cite{S_SCHOLLWOCK201196}. Since spontaneous symmetry breaking does not occur at finite system sizes~\cite{S_sachdevQuantumPhaseTransitions2011}, we run DMRG in the presence of a small symmetry-breaking field, $H_\sigma$, with field strength $g_\sigma$. This ensures that the spin field has a non-zero value at $t=0$, but the separation of scales $g \gg g_\sigma$ allows us to neglect $g_\sigma$ in the scaling collapse of the quench dynamics. For system size $L < 1000$ (Potts model), we fix $g_\sigma = 10^{-4}$, while for $L \geq 1000$ (ANNNI model), we fix $g_\sigma = 10^{-5}$. 

For finite-temperature quenches, we first use the purification method to find near-critical Gibbs ensembles. For real-time dynamics, we again use the purification method and apply the backward time-evolution disentangler, which can reduce the bond dimension~\cite{S_karraschReducingNumericalEffort2013}.

We use the same procedure to assess convergence in the quench dynamics for boundary-CFT, ground state, and finite-temperature initial conditions. First, we choose a maximum simulation time, $Jt_{\max}$. To test convergence, we fix the scaling parameter ($\tau_0$, $g^{-\nu}$, or $\beta$, respectively) and sweep across three values of $L$ and $\chi$, summarized in Table~\ref{table:convergence_params} below. Fixing $\chi$ ($L$) to its maximum value, we compare observable values between the second-largest and largest $L$ ($\chi$) simulations, resulting in a convergence time, $t_{\rm conv}^{(L)}$ ($t_{\rm conv}^{(\chi)}$), where all observable dynamics agree up to a relative error of $5\%$. We identify the maximum convergence time, $t_{\rm conv}\equiv \min(t_{\rm conv}^{(L)},t_{\rm conv}^{(\chi)})$, which corresponds to the timescales shown in the main text.

\begin{table}
    \begin{tabular}{ccccc}
    \toprule
     Initial State & Model & $L$ & $\chi$ & $t_{\max}$  \\
    \midrule
     \multirow{3}{*}{ Boundary-CFT state} & TFIM & $\{50, 100, 200\}$ & \multirow{3}{*}{$\{128, 256, 512\}$} & \multirow{3}{*}{$20$}\\
     & Potts & $\{50, 100, 200\}$ &   & \\
    & ANNNI & $\{100, 200, 400\}$ &  & \\     
    \midrule
    \multirow{2}{*}{ Ground state} 
     & Potts & $\{250, 375, 500\}$ & \{128, 256, 512\} & $20$\\
    & ANNNI & $\{1000, 1500, 2000\}$ & \{128, 256, 384\} & $20$\\     
    \midrule
     \multirow{3}{*}{ Finite-temperature ensemble} & TFIM & \multirow{3}{*}{$\{100, 200, 400\}$} & \{128, 256, 384\} & $12$\\
     & Potts &  & \{128, 256, 384\} & $6$\\
    & ANNNI & & \{128, 192, 256\} & $10$\\     
    \bottomrule
    \end{tabular}
\caption{Parameters used to assess the convergence of MPS quench simulations.}
\label{table:convergence_params}
\end{table}

\section{Details about fitting procedures}

\subsection{Ground state quench: Extracting decay ratios}

\begin{figure*}
\centering
\includegraphics[width=1.0\textwidth]{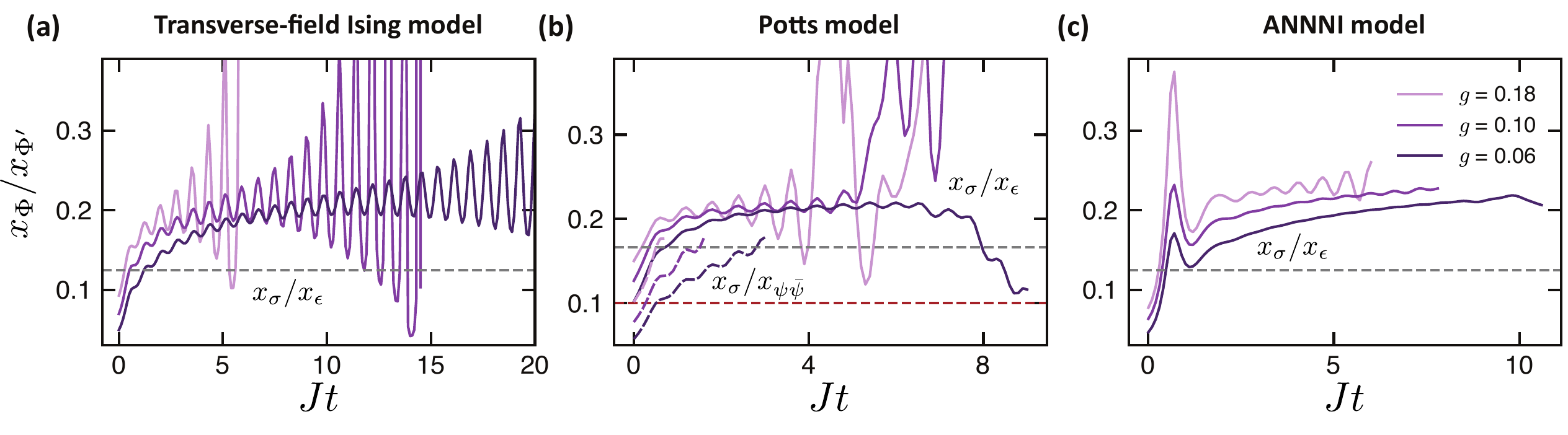}
\caption{The ratio of decay constants in quenches starting from the symmetry-broken ground state in (a) the TFI ($L=500$), (b) Potts ($L=500$, $\chi = 512$), and (c) ANNNI ($L=2000$, $\chi=384$) models. A sliding window of $Jt=1$ is used to fit the dynamics of different primary fields to an exponential decay curve. The horizontal dotted lines indicate universal ratios of the fields' scaling dimensions. In (b), the time-scale is truncated for clarity $x_\sigma/x_{\psi\bar\psi}$ as the fitted ratio diverges from the boundary-CFT expectation of $x_\sigma/x_{\psi\bar\psi} = 1/10$.
}
\label{fig:gsratio}
\end{figure*}

We may fit quench dynamics to an exponential decay curve and examine the ratios of the decay constants, shown in Fig.~\ref{fig:gsratio}. Unlike the boundary-CFT initial condition, we do not observe a plateau in the decay constant that is consistent with the universal ratios in any of the three models.

\subsection{Boundary-CFT quench: Varying the fitting window}

\begin{figure*}
\centering
\includegraphics[width=1.0\textwidth]{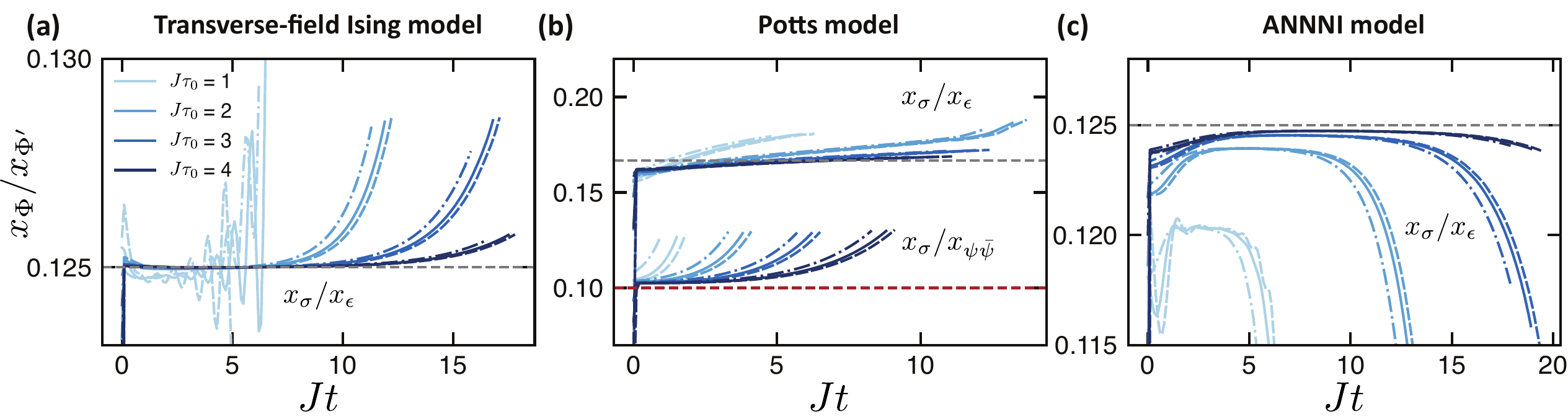}
\caption{The ratio of decay constants in quenches starting from the boundary-CFT state in (a) the TFI ($L=200$), (b) Potts ($L=200$), and (c) ANNNI ($L=400$) models at bond dimension $\chi=512$. The horizontal dotted lines indicate universal ratios of the fields' scaling dimensions. For each value of $J\tau_0$, ratios from three fitting windows are shown: $Jt = 0.5$ (dashed lines), $Jt = 1$ (solid lines), and $Jt = 2$ (dash-dotted lines). 
}
\label{fig:bcftwindowfit}
\end{figure*}

We vary the size of the fitting window in Fig.~\ref{fig:bcftwindowfit} for the boundary-CFT quench dynamics. First, we observe that for all windows, the ratio corresponding to the initial time interval is far from the universal result, indicating an early, nonuniversal time-scale set by the lattice spacing (not shown in the main text). We also observe that the approaches toward and away from the universal ratios have a slight qualitative variation depending on the size of the fitting window. For all three windows, we can identify a plateau consistent with the universal ratios.

\section{Additional data for three-state Potts model quenches}

\begin{figure*}
\centering
\includegraphics[width=1.0\textwidth]{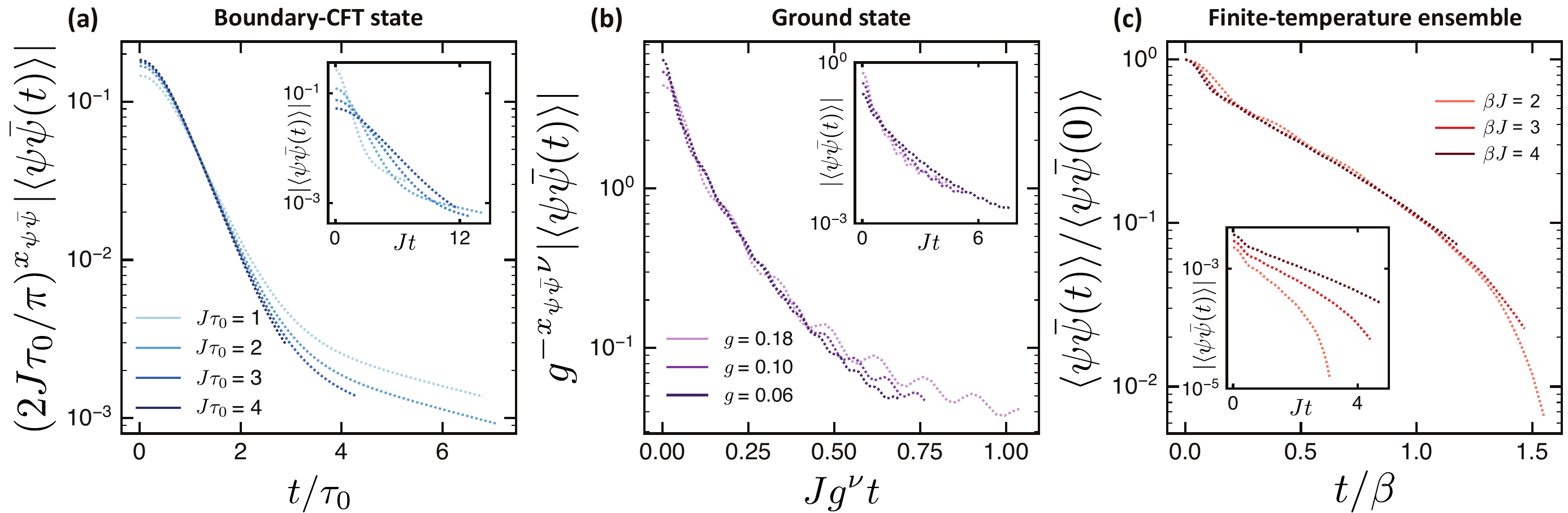}
\caption{The scaling collapse of the dynamics of the parafermion bilinear field, $\psi \bar \psi$, after global quenches in the three-state Potts model, starting from (a) the boundary-CFT state ($L = 400$, $\chi=512$), (b) ground state ($L = 500$, $\chi = 512$), and (c) finite-temperature ensemble with a $\sigma$ perturbation ($L = 400$, $\chi = 384$, $g_\sigma = 1.78\times 10^{-3}$). Insets: The unscaled data.
}
\label{fig:psibarpsi}
\end{figure*}

In Fig.~\ref{fig:psibarpsi}, we show scaling collapses of the parafermion bilinear field, $ \psi \bar \psi$, after global quenches in the three-state Potts model, starting from the three initial conditions. (For completeness, we show the unscaled dynamics for quenches from the boundary-CFT state in App. A.) 

In the ground state quench, we again observe collapse that is consistent with critical exponent $\nu = 5/6$. After normalizing by the initial amplitude of $\langle \psi \bar \psi\rangle$ in the finite-temperature quench, we see a scaling collapse when the time axis is rescaled by $\beta$. However, in contrast to the other examples for the finite-temperature initial condition, we note that this case is not explained by the conformal perturbation theory framework for primary fields, because the three-point structure constant is zero. We hypothesize that the dynamics can still be understood via the overlap between descendant fields in the OPE of the lattice operators, and leave the analytical form for future work.

\section{Analytics}

\subsection{Free fermion scaling limit}
In this section we discuss the origin of the analytical results for quenches to critical points for the transverse-field Ising model quoted in the main text, $H_{\mathrm{TFI}} = -J\sum_i Z_i Z_{i+1} + (1-g)X_i$. In particular, we will focus on deriving results for the $\epsilon$ field through the scaling limit of the lattice observable $Z_i Z_{i+1} - X_i$; the scaling limit of the $Z_i$ has already been calculated in \cite{S_Calabrese_Essler_Fagotti_2011}. Notice that for a transverse-field Ising model with periodic boundary conditions, this simply corresponds to computing the transverse field dynamics, because \begin{align*}
\expct{Z_i Z_{i+1}(t) - X_i(t)}=&\expct{\left(Z_i Z_{i+1}(t) + X_i(t)\right)-2X_i(t)}\\
    =&\expct{\text{Const}-2X_i(t)}\\
    =&\expct{2X_i(\infty)-2X_i(t)}.
\end{align*} The last line follows via the invariance of the critical point under Kramers-Wannier duality, which implies $Z_i Z_{i+1}(t) - X_i(t)$ must vanish as $t \to \infty$. For a quench to the critical point from a model prepared at equilibrium temperature $\beta$ and initial transverse-field perturbation $g$, such transverse-field dynamics have already been calculated by Barouch and McCoy~\cite{S_barouch1970statistical}
\begin{equation}
\expct{Z_i Z_{i+1}(t) - X_i(t)}=\frac{2}{\pi } \int_0^{\pi } \frac{g \cos ^2 \left(\frac{k}{2}\right) \cos  \left(4 J t \sqrt{2-2 \cos  k} \right) \tanh\left(\beta J \sqrt{g^2+2(1+g)(1-\cos k)} \right)}{\sqrt{g^2+2(1+g)(1-\cos k)}} dk
\label{eqn:mccoyfull}
\end{equation}
They took the asymptotic late-time limit, $t\to \infty$, of this expression and found oscillatory behavior at leading order:   
\begin{equation}
\expct{Z_i Z_{i+1}(t) - X_i(t)} \sim -\frac{g}{8  |2-g| \sqrt{\pi } }(Jt)^{-3/2} \tanh\left(\beta J |2-g|\right) \cos \left(8 J t+\frac{\pi }{4}\right),
\label{eq:oscillation}
\end{equation}
as can be observed in Fig.~3 of the main text.

Our goal is to instead calculate the scaling limit of Eqn. ~\ref{eqn:mccoyfull} in order to compare with the CFT results. This amounts to taking the limit $g\to 0$ before taking the late-time limit $t \to \infty$. Keeping only terms to first order in $g$ in Eqn.~\ref{eqn:mccoyfull}  and making the change of variables  $p=\frac{\pi \sqrt{1-\cos k}}{\sqrt{2}}$, we find
\begin{equation}
    \expct{Z_i Z_{i+1}(t) - X_i(t)} \sim \frac{2}{\pi } g \int_0^{\pi } \frac{ \sqrt{\pi ^2-p^2} \cos \left(\frac{ 8 J  t }{\pi } p \right) \tanh \left( \frac{2\beta J}{\pi} p \right) }{\pi \sqrt{p^2+\left(
    \frac{\pi g}{2}\right)^2}} dp
\end{equation}

At late times, the cosine term in the integrand will oscillate rapidly, so we may use the method of stationary phase about $p=0$ to simplify the integrand and extend the upper limit:
\begin{equation}
    \expct{Z_i Z_{i+1}(t) - X_i(t)} \sim \frac{2}{\pi } g \int_0^{\infty } \frac{ \cos \left(\frac{ 8 J  t }{\pi }p\right) \tanh \left( \frac{2 \beta J}{\pi} p \right) }{ \sqrt{p^2+\left(
    \frac{\pi g}{2}\right)^2}} dp.
    \label{eq:statphase}
\end{equation}
As this integral is now in the form of a Fourier transform, it may be easily evaluated. There are two possible cases, depending on the temperature.

\paragraph{Ground state quench} For the ground state quench, $\beta \to \infty$, we utilize
\begin{equation}
    \int_0^{\infty } \frac{1}{\sqrt{p^2+\alpha^2}} \cos (\omega p) dp=K_0\left(\alpha \omega\right),
\end{equation}
where $K_n(x)$ is the modified Bessel function of the second kind. The leading order term in the series expansion of $K_0(x)$ as $x\to \infty$ is $\sqrt{\frac{\pi }{2}}e^{-x}  \sqrt{\frac{1}{x}}$. Finally, we obtain
\begin{equation} \label{eq:asymptote1}
    \expct{\epsilon_\text{I}(t)} \sim \expct{Z_i Z_{i+1}(t) - X_i(t)}\sim{\sqrt[]{\frac{g}{2\pi}}}e^{-4 gJt}{\sqrt[]{\frac{1}{Jt}}},
\end{equation}
as quoted in the main text. We note that this result may also be obtained by exploiting the horizon effect to equate the one-point function squared of $X_i(t)$, decaying as a function of time, with the two-point function in equilibrium, decaying as a function of space. For ground state quenches, the generalized Gibbs ensemble equilibrium two-point function has already been calculated~\cite{S_Calabrese_Essler_Fagotti_2011}.

\paragraph{Finite temperature}
For finite $\beta$, we may simplify the integrand in Eqn.~\ref{eq:statphase} by neglecting the term in the denominator proportional to $g^2$, as unlike for the ground state case the integral converges without this term. Then we may evaluate the integral using the Fourier transform
\begin{equation}
\int_0^{\infty} \frac{\cos(\omega p)\tanh(\gamma p)}{p}dp = - \log \tanh\frac{\pi \omega}{4\gamma}. 
\end{equation}
The leading term of the series expansion of $- \log \tanh x$ as $x \to \infty$ is $2 e^{-2x}$, so finally we recover the exponential decay \begin{equation}
 \expct{\epsilon_\text{I}(t)}\sim\expct{Z_i Z_{i+1}(t) - X_i(t)} \sim \frac{4 g}{\pi}e^{-2\pi t/\beta},
 \label{eq:asymptote2}
\end{equation}
consistent with the boundary-CFT expression under the identification $\beta = 4\tau_0$.

\subsection{Lattice-to-field mapping}
It is useful to take a closer look at the mapping between the critical lattice Hamiltonians and perturbations, which are implemented in numerical simulation, and the underlying continuum CFT Hamiltonians and field perturbations, from which a conformal perturbative analysis yields scaling forms. 

A primary field $\Phi$ scales under a conformal transformation by its scaling dimension~\cite{S_Di_Francesco_Mathieu_Sénéchal_1997}, which is denoted by $\Delta_\Phi$ throughout the supplemental material~\footnote{This notation for the scaling dimension deviates from the main text's convention, $x_\Phi$, in order to avoid confusion with the conformal space-time co-ordinates $\mathbf{x}$.}, and defined by the power-law decay of their two-point correlation functions,
\begin{align}
    \langle \Phi(\mathbf{x}_1)\Psi(\mathbf{x}_2) \rangle = \delta_{\Delta_\Phi,\Delta_\Psi}\left|\mathbf{x}_1-\mathbf{x}_2\right|^{-2\Delta_\Psi},
\end{align}
following customary normalization conventions. The Hamiltonian has units of inverse length $[H] = [J] = -1$. Local lattice operators $O_i$, which are dimensionless, can be written as a linear combination of fields in the CFT~\cite{S_milsted2017extraction},
\begin{align}
    O_i = \sum_\varphi c_{\varphi} a^{\Delta_\varphi}\varphi, 
\end{align} 
where $a$ is the lattice spacing and $c_{\varphi} $ are dimensionless constants. In the continuum limit, a sum over local lattice operators becomes an integral over fields: $\sum_i \to \frac{1}{a}\int dr$. Since any field in the CFT is either a primary field or a descendant thereof~\cite{S_Di_Francesco_Mathieu_Sénéchal_1997}, we consider then, for simplicity, the case where the lattice operator maps to just a primary field, $O_i = c_\Psi a^{\Delta_\Psi}\Psi_i  $. We have
\begin{align}
    H = H_{\mathrm{c}} + g J  \sum_i O_i \to H_{\mathrm{CFT}} + \tilde{g}  \int dr \Psi\left( r \right) \implies \tilde{g} = g c_\Psi J a^{\Delta_\Psi-1}.
\end{align}
Immediately we note that the CFT perturbation strength, $\tilde{g}$, is proportional to the dimensionless lattice perturbation strength, $g$, and, unless the primary field $\Phi$ is a marginal operator (i.e. $\Delta_\Psi = 2$), it must have dimensions, $[\tilde{g}] =  \Delta_\Psi-2$. 

The lattice-to-field mapping and corresponding constants $c_\Psi$ for the TFIM are discussed at length in the literature, see, e.g., \cite{S_sachdevQuantumPhaseTransitions2011,Calabrese_Essler_Fagotti_2011,milsted2017extraction}, and references therein. However, since we do not generically know $c_\Psi$ for arbitrary lattice operators, we do not focus on explicit calculation of the pre-factor in each term of the perturbative expansion in the main text, but instead on the scaling forms with respect to temperature and time, as well as the \emph{ratio} of the different terms for $\mathcal{O}(g)$. We set $a = 1$ from here on out unless explicitly stated. 

\subsection{Conformal perturbation theory asymptotics}
\begin{figure}[tbp!]
    \centering
    \includegraphics[width=0.9\linewidth]{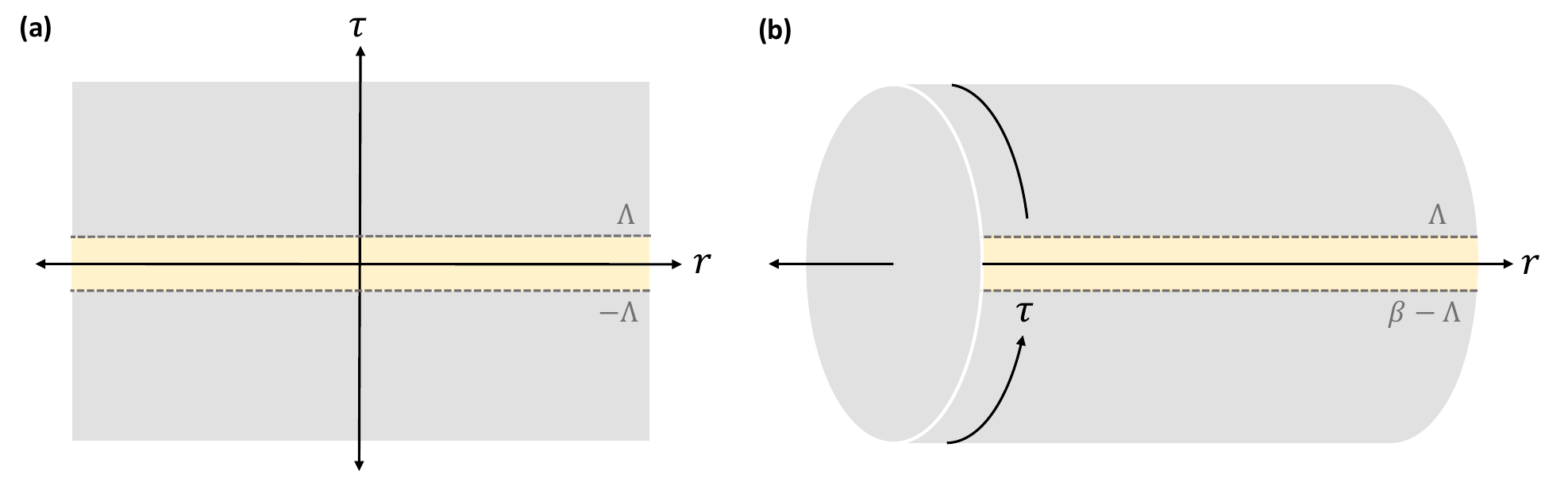}
    \caption{(a) The Euclidean geometry running over all real space $r$ and imaginary time $\tau$, providing no natural infrared (IR) regulator. (b) Considering the CFT at finite temperature is equivalent to putting the Euclidean geometry on a cylinder, where the circumference is in the imaginary time direction $\tau$ and is given by the inverse temperature $\beta \sim T^{-1}$. The finite imaginary time provides an IR regulator. In both geometries, the introduction of a cutoff $\tilde{g}$ allows for the regularization of ultraviolet (UV) divergences; if the calculation is convergent one should be able to safely take $\Lambda \to 0$. }
    \label{fig:geometryboth}
\end{figure}
We are interested in the relaxation behavior of local observables following a shallow global quench to the critical point in $1+1$D, starting from a nearby thermal ensemble. Having discussed the lattice-to-field mapping above, we proceed in the language of CFT. In the conformal perturbative framework~\cite{S_Zamolodchikov_1991,Mussardo_2010}, the quench, characterized by some relevant scalar operator $\Psi$ with $\Delta_\Psi <2$, is modeled as a small deformation of the CFT action $S_{\mathrm{CFT}}$, 
\begin{align}
    S = \int d^2\mathbf{x} \mathcal{L}(\mathbf{x}) = S_{\mathrm{CFT}} - \tilde{g}\int d^2\mathbf{x} \Psi(\mathbf{x}),
\end{align}
where $\mathbf{x}= (r,\tau)$ are the conformal space-time co-ordinates, and $ \mathcal{L}$ is the Lagrangian density of the system. Since the system is at finite temperature, the Euclidean geometry of the CFT is a cylinder of circumference $\beta$ in imaginary time $\tau$, and infinite in real space $r$ (Fig. \ref{fig:geometryboth}). This geometry naturally introduces an infrared (IR) regulator, and is conformally equivalent to flat space~\cite{S_Berenstein_Miller_2014}. Then:
\begin{align}
S=  S_{\text{CFT}}- \tilde{g} \int_{0}^{\beta}d\tau \int_{-\infty}^{\infty} dr\Psi(r,\tau).
\end{align}
Note that the procedure may require the introduction of an ultraviolet (UV) cutoff in imaginary time, $\Lambda$, to regularize any divergences [see Fig. \ref{fig:geometryboth}(b)]\footnote{$\Lambda $ should be distinguished from $a$, which is the UV cutoff corresponding to the lattice spacing}; UV divergences can be dealt with via standard renormalization methods and $\Lambda$ can  be taken to zero after the computation~\cite{S_Mussardo_2010}.  
The partition function takes the form:
\begin{align}
    Z_{\tilde{g}}=\left\langle\exp \left(\tilde{g} \int d^2 \mathbf{x} \Psi(\mathbf{x})\right)\right\rangle_0=\left\langle\sum_{n=0}^{\infty} \frac{1}{n!}\left(\tilde{g} \int d^2 \mathbf{x} \Psi(\mathbf{x})\right)^n\right\rangle_0= 1 +\mathcal{O}(\tilde{g}^2),
    \label{eq:partitionfunction}
\end{align}
where $\langle \ldots\rangle_0 $ denotes correlation functions of the undeformed finite-temperature CFT. Corrections to local operators can then be determined via a perturbative expansion over $n$-point finite-temperature correlators,
\begin{equation}
\begin{aligned}
    \left\langle O\left(\mathbf{x}\right)\right\rangle_{\tilde{g}} & = \frac{1}{Z_{\tilde{g}}} \left\langle O(\mathbf{x}) \exp \left(\tilde{g} \int d^2 \mathbf{x}' \Psi(\mathbf{x}')\right)\right\rangle_0\\
    &= \frac{1}{Z_{\tilde{g}}}\sum_{n=1}^{\infty} \frac{1}{n!}\tilde{g}^n \int  d^2 \mathbf{x}_1 \ldots d^2 \mathbf{x}_n\left\langle \Psi(\mathbf{x}_1)\ldots \Psi(\mathbf{x}_n)O\left(\mathbf{x}\right)\right\rangle_0 \\
    &=  \tilde{g} \int  d^2 \mathbf{x}_1 \left\langle \Psi(\mathbf{x}_1) O\left(\mathbf{x}\right)\right\rangle_0 +
    \frac{1}{2}\tilde{g}^2\int d^2 \mathbf{x}_1 d^2 \mathbf{x}_2\left\langle \Psi(\mathbf{x}_1)\Psi(\mathbf{x}_2) O\left(\mathbf{x}\right)\right\rangle_0 +\mathcal{O}\left(\tilde{g}^3\right)\\
    & =\left\langle O\left(\mathbf{x}\right)\right\rangle_{1}+ \left\langle O\left(\mathbf{x}\right)\right\rangle_{2} +\mathcal{O}\left(\tilde{g}^3\right),
    \label{eq:perteexpn}
\end{aligned}
\end{equation}
where the correlators can be obtained from the operator product expansions (OPE) of the observable $O$ with the quench perturbation $\Psi$, again in the undeformed finite-temperature CFT~\cite{S_Di_Francesco_Mathieu_Sénéchal_1997}. Since the one-point functions of primary fields evaluated in the undeformed CFT are zero, corrections from the renormalized partition function in Eqn.~\ref{eq:partitionfunction} do not manifest until $\mathcal{O}(\tilde{g}^3)$ (when we consider a Taylor expansion in $\tilde{g}$). Note, from Section V B, that $[\tilde{g}] = \Delta_O-2$, and the units of each term in Eqn.~\ref{eq:perteexpn} are $[O(\mathbf{x})] = -\Delta_O$. Assuming translation invariance, expectation values of deformed one-point functions are position-independent and higher-point functions depend only on the separation between fields. The decay of local observables is then obtained via analytic continuation to real time. 

For simplicity, we consider only the dynamics of quench fields and observables which are (the real components of) primary fields. 
\subsubsection{First-order perturbation theory}
At leading order, the expectation value of the observable $\Phi(\mathbf{x})$ at space-time coordinate $\mathbf{x} = (0,\tau) \equiv \tau$, under a quench characterized by $\Psi$, takes the form
\begin{align}
    \langle \Phi(\tau)\rangle_1&= \tilde{g} \int d^2 \mathbf{x'} \left\langle \Psi(\mathbf{x'})\Phi\left( \tau\right)\right\rangle_0 \label{eqn:integralsetup},
\end{align}
where the integrand is the two-point finite-temperature correlation function which is non-zero only if $\Psi = \Phi$ by symmetry constraints, and describes pairs of left- and right-moving quasiparticles~\cite{S_Di_Francesco_Mathieu_Sénéchal_1997},
\begin{align}
    \left\langle \Psi(\mathbf{x'})\Phi\left( \mathbf{x}\right)\right\rangle_0  &= \delta_{\Delta_\Psi \Delta_\Phi} \left(\frac{\pi^2}{\beta^2}\right)^{\Delta_\Phi}\frac{1}{\left\{\sin \left[\frac{\pi}{\beta}(\tau' -\tau)+\frac{i \pi }{v\beta} (r'-r)\right] \sin \left[\frac{\pi}{\beta}(\tau' -\tau)-\frac{i \pi }{v\beta} (r'-r)\right]\right\}^{\Delta_\Phi}}\\
    &= \delta_{\Delta_\Psi \Delta_\Phi}\left(\frac{2\pi^2}{\beta^2}\right)^{\Delta_\Phi}\frac{1}{\left\{ \cosh\left[\frac{2\pi}{v\beta}(r-r')\right]-\cos\left[\frac{2\pi}{\beta}(\tau-\tau')\right]\right\}^{\Delta_\Phi}} \\
    &=  \delta_{\Delta_\Psi \Delta_\Phi}\left(\frac{2\pi^2}{\beta^2}\right)^{\Delta_\Phi} w\left(\mathbf{x'},\mathbf{x}\right)^{-\Delta_\Phi}
    \label{eqn:fintempcorrelator},
\end{align}
where $\delta_{ij}$ denotes the Kronecker delta function, $v = \partial \epsilon_k/\partial k\vert_{k\ll 1}$ is a speed determined by the linear dispersion of the CFT, and where we have re-expressed the correlator in terms of the variable $w(\mathbf{x'},\mathbf{x})$. 

\begin{figure}
  \centering
  \includegraphics[width=.9\linewidth]{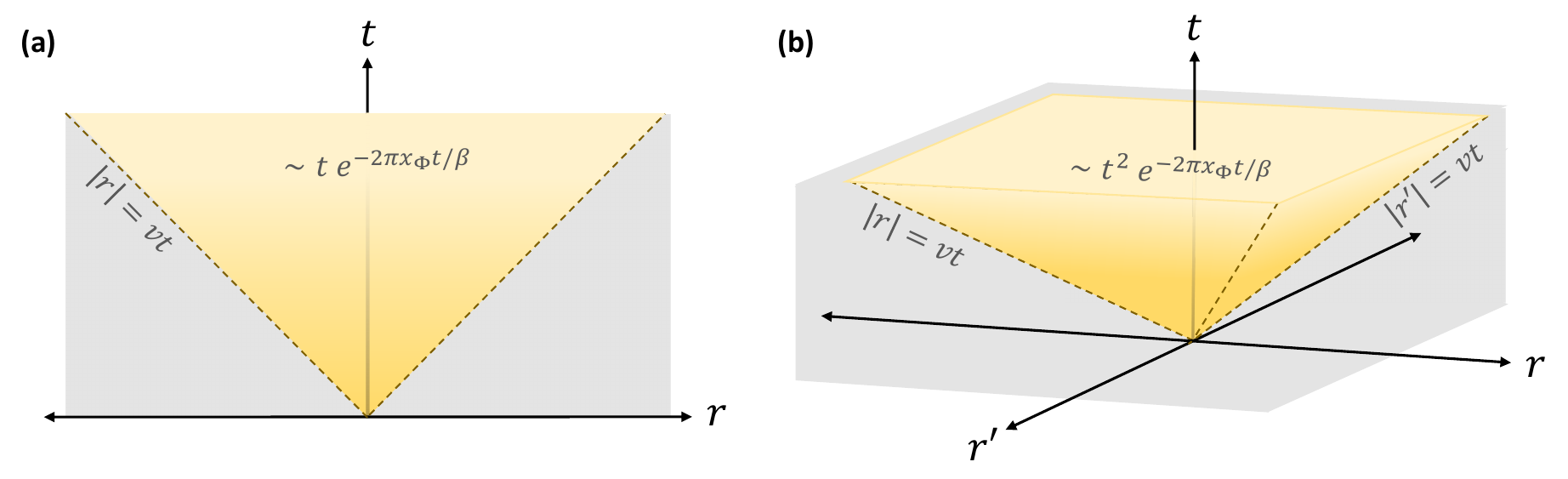}
  \caption{The horizon effect of the forward-spreading light cone(s) of a perturbation in CFT to (a) first order and (b) second order. The initial thermal ensemble acts as a source of quasiparticles which propagate at velocity $v$. Spatially decaying correlations lead to exponential decay of observables with time both inside (yellow) and outside (gray) the (intersecting) light cone(s). Constant spatial correlations can build up within the (intersecting) light cone(s) where space-time points are causally connected, leading to maximal exponentially-damped polynomial growth terms of the form $t^n e^{-2\pi t \Delta_\Phi /\beta}$, for $n$ intersecting light cones. These regimes lead to two qualitatively different scaling forms in the perturbation theory as in Eqns.~\ref{eqn:(A+Bt)} and \ref{eqn:scalingorderg2}. In the first-order correction, both contributions from within the light cone are zero for integer scaling dimension, leading to the purely exponential decay of $\epsilon_\text{I}$ under an $\epsilon_\text{I}$ quench.}
  \label{fig:lightcone}
\end{figure}

To proceed with the calculation, we firstly make use of analytic continuation to real time, $\tau\rightarrow -it$~\footnote{We point out that an equivalent formulation up until the Wick rotation was obtained by Ref.~\cite{S_Berenstein_Miller_2014}, but the following calculation is distinct.}. Secondly, in light of Ref.~\cite{S_calabrese2006time}'s result that local observables exhibit exponential decay for ``large" times $t\gg \tau_0$, we note that our primary region of interest is analogously $2\pi t\gg\beta$ (though we discuss the ``short" time regime $2\pi t\ll\beta$ later). Then, after analytic continuation, and at late times,  the cosine of $w(\mathbf{x},\tau\to-it)\equiv w(\mathbf{x},t)$ becomes exponentially suppressed,
\begin{align}
   w(\mathbf{x'},t) = \cosh\left[\frac{2\pi r'}{v\beta}\right] -\cos \left[\frac{2\pi}{\beta}(i t +\tau')\right] \xrightarrow{2\pi t\gg\beta}  \cosh\left(\frac{2\pi r'}{v\beta}\right)-\frac{1}{2}e^{2\pi (t-i\tau')/\beta},
   \label{eq:w}
\end{align}
so that the first-order correction to the local observable at late times after a quench takes the form,
\begin{equation}
\begin{aligned}
\langle \Phi(t)\rangle_{1} & \simeq \tilde{g} \delta_{\Delta_\Psi \Delta_\Phi}\left(\frac{2 \pi^2}{\beta^2}\right)^{\Delta_\Phi}  \int_{-\infty}^\infty  d r' \int_0^\beta d\tau' \frac{1}{\left\{\cosh \left[\frac{2 \pi r' }{ v \beta}\right]-\frac{1}{2}\exp \left[\frac{2 \pi } {\beta}\left(t-i \tau'\right)\right]\right\}^{\Delta_\Phi}},
\end{aligned}
\end{equation}
which is even over real space. We now make the observation that the first term in Eqn.~\ref{eq:w} dominates when $|r'|\gg v t$, and the second term dominates when $|r'| \ll v t$. That is, we can split the real-space integral into two regions which are separated by a ``horizon" $|r'| = vt$ [see Fig. \ref{fig:lightcone}(a)],
\begin{equation}
\begin{aligned}
\int_{0}^{\infty} d r' &= \int_{0}^{vt} d r'+ \int_{vt}^{\infty} d r' \implies \langle \Phi(t)\rangle = \langle \Phi(t)\rangle_{ \text{LC}}+\langle \Phi(t)\rangle_{ \text{\sout{LC}}}
\end{aligned}
\end{equation}
Physically, this reflects the light-cone (LC) structure of a relativistic CFT: the perturbing field acts as a source of quasiparticles which propagate through space-time at speed $v$~\cite{S_calabrese2005evolution,cardy2016furtherresults}. The form of the correlator $w(\mathbf{x}',t)$ can be approximated accordingly. 
Outside the light cone, where $|r^{\prime}|\gg vt $, the correlations decay exponentially with distance, $\cosh\left(\frac{2 \pi r' }{ v \beta}\right) \sim \frac{1}{2}\exp \left(\frac{2 \pi r' }{ v \beta}\right)$, and the \emph{spatial decay} term in Eqn. \ref{eq:w} dominates:
\begin{equation}
    w(\mathbf{x^{\prime}},t)^{-\Delta_\Phi}\xrightarrow{|r^{\prime}|\gg vt}
    2^{\Delta_\Phi}e^{-\frac{2\pi r'   }{v\beta}\Delta_\Phi} \left[1 -\left(e^{\frac{2\pi}{\beta} (t-i\tau') -\frac{2\pi r'}{v\beta}}-e^{-\frac{4\pi r'}{v\beta}} \right)\right]^{-\Delta_\Phi}.
\label{eq:LCapproxlarger}
\end{equation}
The spatial decay term, $\exp({-2\pi\Delta_\Phi r'/ v\beta})$, is explicitly taken out of the parentheses as the leading contribution. The binomial $(1+y)^{-\Delta_\Phi}$ in Eqn.~\ref{eq:LCapproxlarger} can then be expressed as an algebraic expansion of the sub-leading terms $y$ over an infinite series,  
\begin{equation}
    \left[1 -\left(e^{\frac{2\pi}{\beta} (t-i\tau') -\frac{2\pi r'}{v\beta}}-e^{-\frac{4\pi r'}{v\beta}} \right)\right]^{-\Delta_\Phi}=
    1 +\sum_{k=1}^{\infty}{\Delta_\Phi+k -1\choose k}\left(e^{\frac{2\pi}{\beta} (t-i\tau') -\frac{2\pi r'}{v\beta}}-e^{-\frac{4\pi r'}{v\beta}} \right)^k.
    \label{eq:binomialexp}
\end{equation}
The leading contribution to the behavior of the primary field from outside the light cone comes from the spatial decay term:
\begin{align}
\langle \Phi(t)\rangle_{ 1_\text{\sout{LC},pure decay}}
&\simeq 2 \tilde{g} \delta_{\Delta_\Psi \Delta_\Phi}\left(\frac{2 \pi}{\beta}\right)^{2\Delta_\Phi}  \int_{vt}^\infty  d r' \int_\Lambda^{\beta-\Lambda} d\tau' e^{- 2 \pi r'\Delta_\Phi/ v \beta}\bigg|_{\Lambda\to 0}\nonumber\\
&\simeq  \tilde{g} \delta_{\Delta_\Psi \Delta_\Phi}\left(\frac{2 \pi}{\beta}\right)^{2\Delta_\Phi}   \frac{v\beta^2}{\pi \Delta_\Phi}e^{-2\pi t \Delta_\Phi/\beta}.
\label{eq:decay1}
\end{align}
The $\tau'$-dependent terms in the binomial expansion around $e^{- 2 \pi r'\Delta_\Phi/ v \beta}$ in Eqn.~\ref{eq:LCapproxlarger} vanish, since only positive integer powers of $e^{- 2 \pi i\tau'/\beta}$ appear (see Eqn.~\ref{eq:binomialexp}) and therefore cancel out over the closed imaginary time loop. The first $\tau'$-independent correction term in the binomial expansion is integrated to obtain a sub-leading term of order $\mathcal{O}(e^{-2\pi(2+\Delta_\Phi)t/\beta})$.

On the other hand, inside the light cone where $|r^{\prime}|\ll vt$, the correlations can be expanded around the dominant time decay term that is \emph{spatially constant}:
\begin{equation}
    w(\mathbf{x^{\prime}},t)^{-\Delta_\Phi}\xrightarrow{|r^{\prime}|\ll vt}
    2^{\Delta_\Phi}\left(-1\right)^{-\Delta_\Phi}e^{-\frac{2\pi}{\beta}(t-i\tau^{\prime})\Delta_\Phi}\left[ 1-2\cosh\left(\frac{2\pi r'}{v\beta}\right)e^{-\frac{2\pi}{\beta}(t-i\tau^{\prime})}\right]^{-\Delta_\Phi} ,
\label{eq:LCapproxsmallr}
\end{equation}
where, again, the correction terms can be calculated via a binomial expansion. We see that this spatially constant integrand leads to exponentially damped operator growth (a linear-times-exponential-in-t behavior) which we denote``peak decay", in contrast to ``pure (exponential) decay":
\begin{align}
\langle \Phi(t)\rangle_{1_ \text{LC,peak decay}}&\simeq 2 \tilde{g} \delta_{\Delta_\Psi \Delta_\Phi}\left(\frac{2 \pi}{\beta}\right)^{2\Delta_\Phi}  \int_{0}^{vt}  d r' \int_\Lambda^{\beta-\Lambda} d\tau' \left(-1\right)^{-\Delta_\Phi} e^{-2 \pi \left(t-i \tau'\right)\Delta_\Phi/  \beta}\bigg|_{\Lambda\to 0}\nonumber\\
&\simeq  \tilde{g}\delta_{\Delta_\Psi \Delta_\Phi}\left(\frac{2 \pi}{\beta}\right)^{2\Delta_\Phi}  \frac{v\beta}{\pi \Delta_\Phi} 2t\sin (\pi \Delta_\Phi)e^{-2\pi t\Delta_\Phi /\beta}.
\label{eq:cumulativeterm}
\end{align}
Here, introducing the cutoff $\Lambda$ is crucial for regularizing the branch cut singularities that exist for non-integer scaling dimensions $\Delta_\Phi \neq 1$, since the integrand is a multi-valued function in such cases. In this case, the contribution of the binomial expansion around $e^{- 2 \pi (t-i\tau')\Delta_\Phi/ \beta}$ due to subleading spatially decaying corrections does not vanish around the imaginary time contour, also due to the branch cut for $\Delta_\Phi \neq 1$. We have
\begin{align}
\langle \Phi(t)\rangle_{1_ \text{LC,pure decay}}&\simeq 2\tilde{g} \delta_{\Delta_\Psi \Delta_\Phi}\left(\frac{2 \pi}{\beta}\right)^{2\Delta_\Phi} \sum_{k=1}^{\infty}{\Delta_\Phi\choose k} \frac{2^{k}}{k!} (-1)^{-\Delta_\Phi}  \int_\Lambda d r^{\prime}d \tau^{\prime}  \cosh^k\left(\frac{2\pi r'}{v\beta}\right) e^{-2\pi (\Delta_\Phi+k) (t-i \tau^{\prime})/ \beta }\bigg|_{\Lambda\to0}\nonumber\\
&\simeq \tilde{g} \delta_{\Delta_\Psi \Delta_\Phi}\left(\frac{2 \pi}{\beta}\right)^{2\Delta_\Phi}\frac{ \sin(\pi \Delta_\Phi) }{ 2\pi^2 } v\beta^2 e^{-2\pi \Delta_\Phi t/\beta } \sum_{k=1}^{\infty}{\Delta_\Phi\choose k} \frac{1 }{(\Delta_\Phi+k) k!}\frac{\Gamma\left(-{1\over2} + \frac{1+k}{2}\right)}{\Gamma\left(-{1\over2} + \frac{3+k}{2}\right)}\nonumber \\
&\simeq \tilde{g} \delta_{\Delta_\Psi \Delta_\Phi}\left(\frac{2 \pi}{\beta}\right)^{2\Delta_\Phi}\frac{v\beta^2}{\pi\Delta_\Phi }  \frac{\sin (\pi \Delta_\Phi)}{\pi }\left[  \frac{1}{\Delta_\Phi} - \gamma_E -\pi \csc(\pi \Delta_\Phi) -\frac{\Gamma^{\prime}(1-\Delta_\Phi)}{\Gamma(1-\Delta_\Phi)}\right]e^{-2\pi \Delta_\Phi t/\beta },
\label{eq:decay2}
\end{align}
with corrections again of order $\mathcal{O}(e^{-2\pi(2+\Delta_\Phi)t/\beta})$. 

We see that spatial correlations grow within the light cone where space-time points are causally connected. Constant correlations accumulate over the light cone and lead to exponentially damped operator growth, or ``peak decay" (Eqn. \ref{eq:cumulativeterm}), while spatially decaying correlations lead to purely exponential decay in time (Eqns. \ref{eq:decay1} and \ref{eq:decay2}),
\begin{equation}
\begin{aligned}
\langle \Phi(t)\rangle_{ 1_\text{peak decay}} \sim  \frac{t}{\beta} e^{-2\pi \Delta_\Phi t/\beta },\quad \quad 
\langle \Phi(t)\rangle_{ 1_\text{pure decay}}\sim  e^{-2\pi t \Delta_\Phi/\beta}
\end{aligned}
\end{equation}
Altogether, we have two distinct regimes of behavior,
\begin{align}
\langle \Phi(t)\rangle_1 
 &\simeq \tilde{g} \delta_{\Delta_\Psi \Delta_\Phi}\beta^{2-2\Delta_\Phi}
 \left(A_{\Delta_\Phi} + B_{\Delta_\Phi}\frac{t}{\beta}\right)e^{-2\pi t \Delta_\Phi/\beta}\label{eqn:(A+Bt)}\\
 B_{\Delta_\Phi}&\simeq   v (2\pi)^{2\Delta_\Phi}\frac{2\sin(\pi \Delta_\Phi )}{ \pi \Delta_\Phi}\nonumber\\
 \frac{A_{\Delta_\Phi}}{B_{\Delta_\Phi} }&\simeq\frac{1}{2 \sin(\pi\Delta_\Phi)}\left\{1 + \frac{\sin(\pi\Delta_\Phi)}{\pi}\left[\frac{1}{\Delta_\Phi}-\gamma_E -\pi \csc(\pi \Delta_\Phi) -\frac{\Gamma^{\prime}(1-\Delta_\Phi)}{\Gamma(1-\Delta_\Phi)}\right]\right\},
\label{eqn:scalingorderg}
\end{align}
with corrections of order $\mathcal{O}(e^{-2\pi(2+\Delta_\Phi)t/\beta})$.

A few remarks are in order. Firstly, both contributions coming from within the light cone are proportional to $\sin (\pi \Delta_\Phi) $. Then, since $\sin (\pi \Delta_\Phi) =0$ for integer-valued scaling dimensions, there are no contributions from within the light cone, and the decay is purely exponential. This explains the relaxation dynamics of $\epsilon_\text{I}$ under a finite-temperature $\epsilon_\text{I}$ quench in the main text. Further, since $\sin (\pi \Delta_\Phi) < 0 $ for $1< \Delta_\Phi <2$, the contributions from either side of the horizon have opposite sign: the two competing decay terms will eventually cancel out and the one-point function will cross the $t$-axis before becoming negative, though the amplitude still eventually decays to zero. Finally, we see that a term $\sim( t/\beta) e^{-2\pi \Delta_\Phi t/\beta}$  arises for $\Delta_\Phi \neq 1$. It follows from the causal structure of CFTs: the initial thermal ensemble acts as a source of left- and right-moving pairs of quasiparticles, which are correlated and entangled over lengths $\sim \mathcal{O}\left( vt \right)$. At time $t$, the perturbation of the initial state is felt only from regions within $|r'| < vt$. In this region, the integrand has a term that is spatially constant, leading to a contribution $\sim t/\beta$.

\subsubsection{Second-order perturbation theory}
When the scaling dimensions of the observable and the quench primary fields are not the same, the first order dynamics, $\mathcal{O}(\tilde{g})$, are zero. In that case, we must consider the corrections to second order, for a primary field $\Phi$ under a different primary field $\Psi$ quench: 
\begin{equation}
\begin{aligned}
\left\langle \Phi \left( t\right)\right\rangle_{2} &=  \frac{1}{2}\tilde{g}^2 \int d^2 \mathbf{x'} d^2 \mathbf{x''} \left\langle \Psi(\mathbf{x'})\Psi(\mathbf{x''}) \Phi\left( t\right)\right\rangle_0   \\
&= \frac{1}{2}\tilde{g}^2 \frac{C_{\Psi\Psi\Phi}}{\beta^{\Delta_\Phi+2\Delta_\Psi}} \int d^2 \mathbf{x'} d^2 \mathbf{x''}\frac{w\left(\mathbf{x'},\mathbf{x''}\right)^{\frac{1}{2}\Delta_\Phi-\Delta_\Psi}}{w\left(\mathbf{x'},t\right)^{\frac{1}{2}\Delta_\Phi}w\left(\mathbf{x''},t\right) ^{\frac{1}{2}\Delta_\Phi}}
\label{eq:secondordersetup}
\end{aligned}
\end{equation}
where $C_{\Psi\Psi\Phi}$ is the three-point structure factor of the CFT. Again, we are interested in the $2\pi t\gg \beta$ regime.

There are now two intersecting light cones with respect to  $r'$ and  $r''$, see Fig.~\ref{fig:lightcone}(b). Qualitatively speaking, the regions outside both light cones lead to two distinct rates of exponential relaxation, characterized by the scaling dimensions of twice the quench field $\sim  e^{-4\pi t \Delta_\Psi/\beta} $, and the observable $\sim e^{-2\pi t \Delta_\Phi/\beta}$. Within the two intersecting light cones, in addition to this purely exponential decay, the accumulation of constant spatial correlations result in damped $t$ or $t^2$ terms, with the damping associated only with the scaling dimension of the observable $\sim e^{-2\pi t \Delta_\Phi/\beta}$. 

While the calculation is more involved for intersecting light cones, the procedure follows that of the first-order perturbation. We are primarily focused on the general scaling ansatz with real time $t$ and temperature $\beta^{-1}$; to that end, we focus on the functional form of the polynomial and exponential behavior with real time, and omit explicit calculation of constants. 

We split the spatial integrals up along the horizons $|r^{\prime(\prime)}| = v t$ (without loss of generality taking $r'>0$ and considering $r'' \in (-\infty,\infty)$) and consider the following overlapping regions: (i) outside both light cones, (ii) inside one light cone, and (iii) inside both light cones. We make use of the light-cone approximations for $w(\mathbf{x^{\prime}},t)$ from Eqn.~\ref{eq:LCapproxlarger} and Eqn.~\ref{eq:LCapproxsmallr}, but ignore the contribution of the subleading terms in the binomial expansion, since we assume these will not change the overall functional form (as we saw in the first-order calculation), and only add corrections of order $\mathcal{O}\left(e^{-4\pi (1+\Delta_\Psi)t/\beta},e^{-2\pi (2+\Delta_\Phi)t/\beta}\right)$.
Furthermore, it is useful to consider the approximations for $w(\mathbf{x'},\mathbf{x''})$ when the \emph{quench fields} are spatially ``close" or ``far" apart, meaning:
\begin{align}
    w(\mathbf{x'},\mathbf{x''})=\cosh \left[\frac{2\pi}{v\beta}(r'-r'')\right] -\cos\left[\frac{2\pi}{\beta}(\tau^{\prime}-\tau^{\prime\prime})\right]\to\begin{cases}
        \frac{1}{2}e^{\frac{2\pi}{v\beta}|r'-r''|} \quad &\text{ for }\quad |r'-r''|\gg \frac{\beta}{2\pi}\\
        1 -\cos\left[\frac{2\pi}{\beta}(\tau^{\prime}-\tau^{\prime\prime})\right]\quad &\text{ for }\quad |r'-r''|\ll \frac{\beta}{2\pi}
    \end{cases}    
\end{align}
Outside both or one of the light cones, the contributions are from the following spatial integrals:
\begin{align}
    I_{\text{0LC}}&\sim \int_{vt}^{\infty} dr' \left(\int_{-\infty}^{-vt} dr''  +\int_{vt}^{r'} dr''+ \int_{r'}^{\infty} dr''\right)\label{eq:0LC}\\
    I_{\text{1LC}}&\sim \int_{vt}^{\infty} dr' \int_{-vt}^{vt}dr''+\int_{0}^{vt} dr' \left(\int_{-\infty}^{-vt} dr''  +\int_{vt}^{\infty} dr'' \right)
\end{align}
For instance, the explicit calculation of the quadruple integral in Eqn.~\ref{eq:secondordersetup}, for the first region in Eqn.~\ref{eq:0LC}, is:
\begin{equation}
\begin{aligned}
    &\tilde{g}^2 \frac{C_{\Psi\Psi\Phi}}{\beta^{\Delta_\Phi+2\Delta_\Psi}}2^{\frac{1}{2}\Delta_{\Phi}+\Delta_{\Psi}-1} \int  d\tau' d\tau'' \int_{vt}^{\infty} dr' \int_{-\infty}^{-vt} dr'' \frac{\left(e^{\frac{2\pi}{v\beta}|r'-r''|}\right)^{\frac{1}{2}\Delta_{\Phi}-\Delta_{\Psi}}}{\left(e^{\frac{2\pi}{v\beta}|r'|}\right)^{\frac{1}{2}\Delta_{\Phi}}\left(e^{\frac{2\pi}{v\beta} |r''|}\right)^{\frac{1}{2}\Delta_{\Phi}}}\\
    \simeq &\tilde{g}^2 C_{\Psi\Psi\Phi} \beta^{4- \Delta_\Phi-2\Delta_\Psi}\left(\frac{v }{2\pi\Delta_{\Psi}}\right)^2 2^{\frac{1}{2}\Delta_{\Phi}+\Delta_{\Psi}-1} e^{-4\pi \Delta_{\Psi}t/\beta }.
\end{aligned}
\end{equation}
The correlations outside both or one of the light-cones lead to a scaling form comprising \emph{two} purely exponential decays,
\begin{align}
    \left\langle \Phi \left( t\right)\right\rangle_{2_\text{\sout{2LC}}} \simeq \tilde{g}^2 C_{\Psi\Psi\Phi} \beta^{4-\Delta_\Phi-2\Delta_\Psi} \left(D_{{\Delta_\Phi,\Delta_\Psi}_\text{\sout{2LC}}}e^{-4\pi \Delta_{\Psi} t/\beta}+E_{{\Delta_\Phi,\Delta_\Psi}_\text{\sout{2LC}}}e^{-2\pi \Delta_{\Phi} t/\beta}\right),
\end{align}
where the decay exponents are that of twice the quench field, $2\Delta_\Psi$, and the observable field, $\Delta_\Phi$. 

Even though one would naively think that spatial correlations would accumulate inside even just one of the light cones, because of the more complex form of the three-point correlator --- i.e., due to the term $\cosh(2\pi/v\beta(r'-r''))$ in $w(\mathbf{x'},\mathbf{x''})$ --- the three-point spatial correlations are not constant inside only one light cone. This is because the quench fields must be (approximately) far apart for one field to be inside its light cone and another to be outside; $w(\mathbf{x'},\mathbf{x''})$ still spatially decays and the integral is not constant over either spatial coordinate. Indeed, causality does not necessitate constant correlations but rather facilitates them; a further condition is the proximity of fields. We already saw this in the first-order correction, where there were still purely exponential contributions inside the (single) light cone. 

To meet this condition for two intersecting light cones, one needs to be inside \emph{both} light cones to see (damped) operator growth (i.e. linear or quadratic ``peaked decay"):
\begin{align}
    I_{\text{2LC}}&\sim \int_{0}^{vt} dr' \left(\int_{-vt}^{r'} dr'' +\int_{r'}^{vt} dr'' \right) +\int_{0}^{vt} dr' \int_{0}^{vt} dr''\\
\left\langle \Phi \left( t\right)\right\rangle_{2_\text{2LC}} &\simeq \tilde{g}^2 C_{\Psi\Psi\Phi} \beta^{4-\Delta_\Phi-2\Delta_\Psi} \Bigg[D_{{\Delta_\Phi,\Delta_\Psi}_\text{2LC}}e^{-4\pi  \Delta_\Psi t/\beta} +\Bigg(E_{{\Delta_\Phi,\Delta_\Psi}_\text{2LC}} + F_{\Delta_\Phi,\Delta_\Psi} \frac{t}{\beta} + G_{\Delta_\Phi,\Delta_\Psi}\frac{t^2}{\beta^2}\Bigg)e^{-2\pi  \Delta_\Phi t/\beta}\Bigg].
\end{align}
In this case, there are relaxation terms associated with the observable field $\sim e^{-2\pi t \Delta_\Phi/\beta}$, for which the correlations are constant across one or both spatial integrals. This leads to exponentially damped linear and quadratic operator growth $\sim t, t^2$. The damping term associated with the quench operator $\sim e^{-4\pi t \Delta_\Psi/\beta}$ only arises when the integrand is some exponential function of \emph{both} space directions, and so is purely exponential in time. 
Generically, the scaling form at second order in $\tilde{g}$ is
\begin{equation}
    \langle\Phi (t)\rangle_2 \simeq \tilde{g}^2 C_{\Psi\Psi\Phi} \beta^{4-\Delta_\Phi-2\Delta_\Psi} \Bigg[D_{\Delta_\Phi,\Delta_\Psi} e^{-4\pi  \Delta_\Psi t/\beta}+ \Bigg(E_{\Delta_\Phi,\Delta_\Psi} + F_{\Delta_\Phi,\Delta_\Psi} \frac{t}{\beta} + G_{\Delta_\Phi,\Delta_\Psi}\frac{t^2}{\beta^2}\Bigg)e^{-2\pi  \Delta_\Phi t/\beta}\Bigg],
    \label{eqn:scalingorderg2}
\end{equation}

with corrections of order $\mathcal{O}\left(e^{-4\pi (1+\Delta_\Psi)t/\beta},e^{-2\pi (2+\Delta_\Phi)t/\beta}\right)$. Then, so long as $\Delta_\Psi < \Delta_\Phi/2$, the second order correction is dominated by a purely exponential term set by the scaling dimension of the quench operator --- this is the case for the relaxation dynamics of $\epsilon_\text{I}$ after a finite-temperature $\sigma_\text{I} $ quench.

The picture of $n$-intersecting light cones naturally extends to arbitrary order $\mathcal{O}(\tilde{g}^n)$, with each order contributing up to $\sim \tilde{g}^n (t/\beta)^n e^{-2\pi t\Delta_\Phi/\beta}$ in the scaling function. 

\subsubsection{Ground state conformal perturbation theory and the finite temperature $2\pi t \ll \beta $ regime}
Thus far, we have only considered the finite-temperature quench perturbatively in the --- we discuss now why this is so.  If we instead consider a quench to the critical point from a ``nearby" ground state, the equivalent action is now on the infinite Euclidean plane rather than the cylinder [see Fig.~\ref{fig:geometryboth}(a)]. The relaxation dynamics of the primary field $\Phi$ that constitutes the quench perturbation are now calculated to first order from the integral of the two-point correlation function of the primary field on the plane. This takes the form:
\begin{equation}
\langle \Phi(\tau)\rangle_{\tilde{g}}  = \tilde{g} \int d^2 \mathbf{x}' \left[ \frac{1}{(\tau-\tau' )^2+ r^{\prime 2}}\right]^{ \Delta_\Phi} +\mathcal{O}(\tilde{g}^2),
\label{eq:zerotempcorrs}
\end{equation}
where imaginary time may still be regularized by a UV cutoff $\tilde{g}$: $\int d\tau \sim \int_{-\infty}^{-\Lambda} + \int_{\Lambda}^{\infty}$. Considering, for example, $\epsilon$ in the TFIM, we obtain, after rotation back to Minkowski space:
\begin{equation}
    \langle \epsilon_\text{I}(t)\rangle  \simeq \tilde{g}  \lim_{\tau'\to\infty}  \log \frac{\tau'}{ t}.
\end{equation}
We see that, while the UV divergences can be dealt with via standard renormalization methods, the lack of a natural IR cutoff $\mu $ leads to divergences at zero temperature which are of a different type: they cannot be absorbed in the redefinition of the local quantities and so give rise to non-analytic expressions in the coupling constants~\cite{S_Mussardo_2010}.

Indeed, since the vacuum state of the deformed theory is not adiabatically related to the vacuum states of the conformal theory, the ``nearby" ground state is not so nearby after all. The introduction of a natural IR regulator therefore plays a crucial role in adiabatically connecting the two theories, whether that's by introducing a boundary to the CFT to obtain a slab geometry~\cite{S_calabrese2006time, Calabrese_Cardy_2007, cardy2016furtherresults}, such that $\mu \sim \tau_0$, or by putting the CFT on a cylinder, such that $\mu \sim \beta$.

Moreover, let us consider the relaxation of $\Phi$ under its own quench at finite temperature (see Eqn.~\ref{eqn:integralsetup}), but this time in the $2\pi t \ll \beta$ regime. For $\epsilon$ in the TFIM, we obtain 
\begin{equation}
    \langle \epsilon_\text{I}(t)\rangle \simeq \tilde{g}   \log \frac{\beta}{\pi t},
\end{equation}
analogous to Eqn.~\ref{eq:zerotempcorrs}. As $\beta \to \infty$, the quench depth correspondingly must vanish, $\tilde{g} \sim g \to 0$, in order for the integral to converge; such ground state quenches cannot be treated via a conformal perturbative analysis, as the quench cannot be ``shallow" enough. 

Regardless, while the one-point functions should always be universal in the case of the boundary-CFT quench, it's only in the late time region, $t \gg \tau_0$, that the exponential decay functional form appears generically for all primary fields such that we can extract a universal ratio of scaling dimensions in the ratio of their relaxation rates. It is natural to extend this notion to the finite temperature quench: observables will not ``relax" until they ``notice" the finite cutoff $\mu$ of the geometry they live on. Even at finite $\beta$ with sufficiently shallow quenches, the $2\pi t\ll \beta$ regime does not appear to generically give an exponential or an exponential-times-polynomial decay from which we can extract a universal ratio of scaling dimensions.

\section{Lattice crossover: emergent conformal symmetry and non-linear dispersion}
When quenching to the critical point of a quantum spin chain, local observables generally exhibit scaling behavior that reflects both universal critical behavior (at least in the scaling limit) as well as non-universal lattice effects. The low-lying degrees of freedom are described by a continuum field theory with emergent conformal symmetry, and therefore have linear dispersion,
\begin{align}
    \epsilon_{k\ll 1} \sim v k \implies
    v_g\left(k\ll 1 \right) = \frac{\partial}{\partial k} \epsilon_k\bigg|_{k\ll 1} \sim v .
\end{align}
These modes travel at the same group velocity $v_g \sim v$ and give rise to ballistic, coherent transport and a sharp light-cone effect. 
This light-cone picture relies on Lorentz invariance; however, most 1+1D critical spin chains are not microscopically relativistic, but rather exhibit an emergent Lorentz symmetry only at low energies near criticality. High-energy modes in the lattice may break this emergent symmetry, having some nonlinear dispersion $\epsilon_{k\sim 1} \sim v k + \delta \epsilon(k) $.
High-energy quasiparticles still move ballistically (since the system is integrable) but with momentum-dependent velocities which decrease away from the linear regime. These modes are dispersive: the wave packet will spread out and lose shape over time due to these components traveling at different speeds. 

\begin{figure}
    \centering
    \includegraphics[width=0.4\linewidth]{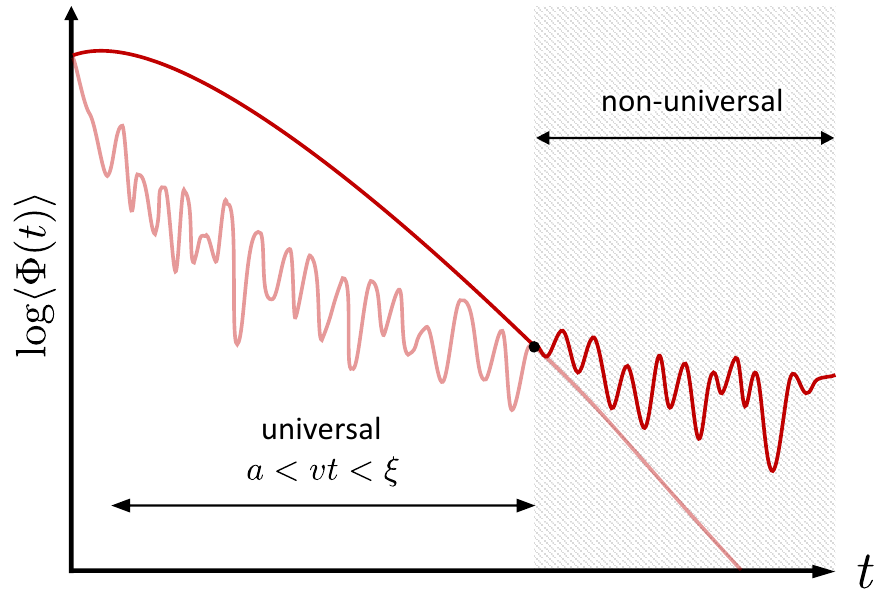}
    \caption{Schematic of the crossover time (black dot) between universal scaling behavior and dispersive lattice effects. The former is governed by the low-energy modes of the model which have emergent conformal symmetry, while the latter are determined by the full energy spectrum of the underlying lattice model, including high-energy modes, which eventually break the conformal invariance.}
    \label{fig:gsgenericcrossover}
\end{figure}

Many different modes can contribute to observable behavior, depending on the model, the nature of the initial state, and the observable itself. In particular, in the case of a global quench, the initial state is highly excited in terms of the post-quench Hamiltonian — it populates many $k$-modes, not just low-energy ones. Then the post-quench operator wavefront will be initially dominated by a coherent light-cone picture with $v_{g,\text{max}}\sim v$, while high-energy modes can lead to oscillatory, dispersive dephasing and velocity spread at long times. Consequently, there may exist a regime in which the universal critical contributions decay faster than the non-universal oscillations, leading to a crossover at a time $t^*$ beyond which the dispersive effects become dominant.
A general scaling ansatz is 
\begin{equation}
    \langle  O(t) \rangle \sim \langle  O(t) \rangle_{\text{univ}} + \langle  O(t) \rangle_{\text{latt}} 
\end{equation}
where the regime $ a/v \ll t \ll t^*$ is governed by universal scaling, and the lattice crossover time $t^*$ occurs when this universal contribution has decayed to the same magnitude as the underlying lattice effects --- see Fig.~\ref{fig:gsgenericcrossover},
\begin{equation}
    \langle  O(t^{*}) \rangle_{\text{univ}} \sim \langle  O(t^{*}) \rangle_{\text{latt}}. 
    \label{eq:crossovertime1}
\end{equation}

Determining $t^*$ requires knowledge of universal scaling as well as dispersive corrections from the underlying lattice dynamics. A quench between integrable Hamiltonians presents the challenge of translating between eigenbases of the two integrable theories, which can have complicated structures described by the Bethe Ansatz. This is a difficult undertaking analytically, with no general formalism currently known~\cite{S_Essler_Fagotti_2016}. Even in the non-interacting (TFIM) case, this proves to be an ambitious undertaking both analytically~\cite{S_Calabrese_Essler_Fagotti_2011,calabrese2012quantum, Granet_Fagotti_Essler_2020} and numerically, as shown in this work. Extensions to interacting integrable models are an area of current research. 

The extent of this work is namely to propose universal scaling ansätze for ground state and finite-temperature critical quenches in these quantum spin chains, and leave parametric determination of crossover times from universal to lattice-dominated behavior to future work. That said, since the behavior of the transverse magnetization has already been obtained~\cite{S_barouch1970statistical}, we can at least calculate these crossover times in that case.
For the ground state quench, we can equate the scaling limit in Eqn.~\ref{eq:asymptote1} to the power-law lattice tail without the oscillation in Eqn.~\ref{eq:oscillation} with $\beta \to \infty$, to obtain the crossover time $Jg t^*$ which scales polynomially in $g$,
\begin{equation}
    Jg t^* \sim -\frac{1}{4}W_{-1}\left(-\frac{g^{3/2}}{|2-g| \sqrt{2}}\right)
\sim a g^{-p},
\end{equation}
where $W_{-1}$ is the product log function, $a \approx 1.37$, and $ p \approx 0.14$. Similarly, equating the scaling limit in Eqn.~\ref{eq:asymptote2} with the power-law tail in Eqn.~\ref{eq:oscillation}, we obtain the crossover time $t^*/\beta $ for the finite-temperature critical TFI quench which scales logarithmically with $\beta J $,
\begin{equation}
    \frac{t^*}{\beta} \sim -\frac{3  }{4 \pi }W_{-1}\left\{-\frac{1}{6\beta J} \left[\frac{\pi^{2}  \tanh (|2-g| \beta J)}{|2-g|\sqrt{2}}\right]^{2/3}\right\}\sim a' \log (b'\beta J),
\end{equation}
with $a' \approx 0.26$, and $ b' \approx 9.88$. 
One may observe that operators $O$ which are more relevant than $\epsilon_\text{I}$ --- that is, more governed by low-lying modes and less sensitive to high-energy corrections --- may have a later lattice crossover,
\begin{align}
    \Delta_O < 1 \implies t^*_O > \beta \log \beta,
\end{align}
while, conversely, there may exist less relevant operators such that $t^*_{O}\leq \beta/2\pi $, for which lattice effects obfuscate the asymptotic universal relaxation regime entirely. 

\section{Integrability breaking, thermalization, and the dual quench protocol} 
To demonstrate the broader applicability of our results, we show that strict Bethe-Ansatz solubility on the lattice is not a requirement for observing the dynamical scaling forms we derive. We support this claim by considering critical quenches in the ANNNI model, which includes (weak) integrability-breaking perturbations. Crucially, although the model is not fully integrable, it remains governed by the underlying integrable Ising CFT for the regime of $\gamma$ and $t$ considered. More generally, we expect integrability-breaking terms to eventually induce a crossover to diffusive dynamics, thereby limiting the window in which universal scaling behavior appears. This ``thermalization" time $t_{\text{therm}}$ should take the form:
\begin{equation}
    t_{\text{therm}} \sim \Delta E^{-a} \gamma^{-2l},
    \label{eqn:thermtime}
\end{equation}
where $\Delta E = E - E_0(g)$ is the energy density injected into the system by the quench, and $a,l$ are positive integers --- see Ref.~\cite{S_Mori_Ikeda_Kaminishi_Ueda_2018} and references therein. Though Fermi's Golden Rule arguments would yield $l = 1$, Ref.~\cite{S_Surace_Motrunich_2023} identifies the self-dual ANNNI as a special case where $l = 2$. The universal regime on the lattice is generally a limited window either way; whether thermalization of the subsystem occurs before or after the crossover to dispersive lattice corrections merely determines which phenomenon constitutes the greatest limiting factor. Then, since $t_{\text{therm}}$ implicitly depends on the integrability-breaking strength, $\gamma$ could be chosen such that thermalization effects were not observed on the timescales we considered. Indeed, our scaling results would still hold even for models that thermalize in a more typical fashion (when $l = 1$).

\begin{figure}
    \centering
    \includegraphics[width=0.4\linewidth]{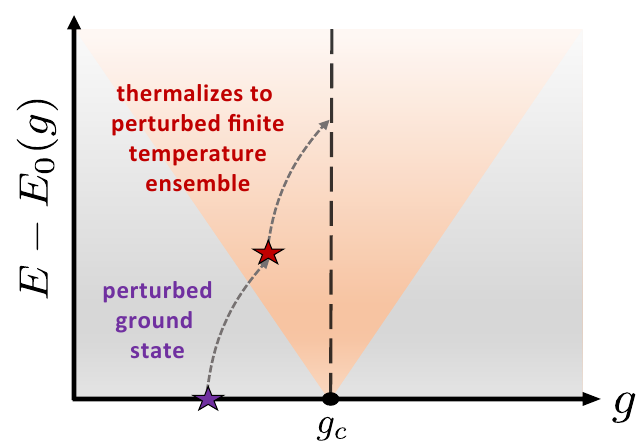}
    \caption{The proposed ``dual quench" protocol for implementing effective finite-temperature critical quenches on a quantum simulator. First, one quenches from a ground state with off-critical tuning $g_0$, which is close to the critical point, to an ensemble at $g_1$, even closer to the critical point. This injects the state with some small amount of energy density $\Delta E=E- E_0(g)$, which, when the Hamiltonian has integrability-breaking terms, should locally ``thermalize", becoming an effective finite temperature ensemble $ \beta^{-1} \sim E- E_0(g)$ for the subsystem after some thermalization time $t_{\text{therm}}$. Finally, the thermalized subsystem ensemble can be used as the initial condition for the finite-temperature quench; the decay of local observables after this second quench should then follow the scaling laws derived from the finite-temperature conformal perturbation theory as in Eqns. \ref{eqn:scalingorderg} and \ref{eqn:scalingorderg2}, for some regime of $\gamma$, 
$t$.}
    \label{fig:dualquench}
\end{figure}
On the other hand, the fact that integrability-breaking lattice terms induce thermalization actually facilitate the simulation of critical quenches from effective ``finite temperature" initial states in isolated quantum simulators. Such a ``dual quench" protocol is outlined in Fig.~\ref{fig:dualquench}.

\end{document}